\documentclass[
  paper=a4,
  BCOR=10mm,
  oneside,
  fontsize=12pt,
  fleqn,
  numbers=noendperiod,
  headings=standardclasses,
  chapterprefix=true,
  version=3.03,
  toc=bibliography,
  listof=totoc,
  parskip=half
]{scrreprt}
\usepackage[utf8]{inputenc}
\usepackage[T1]{fontenc}

\usepackage{color}
\usepackage{amsmath}
\usepackage{tabularx}
\usepackage{graphicx}
\usepackage[ngerman,english]{babel}
\usepackage{pgffor}
\usepackage{listings}
\usepackage{csquotes}
\usepackage{acro}
\DeclareAcroEnding{possessive}{'s}{'s}
\NewAcroCommand\acg{m}{\acropossessive\UseAcroTemplate{first}{#1}}
\NewAcroCommand\acsg{m}{\acropossessive\UseAcroTemplate{short}{#1}}
\NewAcroCommand\aclg{m}{\acropossessive\UseAcroTemplate{long}{#1}}
\usepackage[super]{nth}
\usepackage{scrhack}
\usepackage[toc,page]{appendix}
\usepackage[inline]{enumitem}
\usepackage{booktabs}
\usepackage{rotating}
\usepackage{subcaption}
\usepackage{listings}
\lstdefinestyle{thesis}{
  showstringspaces=false,
  commentstyle=\color{red},
  keywordstyle=\color{blue},
  breaklines=true,
  postbreak=\mbox{\textcolor{red}{$\hookrightarrow$}\space},
  numbers=left,
  stepnumber=1,
}
\usepackage[
	pdftitle={Honeypot Implementation in a Cloud Environment},
	pdfauthor={Stefan Machmeier},
	pdfsubject={Masterthesis},
	pdfcreator={pdflatex, LaTeX with KOMA-Script},
	pdfpagemode=UseOutlines,
	pdfdisplaydoctitle=true, 		
	pdflang={english}, 
]{hyperref}

\usepackage[numbers]{natbib}
\usepackage{verbatim}
\usepackage{fancyvrb}
\usepackage{fvextra}

\usepackage{epigraph} 

\DeclareAcronym{nist}{short=NIST, long=National Institute of Standards and Technology}
\DeclareAcronym{saas}{short=SaaS, long=Software-as-a-Service}
\DeclareAcronym{iaas}{short=IaaS, long=Infrastructure-as-a-Service}
\DeclareAcronym{paas}{short=PaaS, long=Platform-as-a-Service}
\DeclareAcronym{http}{short=HTTP, long=Hypertext Transfer Protocol}
\DeclareAcronym{daas}{short=DaaS, long=Data-as-a-Service}
\DeclareAcronym{haas}{short=HaaS, long=Hardware-as-a-Service}
\DeclareAcronym{dtk}{short=DTK, long=Deception Toolkit}
\DeclareAcronym{cert}{short=CERT, long=Computer Emergency Response Team}
\DeclareAcronym{os}{short=OS, long=operating system}
\DeclareAcronym{ids}{short=IDS, long=intrusion detection system}
\DeclareAcronym{ips}{short=IPD, long=intrusion prevention system}
\DeclareAcronym{nsm}{short=NSM, long=network security monitoring}
\DeclareAcronym{iocta}{short=IOCTA, long=Internet Organised Crime Threat Assessment}
\DeclareAcronym{dmz}{short=DMZ, long=demilitarized zone}
\DeclareAcronym{eu}{short=EU, long=European Union}
\DeclareAcronym{vm}{short=VM, long=virtual machine}
\DeclareAcronym{vmm}{short=VMM, long=virtual machine monitors}
\DeclareAcronym{adb}{short=ADB, long=Android Debug Bridge}
\DeclareAcronym{asa}{short=ASA, long=Adaptive Security Appliance}
\DeclareAcronym{dos}{short=DoS, long=Denial of Service}
\DeclareAcronym{adc}{short=ADC, long=Application Delivery Controller}
\DeclareAcronym{ics}{short=ICS, long=Industrial Control System}
\DeclareAcronym{scada}{short=SCADA, long=Supervisory Control and Data Acquisition}
\DeclareAcronym{ddos}{short=DDoS, long=Distributed Denial of Service}
\DeclareAcronym{ntp}{short=NTP, long=Network Time Protocol}
\DeclareAcronym{dns}{short=DNS, long=Domain Name System}
\DeclareAcronym{ssdp}{short=SSDP, long=Simple Service Discovery Protocol}
\DeclareAcronym{chargen}{short=CHARGEN, long=Character Generator Protocol}
\DeclareAcronym{udp}{short=UDP, long=User Datagram Protocol}
\DeclareAcronym{dicom}{short=DICOM, long=Digital Imaging and Communications in Medicine}
\DeclareAcronym{rdp}{short=RDP, long=Remote Desktop Protocol}
\DeclareAcronym{aws}{short=AWS, long=Amazon Web Services}
\DeclareAcronym{gcp}{short=GCP, long=Google Cloud Platform}
\DeclareAcronym{fhir}{short=FHIR, long=Fast Healthcare Interoperability Resources}
\DeclareAcronym{ipp}{short=IPP, long=Internet Printing Protocol}
\DeclareAcronym{acl}{short=ACL, long=Access Control List}
\DeclareAcronym{belwue}{short=BelWÜ, long=Baden-Württembergs extended LAN}
\DeclareAcronym{bpp}{short=BPP, long=Binary Packet Protocol}
\DeclareAcronym{mac}{short=MAC, long=Message Authentication Code}
\DeclareAcronym{soho}{short=SOHO, long=small office/home office}
\DeclareAcronym{snmp}{short=SNMP, long=Simple Network Management Protocol}
\DeclareAcronym{cve}{short=CVE, long=Common Vulnerabilities and Exposures}
\DeclareAcronym{vpn}{short=VPN, long=virtual private network}
\DeclareAcronym{ident}{short=IDENT, long=Identification Protocol}
\DeclareAcronym{smtp}{short=SMTP, long=Simple Mail Transfer Protocol}
\DeclareAcronym{ncp}{short=NVP, long=Network Control Protocol}
\DeclareAcronym{tcp}{short=TCP, long=Transmission Control Protocol}
\DeclareAcronym{dfn}{short=DFN , long=German National Research and Education Network}
\DeclareAcronym{ssh}{short=SSH, long=Secure Shell}
\DeclareAcronym{mitm}{short=MITM, long=man-in-the-middle}
\DeclareAcronym{ftp}{short=FTP, long=File Transport Protocol}
\DeclareAcronym{smb}{short=SMB, long=Server Message Block}
\DeclareAcronym{api}{short=API, long=Application Programming Interface}
\DeclareAcronym{ip}{short=IP, long=Internet Protocol}
\DeclareAcronym{gpu}{short=GPU, long=graphics processing unit}
\DeclareAcronym{voip}{short=VoIP, long=Voice over Internet Protocol}
\DeclareAcronym{gb}{short=GB, long=gigabyte}
\DeclareAcronym{vcpu}{short=vCPU, long=virtual central processing unit}
\DeclareAcronym{cpu}{short=CPU, long=central processing unit}
\DeclareAcronym{ram}{short=RAM, long=random-access memory}
\DeclareAcronym{xml}{short=XML, long=Extensible Markup Language}
\DeclareAcronym{usb}{short=USB, long=Universal Serial Bus}
\DeclareAcronym{imap}{short=IMAP, long=Internet Message Access Protocol}
\DeclareAcronym{tls}{short=TLS, long=Transport Layer Security}
\DeclareAcronym{nla}{short=NLA, long=Network Level Authentication}
\DeclareAcronym{asn}{short=ASN, long=Autonomous System Number}
\DeclareAcronym{as}{short=AS, long=Autonomous System}
\DeclareAcronym{cgi}{short=CGI, long=Common Gateway Interface}
\DeclareAcronym{vnc}{short=VNC, long=Virtual Network Computing}
\DeclareAcronym{vlan}{short=VLAN, long=Virtual Local Area Network}
\DeclareAcronym{icmp}{short=ICMP, long=Internet Control Message Protocol}
\DeclareAcronym{pop}{short=POP, long=Post Office Protocol}
\DeclareAcronym{sip}{short=SIP, long=Session Initiation Protocol}
\DeclareAcronym{nat}{short=NAT, long=network address translation}
\DeclareAcronym{rdbms}{short=RDBMS, long=relational database management system}
\DeclareAcronym{dsa}{short=DSA, long=Digital Signature Algorithm}

\newcommand{\ipAddress}[1]{%
    {\fontfamily{pcr}\selectfont%
        #1%
    }%
}

\newcommand{\Autoref}[1]{%
  \begingroup%
  \def\chapterautorefname{Chapter}%
  \def\sectionautorefname{Section}%
  \def\subsectionautorefname{Subsection}%
  \autoref{#1}%
  \endgroup%
}

\usepackage{numprint}
\npthousandsep{,}

\title{Honeypot Implementation in a Cloud Environment}
\author{Stefan Machmeier}
\date{01. February 2022}

\begin{document}

\thispagestyle{empty}
\begin{center}
  \renewcommand{\baselinestretch}{2.00}
  \Large%\sffamily
  Faculty of Mathematics and Computer Science\\
  \large Heidelberg University
  \par\vfill\normalfont
  Master thesis\\
  in Computer Science\\
  submitted by\\
  Stefan Machmeier\\
  born in Heidelberg\\
  2022
\end{center}
\newpage

\thispagestyle{empty}
\begin{center}
  \renewcommand{\baselinestretch}{2.00}
  \Large\bfseries%\sffamily
    Honeypot Implementation in a Cloud Environment
  \par
  \vfill
  \large\normalfont
  This Master thesis has been carried out by Stefan Machmeier\\
  at the\\
  Engineering Mathematics and Computing Lab\\
  under the supervision of\\
  Prof. Dr. Vincent Heuveline
  %% additionally insert second supervisor here if carrying out an
  %% external diploma thesis. Reduce vspace in L. 44 accordingly.
\end{center}\par
\vspace{5\baselineskip}

% reset baselinestretch
\renewcommand{\baselinestretch}{1.00}\normalsize

\pagenumbering{Roman}

\thispagestyle{empty}
\begin{center}
    \begin{minipage}[c][0.48\textheight][b]{0.9\textwidth}
        \small
        \begin{center}
            \textbf{Abstract}
        \end{center}\par
        \vspace{\baselineskip}
        In this age of digitalization, Internet services face more attacks than ever.
        An attacker's objective is to exploit systems and use them for malicious purposes.
        Such efforts are rising as vulnerable systems can be discovered and compromised through Internet-wide scanning.
        One known methodology besides traditional security leverages is to learn from those who attack it.
        A honeypot helps to collect information about an attacker by pretending to be a vulnerable target.
        Thus, how honeypots can contribute to a more secure infrastructure makes an interesting topic of research.
        This thesis will present a honeypot solution to investigate malicious activities in heiCLOUD and show that attacks have increased significantly.
        To detect attackers in restricted network zones at Heidelberg University, a new concept to discover leaks in the firewall will be created.
        Furthermore, to consider an attacker's point of view, a method for detecting honeypots at the transport level will be introduced.
        Lastly, a customized OpenSSH server that works as an intermediary instance will be presented to mitigate these efforts.
        \end{minipage}\par
    \vfill
\end{center}

\thispagestyle{empty}
\begin{center}
    \begin{minipage}[c][0.48\textheight][b]{0.9\textwidth}
        \small
        \begin{center}
            \textbf{Zusammenfassung}
        \end{center}\par
        \vspace{\baselineskip}
        Heutzutage sind Dienste, die über das Internet zugänglich sind, mehr Angriffen ausgesetzt als je zuvor.
        Das Ziel von Angreifern ist es, Systeme auszunutzen und sie für ihre eigenen bösartigen Zwecke zu verwenden.
        Derartige Bemühungen nehmen zu, da verwundbare Systeme durch internetweites Scannen entdeckt und kompromittiert werden können.
        Neben den traditionellen Sicherheitsmaßnahmen besteht auch die Möglichkeit, von den Angreifern zu lernen.
        Ein Honeypot hilft dabei, Informationen über Angreifer zu sammeln, indem er vorgibt, ein verwundbares Ziel zu sein.
        Daher ist es eine interessante Forschungsfrage, wie Honeypots zu einer sichereren Infrastruktur beitragen können.
        In dieser Arbeit wird eine Honeypot-Lösung zur Untersuchung von Böswillige-Aktivitäten in der heiCLOUD vorgestellt und gezeigt, dass die Angriffe aus dem Internet erheblich zugenommen haben.
        Des Weiteren wird versucht, Angreifer in eingeschränkten Netzwerkzonen der Universität Heidelberg zu entdecken.
        Die Ergebnisse zeigen, dass die Firewall Lücken aufweist und Angreifer in der Lage waren gewisse Bereiche zu scannen.
        Zusätzlich wird die Sichtweise eines Angreifers eingenommen und eine Methode zur Erkennung von Honeypots auf Transportebene vorgestellt.
        Abschließend wird ein angepasster OpenSSH-Server vorgestellt, der als Zwischeninstanz fungiert, um diese Bemühungen zu verhindern.
    \end{minipage}\par
    \vfill
\end{center}

\thispagestyle{empty}
\setlength{\parindent}{0em}

\chapter*{Erklärung}
\thispagestyle{empty}

\vspace{3\baselineskip}
Ich versichere hiermit, dass ich die vorliegende Arbeit selbständig verfasst und keine anderen als die angegebenen Hilfsmittel benutzt habe. 
Sowohl inhaltlich als auch wörtlich entnommene Inhalte wurden als solche kenntlich gemacht.

Die Arbeit ist in gleicher oder vergleichbarer Form noch bei keiner anderen Prüfungsbehörde eingereicht worden.
\vspace{5\baselineskip}

\begin{tabular}[t]{c}
    Heidelberg, den 21.01.2022
\end{tabular}
\hfill
\begin{tabular}[t]{c}
    \rule{15em}{0.4pt} \\ Stefan Machmeier
\end{tabular}

\chapter*{Acknowledgements}
\thispagestyle{empty}

The research included in this thesis could not have been performed if not for many individuals' assistance, patience, and support. 
 
First and foremost, I am deeply grateful to my supervisor, Prof. Dr. Vincent Heuveline for his valuable and constructive input.
Without his guidance and mentorship,
I would not have been able to finish this thesis.
Moreover, I grew as a researcher, and I am immensely grateful for the opportunity to continue my research as a future Ph.D. candidate under his supervision. 

I want to extend my gratitude towards Stefan Steiger and Olaf Pichler from the Computing Centre at Heidelberg University.
Thank you for offering insightful comments and brilliant suggestions when the task got challenging.
They always had time and provided me with ample support no matter what happened.

I am indebted to Joachim Peeck for generously agreeing to examine my results and providing valuable inputs.
His timely advice and scientific knowledge helped me understand essential parts of the topic assisted me to a great extent in accomplishing this task.

Lastly, I could not have completed this thesis without the support of my girlfriend, Carmen.
Thank you for being so patient in providing emotional support and stimulating discussions during my research.

\tableofcontents
\newpage

% Acronyms
\cleardoublepage
\printacronyms

% List of figures
\cleardoublepage
\listoffigures

% List of tables 
\cleardoublepage
\listoftables

% List of listings
\cleardoublepage
\lstlistoflistings

\cleardoublepage
\pagenumbering{arabic}

%% Put your contents here

  \IfFileExists{chapters/introduction}{
    \chapter{Introduction}

Recently, Europol\footnote{An agency that fights against terrorism, cybercrime, and other threats \cite{europol2021}} raised awareness of new cyber threats related to the ongoing pandemic.
As stated in their yearly \ac{iocta} report, scanning of corporate infrastructures has been skyrocketing within the last 12 months by ransomware groups, respectively increasing malware usage.
Attackers use scans to find potential vulnerabilities in remote desktop sharing software, or \acp{vpn} in order to deploy malware and blackmail companies \cite{iocta2020}.
The rapid increase dates back to the pandemic and the shift to home office, forcing companies to adapt their infrastructures quickly.
Such changes come with the downside of adding new threats to an organization.
The latest incident at the SRH University Heidelberg points out the obstacles institutions face when ransomware groups have access and exploit various parts of the infrastructure with malware.
An unknown group infected systems with malware and distributed internal data in the darknet.
Such incidents emphasize the rise of malicious activities.

Especially in cloud computing, controlling access to services is becoming a stricter challenge due to access to large data sets and computing resources.
Besides traditional security measures such as firewalls or intrusion detection systems, one known methodology to strengthen infrastructures is learning from those who attack them.
Honeypots are a proper instrument to gather information about attackers.
It is \enquote{a security resource whose value lies in being probed, attacked, or compromised} \cite{Spitzner2003}.
Collecting attacks can reveal shell-code exploitation or bot activity.
In retrospect, this would help to harden infrastructures before proper damage occurs.
For a cloud provider, it is crucial to know whether and how attacks on its service can be prevented.
Considering the Global Security Report by Trustwave, the number of attacks doubled in 2019 and increased by $20\%$ in 2020 \cite{fahim2020}, respectively putting cloud providers to the third most targeted environments for cyberattacks, behind corporate and internal networks.

The Heidelberg University offers its own cloud service, called heiCLOUD.
It enables users to maintain and control computational resources easily. 
Thus, it is interesting to elaborate on the value of honeypots for this cloud solution.
This thesis tries to answer the general research question of whether honeypots can contribute to a more secure infrastructure in a cloud environment.
This includes deploying a honeypot solution in heiCLOUD and presenting the results.
Prior to that, an insight into a recent study investigating honeypots for the cloud providers AWS, GCP, and Microsoft Azure is given.
These findings help to validate the results in heiCLOUD.
In addition, the university network will be investigated to find potential leaks in the stateless firewall.
Therefore, a concept is created using the BSI's honeypot-like detection tool MADACT and deployed on desktop computers inside the university building.
Furthermore, to consider an attacker's point of view, this thesis introduces a recent work to detect honeypots on the transport level.
Lastly, a solution to mitigate these efforts will be presented.

This thesis includes six chapters.
After the introduction, \autoref{chap:background} outlines the background knowledge that is needed to comprehend the upcoming experiments.
It gives the reader a profound understanding of cloud computing, honeypots, and virtualization.
\Autoref{chap:cloud-security}, \textit{Analyze Honeypot Attacks in the Cloud}, presents the status quo of malicious activities in heiCLOUD.
In the beginning, it shows the results that \citet{Kelly2021} claim for AWS, GCP, and Microsoft Azure.
Next, it gives an insight into the T-Pot solution used to collect the data and shows the results after collecting them for three weeks.
Furthermore, \autoref{chap:concept}, \textit{Catching Attackers in Restricted Network Zones}, investigates the university network in which the new concept is deployed for three weeks.
It shows that the concept was able to adapt the firewall, thus, improving the network security at the university.
\Autoref{chap:fingerprinting}, \textit{Mitigate Fingerprint Activities of Honeypots}, presents two experiments.
First, it describes the preliminary work to detect honeypots and finishes with an experiment to prove this assumption.
Next, it drafts the counterpart of mitigating this activity, also closing up with an experiment.
Lastly, chapter 6 completes this thesis with a conclusion that summarizes the results and describes future work in this regard.

  }
  {
  }

  \IfFileExists{chapters/background}{
    \chapter{Background}
\label{chap:background}

\epigraph{A honeypot is a security resource whose value lies in being probed, attacked, or compromised.}{\textit{Lance Spitzner}}

Using honeypots in a cloud environment merges two varying principles.
This chapter introduces the fundamental knowledge needed to comprehend the upcoming experiments.
If the reader has a profound understanding of cloud computing, honeypots, and virtualization, he can skip this chapter.

% ====================================================================================================================
% Virtualization
\section{Virtualization}

Virtualization, often referred to as \acp{vm}, is defined by \citet{kreuter2004} as \enquote{an abstraction layer or environment between hardware component and the end-user}.
A \ac{vm} runs on top of the \acg{os} core components.
Through an abstraction layer, the \acl{vm} is connected with the real machine by hypervisors or \ac{vmm}.
Hypervisors can use real machine hardware components but also support \aclg{vm} operating systems and configurations.
Both are similar to emulators, which are defined by \citet{lichstei1969} as a \enquote{process whereby one computer is set up to permit the execution of programs written for another computer}.
This allows managing multiple \acp{vm} with real machine resources.
There are three different types of virtualization,
\begin{enumerate*}[label=(\roman*)]
    \item software virtual machines,
    \item hardware virtual machines, and
    \item virtual OS/containers
\end{enumerate*}.
Software virtual machines manage interactions between the host and guest operating systems \cite{daniels2009}.
Hardware virtual machines offer direct and fast access to the underlying resources \cite{daniels2009}.
It uses hypervisors, modified code, or \acp{api}.
Lastly, virtual OS/container partitions the host operating system into containers or zones \cite{daniels2009}.

% ====================================================================================================================
% Cloud Computing
\section{Cloud Computing}
\label{sec:cloud-computing}

Cloud Computing has become a buzzword these days.
It has been used by various large companies such as Google and Amazon.
However, the term \enquote{cloud computing} dates back to late 1996, when a small group of technology executives of Compaq Computer framed new business ideas around the Internet \cite{regalado2020}.
Starting from 2007, cloud computing evolved into a serious competitor and outnumbered the keywords' \enquote{virtualization}, and \enquote{grid computing} as reported by Google trends \cite{Wang2010}.
Shortly, various cloud providers become publicly available, each with its strengths and weaknesses.
For example IBM's Cloud\footnote{\url{https://www.ibm.com/cloud}}, Amazon Web Services\footnote{\url{https://aws.amazon.com/}}, and Google Cloud\footnote{\url{https://cloud.google.com/}}.
So, why are clouds so attractive in practice?

\begin{itemize}
    \item It offers major advantages in terms of cost and reliability.
    When demand is needed, consumers do not have to invest in hardware when launching new services.
    Pay-as-you-go allows flexibility.
    \item Consumers can easily scale with demand.
    When more computational resources are required due to more requests, scaling up instances in conjunction with a suited price model is straightforward.
    \item Geographically distributed capabilities supply the need for worldwide scattered services.
\end{itemize}

% ====================================================================================================================
% Definition of cloud computing
\subsection{Definition of Cloud Computing}

According to the definition by Brian Hayes, cloud computing is \enquote{a shift in the geography of computation} \cite{hayes2008}.
Thus, the computational workload is moved away from local instances towards services and data centers that provide the user's needs \cite{Armbrust2010}.

The \ac{nist} defines cloud computing as \enquote{a model for enabling ubiquitous, convenient, on-demand network access to a shared pool of configurable computing resources (e.g., networks, servers, storage, applications, and services) that can be rapidly provisioned and released with minimal management effort or service provider interaction} \cite{Mell2011}.
\ac{nist} not only reflects the geographical shift of resources such as data centers but also mentions on-demand usage that contributes to flexible resource management.
Moreover, \ac{nist} composes the term into five essential characteristics, three service models (see \autoref{subsec:cloud-service}), and four deployment models (see \autoref{subsec:cloud-deployment}) \cite{Mell2011}.

\textit{On-demand-self-service} refers to the unilateral provision computing capabilities.
Consumers can acquire server time and network storage on demand without human interaction.

\textit{Broad network access} characterizes the access of capabilities of the network through standard protocols such as \ac{http}.
Heterogeneous thin and thick client platforms should be supported.

\textit{Resource pooling} allows the provider's computing resources to be pooled across several consumers.
Different physical and virtual resources are assigned on-demand with a multi-tenant model.
Other aspects such as location are independent and cannot be controlled on a low-level by consumers.
Moreover, high-level access to specify continent, state, or datacenter can be available.

\textit{Rapid elasticity} offers consumers to extend and release capabilities quickly.
Further automation to quickly increase resources when demand surges can be supported at any time, regardless of limit or quantity.

\textit{Measured service} handles resources in an automated and optimized manner.
It uses additional metering capabilities to trace storage, processing, bandwidth, and active user accounts.
This helps to monitor and control resource usage.
Thus, contributing to transparency between provider and consumer.

% ====================================================================================================================
% Service models
\subsection{Service models}
\label{subsec:cloud-service}

Service models are categorized by \ac{nist} into three basic models based on usage and abstraction level.
\autoref{fig:cloud-functionalities} shows the connection between each model whereas cloud resource are defined in \autoref{subsec:cloud-deployment}.
\ac*{iaas} builds with a vast range of functionalities the foundation of service models.
Each model on top represents a user-friendly abstraction with derated capabilities.

\begin{figure}
    \centering
    \includegraphics[width=0.4\textwidth]{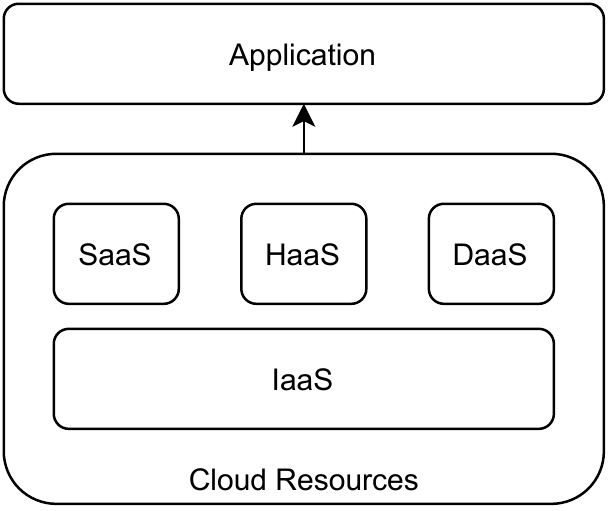}
    \caption[Abstract visualization of service models]{
        Abstract visualization of service models.
        The container \enquote{cloud resources} represents the depth of functionalities.
        Therefore, \ac{iaas} offers the most functionalities, whereas the others have a user-friendly abstraction.
    }
    \label{fig:cloud-functionalities}
\end{figure}

\ac{saas} is a high-level abstraction to consumers.
Controlling the underlying infrastructure is not supported.
Providers often use a multi-tenancy system architecture to organize each consumer's application in a different environment.
It helps to employ scaling with respect to speed, security availability, disaster recovery, and maintenance \cite{Mell2011}.
The main objective of \ac{saas} is to host a consumer's software or application that can be accessed over the Internet using either a thin or rich client \cite{Dillon2010}.
Users can apply custom configuration settings \cite{Mell2011}.

\ac{paas} pivots on the full \enquote{Software Lifecycle} of an application whereas \ac{saas} distinct on hosting complete applications.
\ac{paas} offers ongoing development and includes programming environment, tools, configuration management, and other services.
In addition, the underlying infrastructure is not managed by the consumer \cite{Mell2011}.

\ac{iaas} offers a low-level abstraction to consumers with the ability to run arbitrary software regardless of the operating system or application.
In contrast to \ac{saas}, IT infrastructure capabilities (such as storage and networks) can be used.
It strongly depends on virtualization due to the integration or decomposition of physical resources \cite{Mell2011}.

\ac{daas} serves as a virtualized data storage service on demand.
Motivations behind such services could be upfront costs of on-premise enterprise database systems \cite{Dillon2010}.
Mostly they require \enquote{dedicated server, software license, post-delivery services, and in-house IT maintenance} \cite{Dillon2010} whereas \ac{daas} costs solely what consumers need.
When dealing with a tremendous amount of data, file systems and \acp{rdbms} often lack performance.
\ac{daas} outruns such weak links by employing a table-style abstraction that can be scaled \cite{Dillon2010}.

\ac{haas} offers IT hardware or datacenters to buy as a pay-as-you-go subscription service.
The term dates back to 2006 when hardware virtualization became more powerful.
It is flexible, scalable, and manageable \cite{Wang2010}.

% ====================================================================================================================
% Deployment models
\subsection{Deployment models}
\label{subsec:cloud-deployment}

Deployment models are categorized by \ac{nist} into four basic models.
Each differs in data privacy, location, and manageability \cite{Mell2011}.

With private clouds, users have the highest control regarding data privacy and utilization.
Such clouds are mostly deployed within a single organization, managed by in-house teams or third-party suppliers.
In addition, it can be on- or off-premise.
Within private clouds, consumers have full control of their data.
Especially for European data privacy laws, it is not negligible when data is stored abroad, and thus, under the law of foreign countries.
However, its popularity has not been diminished due to the immense cost of switching to public clouds \cite{Dillon2010, Mell2011}.

Community clouds can be seen as a conglomerate of multiple organizations that merge their infrastructure with respect to a commonly defined policy, terms, and conditions beforehand \cite{Mell2011}.

Public clouds represent the most used deployment models.
Contrary to private ones, public clouds are fully owned by service providers such as businesses, academics, or government organizations.
Consumers do not know where their data is distributed.
In addition, contracts underlie custom policies \cite{Mell2011}.

A hybrid cloud mixes two or more cloud infrastructures, such as private and public clouds.
However, each entity keeps its core element.
Hybrid clouds define \enquote{standardized or proprietary technology to enable data and application portability}\cite{Mell2011}.

% ====================================================================================================================
% Honeypots
\section{Honeypots}

The term \enquote{honeypot} has been established for more than two decades.
1997 was the first time that a free honeypot solution became public.
\ac{dtk}, developed by Fred Cohen, released the first honeypot solution.
However, the earliest drafts of honeypots are from 1990/91 and built the foundation for Fred Cohen's \ac*{dtk}.
Clifford Stoll's book \enquote{The Cuckoo's Egg}\cite{stroll2000}, and Bill Cheswick's whitepaper \enquote{An Evening With Berferd}\cite{Cheswick92} describe concepts that are considered nowadays as honeypots \cite{Spitzner2003}.
A honeypot itself is a security instrument that collects information on buzzing attacks.
It disguises itself as a system or application with weak links, so it gets exploited and gathers knowledge about the adversary.
In 2002, a Solaris honeypot helped to detect an unknown \verb|dtspcd| exploit.
Interestingly, a year before in 2001, the Coordination Center of \acs{cert}\footnote{\acl{cert} is an expert group that handles computer security incidents\cite{cert2021}} shared their concerns regarding the dtspcd.
Communities were aware that the service could be exploited to get access and remotely compromise any Unix system.
However, such an exploit was not known during this time, and experts did not expect any in the near future.
Luckily, early instances based on honeypot technologies could detect new exploits and avoid further incidents.
Such events emphasize the importance of honeypots.

% ====================================================================================================================
% Definition of a honeypot
\subsection{Definition of Honeypots}

Many definitions for honeypots circulate through the web that causing confusion and misunderstandings.
In general, the objective of a honeypot is to gather information about attacks or attack patterns \cite{NawrockiWSKS2016}.
Thus, contributing as an additional source of security measure.
See \autoref{subsec:honeypot-security-concept} for a detailed view regarding honeypots in the security concept.
As \citet{Spitzner2003} has listed, the most misleading definitions: a honeypot is a tool for deception, it is a weapon to lure adversaries or a part of an intrusion detection system.
In order to get a basic understanding, this section wants to exhibit some key definitions.
\citet{Spitzner2003} defines honeypots as a \enquote{security resource whose value lies in being probed, attacked, or compromised}.
Independent of its source (e.g., server, application, or router), he expects the instance to be probed, attacked, and eventually exploited.
If a honeypot does not match this behavior, it will not provide any value.
It is essential to mention that honeypots do not have any production value; thus, any communication that is acquired is suspicious by nature \cite{Spitzner2003}.
In addition, \citet{Spitzner2003} points out that honeypots are not bound to solve a single problem; hence, they function as a generic perimeter and fit into different situations.
Such functions are attack detection, capturing automated attacks, or alert/warning generators.
\autoref{fig:honeypot-example} shows an example of how honeypots could be used in an IT infrastructure.

\begin{figure}
    \centering
    \includegraphics{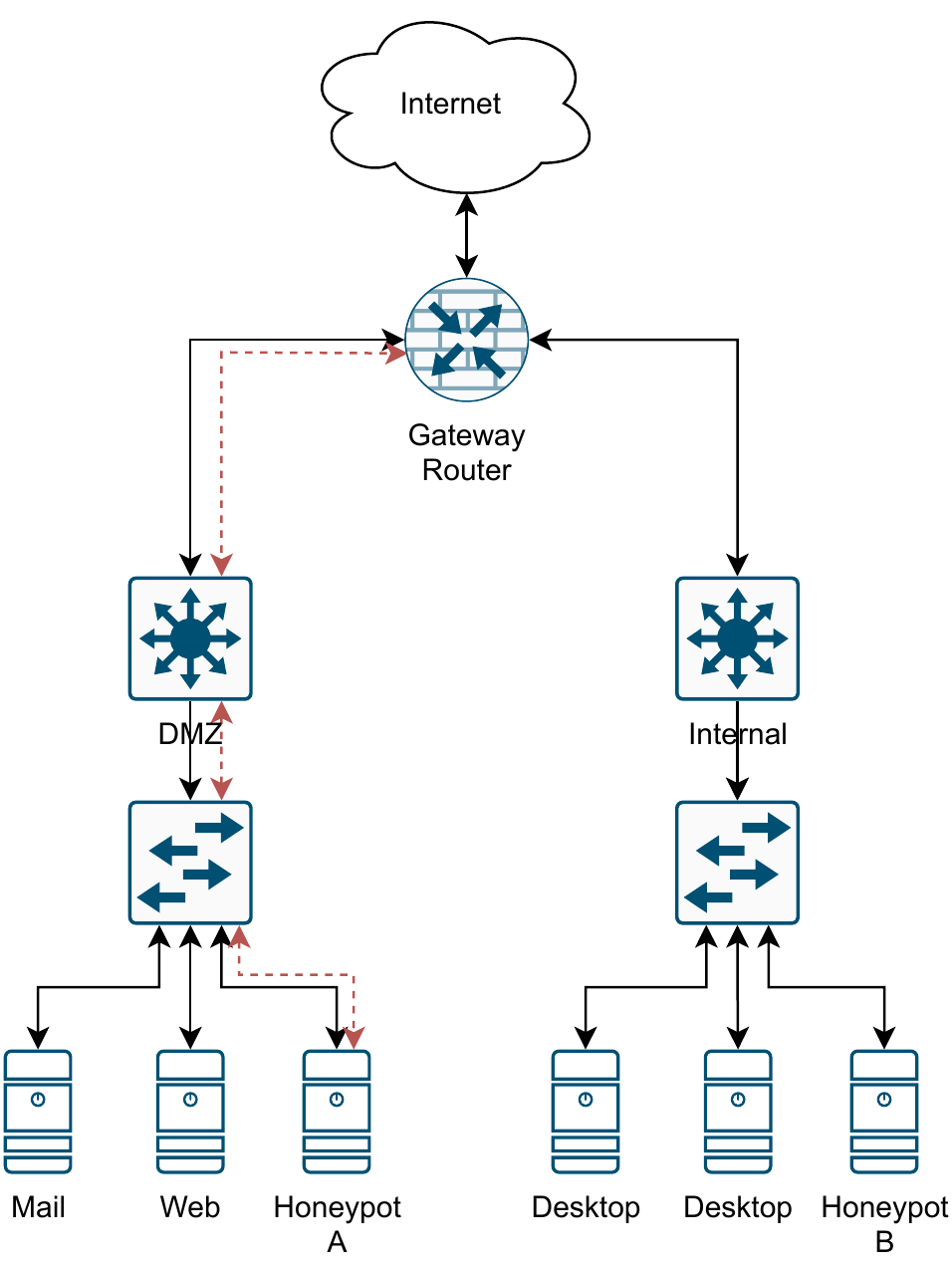}
    \caption[Example of honeypots in a simplified network]{
        Example of honeypots in a simplified network (derived from \cite{Spitzner2003}).
        Each of the \acp{dmz} and internal networks are separated by a router and a Layer-3 switch.
        In each network a honeypot is available (honeypot A, B).
        The red path symbolizes the path of an attacker coming from the gateway router.
    }
    \label{fig:honeypot-example}
\end{figure}

In general, he differentiates two types of honeypots
\begin{enumerate*}[label=(\roman*)]
    \item production honeypots
    \item research honeypots
\end{enumerate*}.
This categorization has its origin from Mark Rosch, a developer of Snort, during his work at GTE Internetworking \cite{vetterl2020}.

Production honeypots are the most common type of honeypots that people would think of.
The objective is to protect production environments and mitigate the risk of attacks.
Usually, production honeypots are easy to deploy within an organization.
Mostly, low-interaction honeypots are chosen due to a significant risk reduction, so adversaries cannot exploit honeypots to attack other systems \cite{Spitzner2003}.
The downside of a low-interaction honeypot is a lack of information, which means only standard information like the origin of attacks or what exploits have been used can be collected \cite{Mokube2007}.
On the contrary, insides about the communication of attackers or deployment of such attacks are unlikely to obtain, whereas research honeypots fulfill this objective \cite{Spitzner2003}.

Research honeypots are used to learn more in detail about attacks.
The objective is to collect information about clandestine organizations, new tools for attacks, or the origin of attacks \cite{Spitzner2003,Mokube2007}.
Research honeypots are unlikely suitable for production environments due to a higher risk increase.
Facing an increase in deployment complexity and maintenance does not attract production usage either \cite{Spitzner2003}.

It is worth mentioning that there is no exact line between research or production honeypots.
A possible use case is a honeypot that functions as a production or a research honeypot.
Due to the dynamic range in which they are applicable, it is difficult to distinguish them.

In addition, \citet{Provos2003} adds a differentiation for the virtual honeypot framework and splits it into the following types:

\begin{itemize}
    \item Physical honeypots are \enquote{real machines on the network with its own \ac{ip} address} \cite{Provos2003}
    \item Virtual honeypots are \enquote{simulated by another machine that responds to network traffic sent to the virtual honeypot} \cite{Provos2003}
\end{itemize}

% ====================================================================================================================
% Level of interaction
\subsection{Level of Interaction}
\label{subsec:interaction-honeypots}

When building and deploying a honeypot, the depth of information has to be defined beforehand.
Should it gather unauthorized activities, such as an \textsc{nmap} scan?
Do you want to learn about buzzing tools and tactics?
Each depth brings a different level of interaction because some information depends on more actions of adversaries.
Therefore, honeypots differ in their level of interaction.

Low-interaction honeypots provide the lowest level of interaction between an attacker and a system.
Only a small set of services like \ac{ssh}, Telnet, or \ac{ftp} are supported, contributing to the deployment time.
In terms of risk, a low-interaction honeypot does not give access to the underlying \ac{os} which makes it safe to use in a production environment \cite{Spitzner2003}.
For example, using an \ac{ssh} honeypot with emulated services allows attackers to log in and execute commands by brute force or guesswork.
The adversary will never gain more access because it is not a real \ac{os}.
However, safety comes with the downside of less information.
The collection is limited for the statistical purpose such as
\begin{enumerate*}[label=(\roman*)]
    \item time and data of attack
    \item source \ac{ip} address and source port of the attack
    \item destination \ac{ip} address and destination port of the attack
\end{enumerate*} \cite{Spitzner2003,Mokube2007}.
The transactional information can not be collected \cite{Spitzner2003}.

A medium-interaction honeypot offers more sophisticated services with a higher level of interaction.
It is capable of responding to specific activities.
For example, a Microsoft IIS Web server honeypot could respond in a way that a worm is expecting.
The worm would get emulated answers and could be able to interact with it in more detail.
In this way, more severe information about the attack can be gathered, including privilege assessment, toolkit capture, and command execution \citet{Spitzner2003}.
In comparison, medium-interaction honeypots allocate more time to install and configure \cite{Spitzner2003,Mokube2007}.
Also, more security checks have to be performed due to a higher interaction level than low-interaction honeypots \cite{Spitzner2003}.

High-interaction honeypots represent a real \ac{os} to provide a full set of interactions to attackers \cite{Spitzner2003}.
They are so powerful because other production servers do not differ much from high-interaction honeypots.
They represent real systems in a controlled environment \cite{Spitzner2003,Mokube2007}.
The amount of information is tremendous.
It helps to learn about
\begin{enumerate*}[label=(\roman*)]
    \item new tools
    \item finding new bugs in the \ac{os}
    \item the black hat community
\end{enumerate*} \cite{Spitzner2003}.
However, the risk of such a honeypot is extremely high.
It needs severe deployment and maintenance processes; thus, it is time-consuming.

% ====================================================================================================================
% security concepts
\subsection{Security concepts}
\label{subsec:honeypot-security-concept}

Security concepts are classified by \citet{Schneier2004} in prevention, detection, and reaction.
Prevention includes any process that
\begin{enumerate*}[label=(\roman*)]
    \item discourages intruders and
    \item hardens systems to avoid any breaches
\end{enumerate*}.
Detection scrutinizes the identification of attacks that threatens the systems'
\begin{enumerate*}[label=(\roman*)]
    \item confidentiality
    \item integrity and
    \item availability
\end{enumerate*}.
Reaction treats the active part of the security concept.
When attacks are detected, it conducts reactive measures to remove the threat.
Each part is designed to be sophisticated so that all of them contribute to a secure environment \cite{NawrockiWSKS2016}.

Honeypots contribute to the security concept like firewalls, or \acp{ids}.
However, honeypots add only a small value towards prevention because security breaches cannot be identified.
Moreover, attackers would avoid wasting time on honeypots and go straight for production systems instead.

Detection is one of the strengths of honeypots.
Attacks often vanish in the sheer quantity of production activities.
If any connection is established to a honeypot, it is suspicious by nature.
In conjunction with an alerting tool, attacks can be detected.

Honeypots strongly supply reaction tools due to their clear data.
It is difficult to find attacks for further data analysis in production environments.
Often data submerge with other activities, which complicates the process of reaction \cite{NawrockiWSKS2016}.
\citet{NawrockiWSKS2016} distinguish honeypots from other objectives such as firewall or log-monitoring.

\begin{table}
    \centering
    \caption[Distinction between security concepts]{
        Distinction between security concepts based on areas of operations (derived from \cite{NawrockiWSKS2016}).
    }
    \begin{tabular}{l|lll}
        \toprule
        \textbf{Objective}        & \textbf{Prevention} & \textbf{Detection} & \textbf{Reaction} \\ \hline
        Honeypot                  & +                   & ++                 & +++               \\
        Firewall                  & +++                 & ++                 & +                 \\
        Intrusion Detection Sys.  & +                   & +++                & +                 \\
        Intrusion Prevention Sys. & ++                  & +++                & ++                \\
        Anti-Virus                & ++                  & ++                 & ++                \\
        Log-Monitoring            & +                   & ++                 & +                 \\
        Cybersecurity Standard    & +++                 & +                  & +                 \\
        \bottomrule
    \end{tabular}
    \label{tab:honeypots-security-concepts}
\end{table}

% ====================================================================================================================
% Value of honeypots
\subsection{Value of Honeypots}

To assess the value of honeypots, this section looks at their advantages and disadvantages \cite{Mokube2007,Kaur2014,Spitzner2003}.

\subsubsection{Advantages}

\begin{itemize}
    \item \textit{Data Value}: Collected data is often immaculate and does not contain noise from other activities.
          Thus, reducing the total data size and speeding up the analysis.
    \item \textit{Resources}: Firewalls and \ac{ids} are often overwhelmed by the gigabits of traffic, thus, dropping network packets for analysis.
          This results in far less effective detection of malicious network activities.
          However, honeypots are independent of resources because they only capture their activities.
          Due to resource limitations, expensive hardware is not needed.
    \item \textit{Simplicity}: A honeypot does not require complex algorithms or databases.
          If a honeypot is too complex, it will lead to misconfigurations, breakdowns, and failures.
          The challenging research honeypots might come with an inevitable increase in complexity in maintenance.
    \item \textit{Return on Investment}: Capturing attacks immediately informs users that suspicious activities occur on the infrastructure.
          This helps to demonstrate their value and contributes to new investments in other security measurements.
\end{itemize}

In addition, \citet{NawrockiWSKS2016} listed four more advantages of honeypots:

\begin{itemize}
    \item \textit{Independent of Workload}: Honeypots only process traffic directed to them.
    \item \textit{Zero-Day-Exploit Detection}: It helps to detect unknown strategies and zero-day-exploits.
    \item \textit{Flexibility}: Well-adjusted honeypots for various specific tasks are available.
    \item \textit{Reduced False Positives and Negatives}: Any traffic or connection to a honeypot is suspicious.
          Client-honeypots verify such attacks based on system state changes.
          This results in either false positive or false negatives.
\end{itemize}

\subsubsection{Disadvantages}

\begin{itemize}
    \item \textit{Narrow Field of View}: Only direct attacks on honeypots can be investigated, whereas attacks on the production system are not detected.
    \item \textit{Fingerprinting}: A honeypot often has a certain fingerprint that attackers can identify.
          Especially commercial ones can be detected by their responses or behaviors.
    \item \textit{Risk to the Environment}: Using honeypots in an environment always increases risk.
          However, it depends on the level of interaction.
\end{itemize}

\subsection{Honeynets}

Instead of having single honeypots that can be attacked, a honeynet offers a complete network of standard production systems such as you would find in an organization \cite{sokol2017}.
Those systems are high-interaction honeypots, thus, allowing them to fully interact with the \ac{os} and applications.
The key idea is that an adversary can probe, attack, and exploit these systems so that the maintainer can derive interaction within this network \cite{Spitzner2003,sokol2017}.
It should be mentioned that a honeynet has to be protected by firewalls.
For example, \autoref{fig:honeynet-example} represents such a honeynet within an organization.

\begin{figure}
    \centering
    \includegraphics{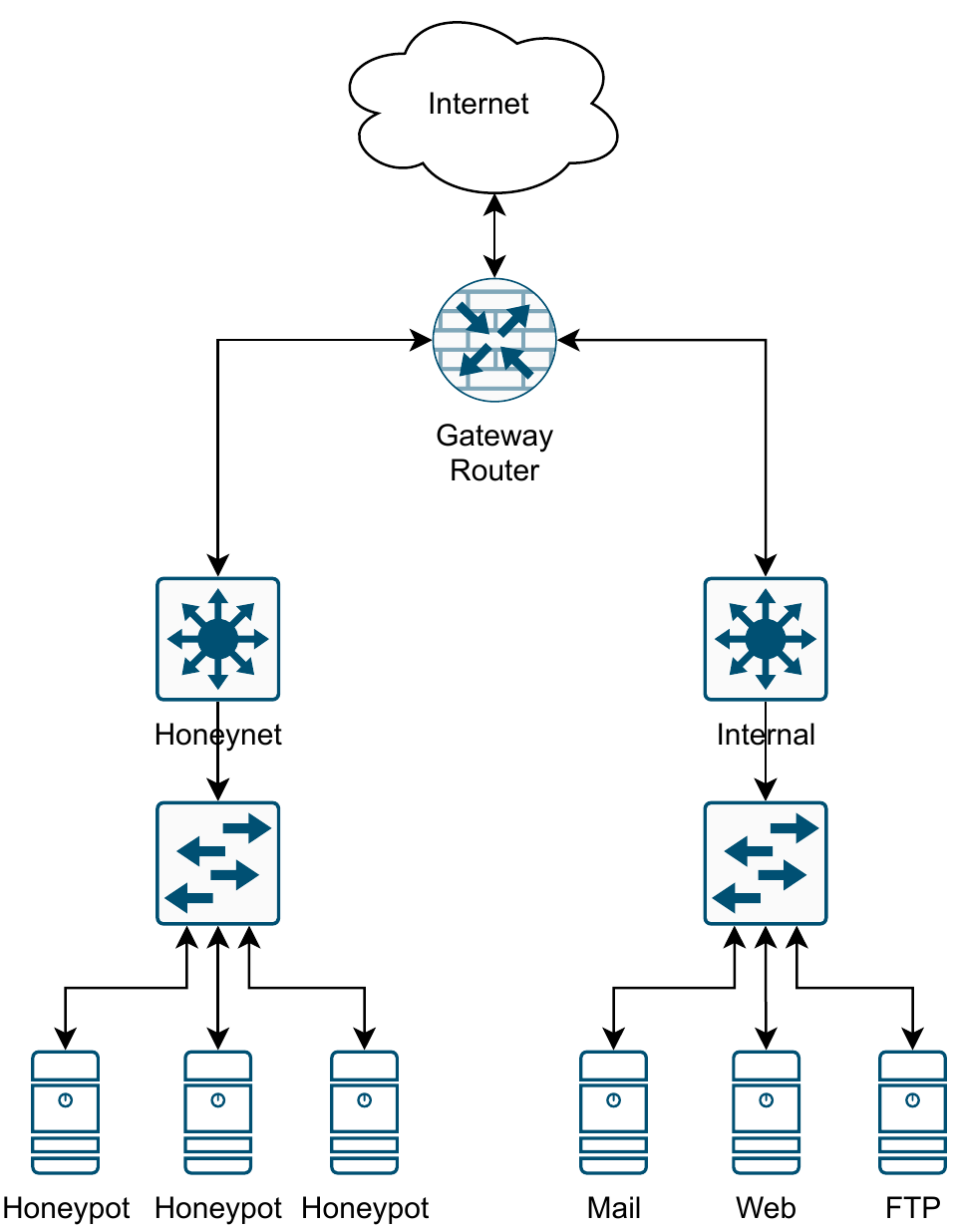}
    \caption[Example of honeynets in a simplified network]{
        Example of honeynets in a simplified network (derived from \cite{Spitzner2003}).
        This network presents the honeynet consisting of several other honeypots on the left.
        On the right, the network presents a common subnet consisting of mail, web, and FTP server.
    }
    \label{fig:honeynet-example}
\end{figure}

Compared to a traditional honeypot, the most significant value of honeynets is the usage of proper production systems.
Black hats often do not know that they attack a honeynet, thus, adding value to prevention.
However, the downsides are the high complexity and maintenance needed to keep a honeynet running \cite{Spitzner2003}.

% ====================================================================================================================
% Legal issues
\subsection{Legal Issues}

Considering questions related to legal issues of honeypots can easily exceed this thesis.
In this regard, this section restricts the study to the country the author resides in.
Thus, only the \ac{eu} regulations, EU directives, and international agreements are considered.
Honeypots collect
\begin{enumerate*}[label=(\roman*)]
    \item content data that is used for communication, and
    \item transactional data that is used to establish the connection
\end{enumerate*}.
\citet{sokol2017} studied the legal conditions for data collection and data retention.
They have concluded that administrators of honeypots have a legal ground of legitimate interest to store and process personal data, such as IP addresses.
Moreover, for production honeypots, the legitimate interest is to secure services.
Regarding the length of data retention, the principle of data minimization has to be considered, which means there is no clear answer.
Any published data of research honeypots needs to be anonymized.

  }
  {
  }

  \IfFileExists{chapters/cloud_security}{
    \chapter{Analyze Honeypot Attacks in the Cloud}
\label{chap:cloud-security}

Attacks from the Internet often originate from bots.
A bot, short for \enquote{robot}, is an automated process that interacts with different network services.
Despite good intentions, bots can be used for malicious purposes.
Mostly, bots try to self-propagate malware across the Internet and try to capture hosts that merge into a botnet \cite{Feily2009}.
Recently, Universities in Germany received more cyberattacks than ever, respectively increasing their costs for damage repairs.
Honeypots are a good solution to catch attackers and learn from their exploits.
However, it is not clear whether honeypots are an appropriate countermeasure to prevent such damage in the age of bots.
Following the rise of cyberattacks, this chapter introduces a method to collect and analyze cyberattacks in a cloud environment.
It further proposes an answer if honeypots are helpful to detect bot activities.
All IP addresses have been redacted.

\section{Introduction}

As previously mentioned in \autoref{sec:cloud-computing}, using cloud resources is becoming the go-to option for new services and applications.
\citet{Kelly2021} thoroughly investigated honeypots on Azure, \ac{aws}, and \ac{gcp}.
Consequently, this chapter presents their results briefly to compare them with the ones heiCLOUD achieves.
The results are collected by T-Pot version $20.06.0$ for three weeks.
In addition, \citet{Kelly2021} considered different server geographical locations.
They have collected data from East US, West Europe, and Southeast Asia.
\autoref{tab:overview-cloud-security} shows the results presented by \citet{Kelly2021}.
Dionaea (a honeypot to capture malicious payload), Cowrie (\ac{ssh} and Telnet honeypot), and Conpot (industrial honeypot for \acs{ics} and \acs{scada}) are the most attacked honeypots in comparison to the others.
Regarding \ac{aws}, Dionaea accounts for $91\%$ of the total attacks, Glutton and Cowrie are minor with $5\%$, and $2\%$.
Interestingly, Cowrie reported several attacks related to the COVID-19 pandemic to enable social engineering methods.
In contrast to \ac{aws}, Cowrie logged the majority of attacks with $51\%$ on \ac{gcp}.
Besides several automated attacks trying to log in with default credentials, adversaries tried to gather information about the \acs{gpu} architecture, scheduled tasks, and privilege escalation.
Microsoft Azure reflects nearly the same results as the other two cloud providers beforehand.

\begin{table}[htbp]
    \centering
    \caption[Overview of attacks on cloud providers]{
        Overview of attacks on cloud providers.
        For a better overview, only the three most attacked honeypots are listed.
        The remaining honeypots are listed in the column named "others".
    }
    \begin{tabularx}{\linewidth}{l|XXXX|l}
        \toprule
        \textsc{Provider} & \multicolumn{4}{c|}{\textsc{Honeypot}} & \textsc{In Total}                                                           \\
                          & \textbf{Dionaea}                       & \textbf{Cowrie}   & \textbf{Glutton} & \textbf{others}  &                   \\
        \hline
        \acl{aws}         & \numprint{228075}                      & \numprint{4503}   & \numprint{11878} & \numprint{3688}  & \numprint{248144} \\
        \acl{gcp}         & \numprint{162570}                      & \numprint{297818} & \numprint{84375} & \numprint{36403} & \numprint{581116} \\
        Microsoft Azure   & \numprint{308102}                      & \numprint{9012}   & \numprint{17256} & \numprint{6365}  & \numprint{340735} \\
        \bottomrule
    \end{tabularx}
    \label{tab:overview-cloud-security}
\end{table}

The overall results show an average ratio of \numprint{55000} attacks per day, summing up to roughly $1.17$ million in total.
Similar results for different regions could have been reproduced.
Their results clearly show the Europe, US, and Asia disparity.
An important question that \citet{Kelly2021} answered is if attackers target services on cloud providers based on the cloud providers' market share.
The study could not confirm this assumption because Google Cloud received most of the attacks with the smallest market share.
In total, most of the attacks are originated from Vietnam, Russia, the United States, and China.
Due to technologies such as \ac{vpn} or Tor, the geolocation only indicates the last node so that location data might be distorted.
Across all providers, roughly $80\%$ of the source \ac{ip} addresses had a bad reputation (identified by Suricata) and could have been filtered by the organization.
The operating devices used for attacking the services are mostly Windows 7 or 8 and different Linux kernels and distributions.
Windows devices target vulnerabilities in remote desktop sharing software.
Such vulnerabilities are
\begin{enumerate*}[label=(\roman*)]
    \item CVE-2006-2369\cite{CVE-2006-2369} (RealVNC) in the US region,
    \item CVE-2001-0540\cite{CVE-2001-0540} (\ac{rdp}) in EU and Asia regions,
    \item CVE-2012-0152\cite{CVE-2012-0152} (\ac{rdp}) in the Asia region, and
    \item CVE-2005-4050\cite{CVE-2005-4050} (\ac{voip}) in EU region
\end{enumerate*}.
In addition, attackers were also capable of disguising any fingerprinting activity of P0f.

This chapter compares the findings \citet{Kelly2021} claimed in the paper \enquote{A Comparative Analysis of Honeypots on Different Cloud Platforms} with ours using the Heidelberg University's cloud solution.
First, a short introduction of heiCLOUD is given, followed by a closer lookup of the T-Pot used to acquire data.
Lastly, it presents the results and does a thorough comparison closing up with a discussion based on a technical report of Cambridge University.

\section{Methodology}

The foremost goal is to track as many attacks as possible.
\autoref{fig:concept} sketches the concept to achieve this goal to gather various attacks from the Internet.
Honeypots should be deployed on a single instance, and their data or log files are stored in a database.
The attacks are analyzed with the help of data visualization tools.
For security reasons, honeypots should run in a virtualized environment to avoid harming the host system.
The host machine runs on a Debian distribution.
The instance runs on heiCLOUD, a cloud service provided by Heidelberg University.
It is capable of 16 \acs{gb} of \acs{ram}, 8 \acsp{vcpu}, and volatile memory of 30 \acs{gb}.
In addition, it mounts a 125 \acs{gb} permanent volume to store the data securely.
In the very early stage of this chapter, different approaches to achieve this goal have been compared.
For example, native implementation approaches, additional frameworks, and ready-to-use solutions have been evaluated.
However, the T-Pot, developed by Telekom, offers a profoundly ready-to-use solution with significant advantages.
It combines several honeypots with various analytic tools to trace the newest attacks.
Furthermore, it helps to compare the findings with the ones \citet{Kelly2021} claim.

\begin{figure}[htbp]
    \centering
    \includegraphics{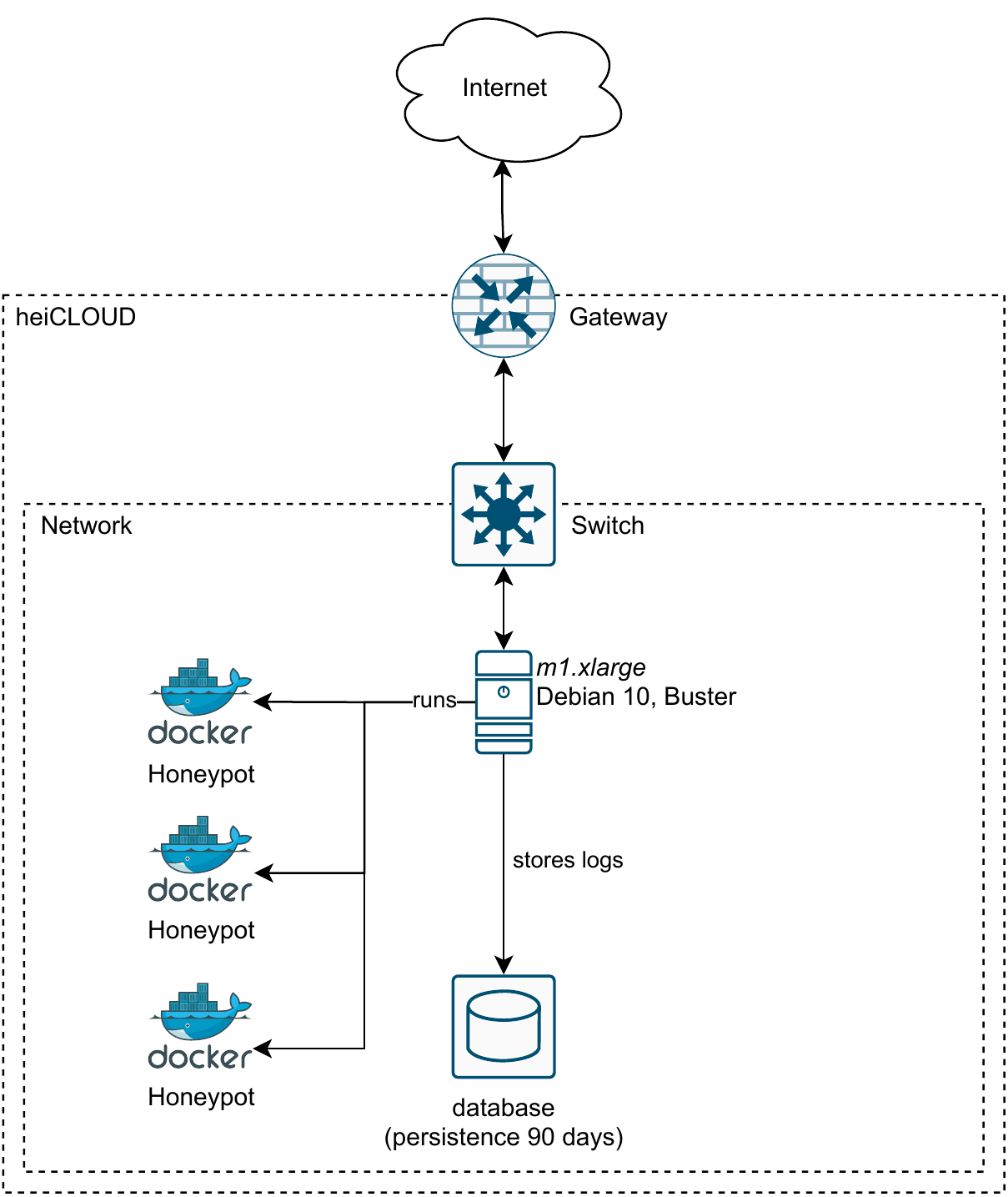}
    \caption[Concept to collect honeypot attacks]{
        Concept to collect honeypot attacks.
        The instance size is referred to the available resources of OpenStack.
        The network is an encapsulated subnet with a switch for incoming and outgoing connections.
        The database is independent of the instance and could run on a separate host.
    }
    \label{fig:concept}
\end{figure}

Running the instance and exposing it to the Internet needs some adjustments beforehand.
Therefore, a virtual network with subnet \ipAddress{192.168.145.0/24} has been created wherein the \ac{ip} address \ipAddress{192.168.145.4} is assigned to the instance.
The instance is accessible from the outside with a floating \ac{ip} address \ipAddress{10.20.30.123}.
Access rules are similar to a stateless firewall, and thus, do not block any attacks.
Ports $1-64000$ are exposed and can be attacked by anyone.
Ports higher than $64000$ are only accessible through the university network \ipAddress{10.20.30.123/16} or eduroam \ipAddress{10.20.31.123/16} and should provide a basic authentication with username and password.

\subsection{heiCLOUD}
\label{subsec:heicloud}

University Computing Center Heidelberg offers a \enquote{\ac{iaas} specially tailored for higher education and research institutions}\cite{urz2021} called heiCLOUD.
It supplies multiple departments at Heidelberg University with storage, virtual machines, or network components.
In addition, heiCLOUD is a DFN\footnote{German National Research and Education Network is the communications network for Science and research in Germany} member and offers others to use their services.
As stated on their information website\cite{heicloud2021}, it 
\begin{enumerate*}[label=(\roman*)]
    \item is capable of freely manageable IT resources,
    \item beholds a stable and fast connection,
    \item ensures high availability and scalability,
    \item has freely selectable VM operating systems, and
    \item has a transparent payment model
\end{enumerate*} \cite{heicloud2021}.
Users can easily create their network areas and manage their space individually based on the open-source application OpenStack.
Unlike well-known cloud providers, heiCLOUD servers are located within Germany, thus, abide by the European data privacy law.
HeiCLOUD has never considered implementing honeypots for additional cybersecurity measurements.

\subsection{T-Pot}
\label{subsec:tpot}

To be able to compare the results with \citet{Kelly2021}, the same approach to capture recent cyberattacks is used.
The T-Pot solution, a mixture of Telekom and Honeypot, stands out with its sheer quantity of various honeypots.
It requires at least $8$ \ac{gb} of \ac{ram} and a minimum of $128$ \ac{gb} of hard drive storage.
Based on a Debian 10 Buster distribution, it relies on Docker to run their services \cite{docker2021}.
T-Pot has to be deployed in a reachable network where intruders are expected.
Either \ac{tcp} and \ac{udp} traffic are forwarded without filtering to the network interface, or it runs behind a firewall with forwarding rules.
Specified ports for attackers are $1$-$64000$; higher ports are reserved for trusted IPs; thus, a reverse proxy asks for basic authentication.
All daemons and tools run on the same network interface, but some are encapsulated in their own Docker network.
The lightweight virtualization technology Docker uses containers to run on the host system \cite{combe2016}.
Unlike virtual machines, Docker reduces overhead with the downside of a greater attack surface.
To mitigate attacks, Docker wraps containers in an isolated environment.
This is achieved by restricting the kernel namespace and control groups (\verb|cgroups|) \cite{combe2016}.
\autoref{fig:overview-tpot} visualizes the technical concept of T-Pot.
Each service has dedicated ports or port ranges that are exposed.
Attackers can communicate either with \ac{tcp} or \ac{udp}.
All honeypots and tools create log files used to get any knowledge about attackers.
In order to view and trace current attacks, T-Pot uses the ELK stack.
ELK is the acronym of Elasticsearch, Logstash, and Kibana \cite{elastic2021}.
The search engine Elasticsearch is based on the Lucene library.
It is multitenant-capable and offers full-text search via \ac{http}.
Logstash is used to feed Elasticsearch.
In general, it offers an open server-side data processing pipeline that helps to send data from multiple sources to an Elasticsearch node.
Kibana is the primary data visualization tool.
It enables users to create plots and dashboards, crawl Elasticsearch, and trace the system's health.
All logs of the honeypots and tools are forwarded to the search engine Elasticsearch by Logstash.
The ELK stack is not directly exposed to the Internet; thus, authentication is unnecessary.
Users can monitor all log files with Kibana by pre-defined dashboards or custom search queries.
In addition, T-Pot features different services types, namely
\begin{enumerate*}[label=(\roman*)]
    \item standard,
    \item sensor,
    \item industrial,
    \item collector,
    \item next generation, and
    \item medical
\end{enumerate*}.
Each service type has a different set of honeypots and tools tailored to its core idea.
T-Pot feeds their data to an external Telekom service; however, this data submission can be turned off.
The latest version, $20.06.0$, has been used in this chapter.
Newer versions might be available by the end of this study and could differ from this.

\begin{sidewaysfigure}[htbp]
    \centering
    \includegraphics[width=\textwidth]{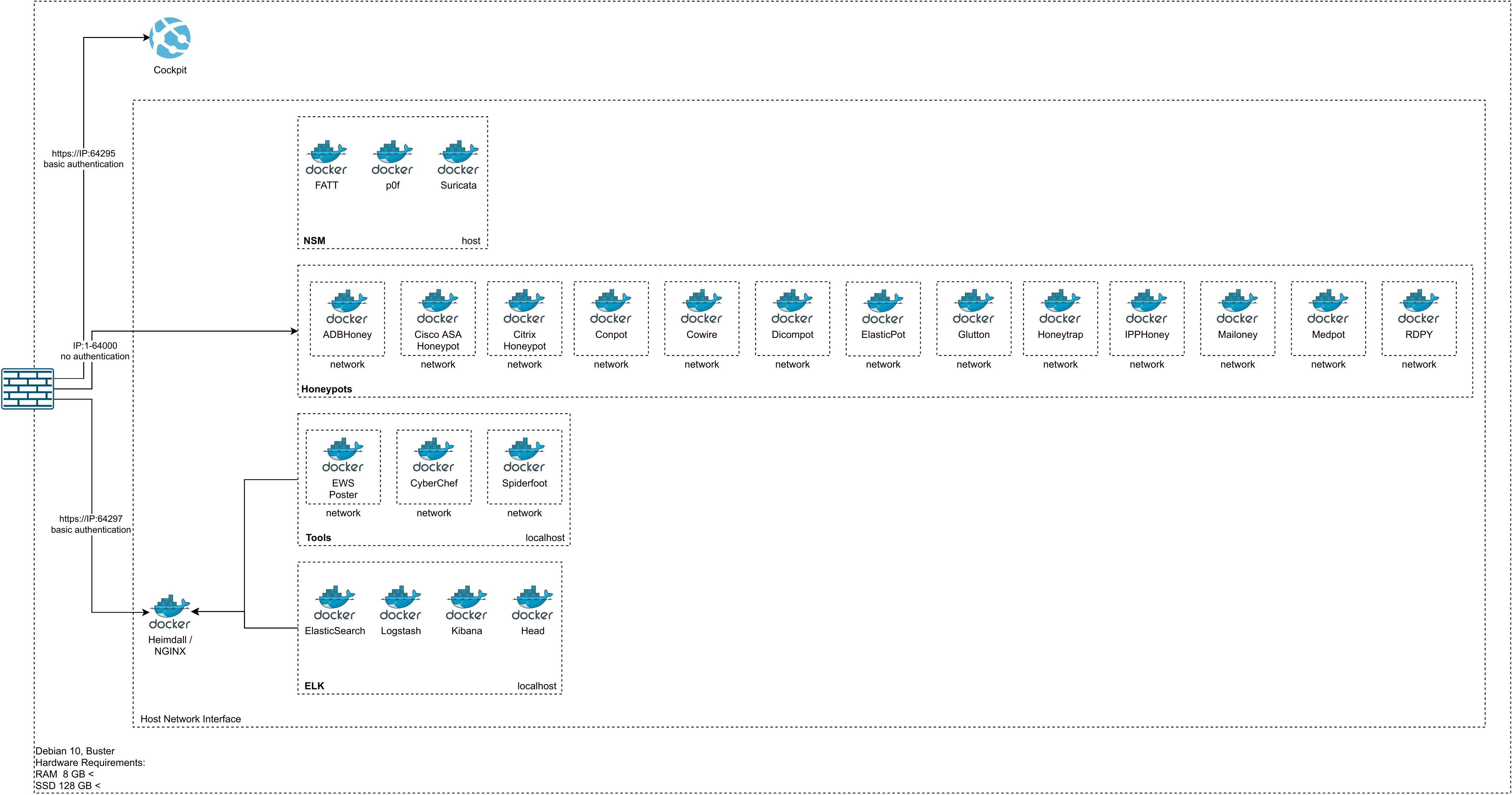}
    \caption[T-Pot architecture]{
        T-Pot architecture derived from \cite{tpot2021}.
        Honeypots are encapsulated in their network.
        NSM runs on the host network, and thus, receives every packet.
        ELK and tools run on localhost and are accessible through NGINX.
        The Cockpit application is a web-based graphical interface for servers.
    }
    \label{fig:overview-tpot}
\end{sidewaysfigure}

\subsubsection{Honeypots}
\label{subsubsec:honeypots}

T-Pot consists of 20 honeypots.
Albeit the sheer quantity of it, a short explanation is given.
In addition, \autoref{tab:overview-honeypots} gives a quick overview of all available honeypots in conjunction with
\begin{enumerate*}[label=(\roman*)]
    \item the port they are running on,
    \item their interaction level, and
    \item a short description
\end{enumerate*}.

\textbf{ADBHoney} \cite{adbhoney2021} is a low-interaction \ac{adb} honeypot over \ac{tcp}/\ac{ip}.
The importance of it lies in the \ac{adb} protocol that is used for debugging and pushing content to an Android device.
However, unlike a \ac{usb} connection, it does not support any kind of ample mechanisms of authentication and protection.
By exposing the \ac{adb} service over any port, an adversary could connect and exploit it.
ADBHoney is designed to catch malware that has been pushed onto devices.

\textbf{Cisco \ac{asa}} \cite{cymmetria2018} is a low-interaction honeypot that detects CVE-2018-0101\cite{CVE-2018-0101}.
It is a vulnerability that could allow an unauthenticated, remote attacker to cause a reload of the affected system and remotely execute code.
This can be achieved by flooding a webvpn-configured interface with crafted \ac{xml} packets.
Consequently, the attacker obtains full control by executing arbitrary code.

\textbf{Citrix \ac{adc} honeypot} \cite{citrixhoneypot2020} detects and logs CVE-2019-19781\cite{CVE-2019-19781} scans and exploitation attempts.
This vulnerability allows adversaries to perform directory traversal attacks.
Files are accessible by path strings to denote the file or directory.
In addition, some file systems include special characters to traverse the hierarchy easily.
Attackers take advantage of it by combining special characters to get access to restricted areas. \cite{flanders2019}

\textbf{Conpot} \cite{conpot2021} is a low-interaction industrial honeypot for \acf{ics}, and \acf{scada}.
It provides a variety of different standard industrial control protocols.
An adversary should be tricked by the complex infrastructure and lured into attacks.
In addition, a custom human-machine interface can be connected to increase the attack surface.
By randomly delaying the response time, Conpot tries to emulate a real machine handling a certain amount of load.

\textbf{Cowrie} \cite{cowrie2021} is a medium- to high-interaction \ac{ssh} and Telnet honeypot.
It offers to log brute-force attacks and shell interactions with attackers.
In medium-interaction mode, Cowrie emulates a Unix shell in Python, whereas in high-interaction mode, it proxies all commands to another system.

\textbf{DDoSPot} \cite{ddosspot2021} is a low-interaction honeypot to log and detect \ac{udp}-based \ac{ddos} attacks.
It is a platform used to support various plugins for different honeypot services and servers.
Currently, it supports \ac{dns}, \ac{ntp}, \ac{ssdp}, \ac{chargen}, and random/mock \ac{udp} server.

\textbf{Dicompot} \cite{dicompot2021} is a low-interaction honeypot for the \ac{dicom} protocol.
As with other honeypots before, it mocks a \ac{dicom} server in Go to collect logs and detect attacks.

\textbf{Dionaea} \cite{dionaea2021} is a medium-interaction honeypot that tries to capture malware copies by exposing services.
It supports various protocols such as \ac{ftp}, \ac{smb}, and \ac{http}.
Several modules can be integrated to work with Dionaea for further malware results, such as VirusTotal.

\textbf{Elasticpot} \cite{elasticpot2021} is a low-interaction honeypot for Elasticsearch, a search engine based on the Lucene library.

\textbf{Glutton} \cite{glutton2021} is a generic low-interaction honeypot that works as a \ac{mitm} for \ac{ssh} and \ac{tcp}.
However, lacking documentation does not provide a deeper insight into this honeypot.

\textbf{Heralding} \cite{heralding2021} is a credential catching honeypot for protocols like \ac{ftp}, Telnet, \ac{ssh}, \ac{http}, or \ac{imap}.

\textbf{HoneyPy} \cite{honeysap2021} is a low to medium-interaction honeypot that supports several protocols such as UDP or \ac{tcp}.
New protocols can be added by writing a custom plugin for them.
HoneyPy gives the freedom of quickly deploying and extending honeypots.

\textbf{HoneySAP} \cite{honeysap2021} is a low-interaction honeypot tailored for SAP services.

\textbf{Honeytrap} \cite{honeytrap2021} is a low-interaction honeypot network security tool.
As stated by \citet*{honeytrap2021}, Honeytrap is vulnerable to buffer overflow attacks.

\textbf{IPPHoney} \cite{ipphoney2021} is a low-interaction \ac{ipp} honeypot.

\textbf{Mailoney} \cite{mailoney2021} is a low-interaction \ac{smtp} honeypot written in Python.

\textbf{MEDpot} \cite{medpot2021} is a low-interaction honeypot focused on \ac{fhir}.
It is a standard description data format to transfer and exchange medical health records.

\textbf{RDPY} \cite{rdpy2021} is a low-interaction honeypot of the Microsoft \ac{rdp} written in Python.
It features client and server-side, and it is based on the event-driven network engine Twisted.
It supports authentication over \ac{tls} and \ac{nla}.

\textbf{SNARE and TANNER} \cite{snare2021, tanner2021} is a honeypot project.
SNARE is an abbreviation for Super Next-generation Advanced Reactive honEypot.
It is a successor of Glastopf, a web application sensor.
In addition, it supports the feature of converting existing web pages into attack surfaces.
TANNER \cite{tanner2021} can be seen as SNARES' brain.
Whenever a request has been sent to SNARE, TANNER decides how the response should be.

\subsubsection{Tools}

T-Pot integrates tools to screen network traffic and block DoS attacks.

\textbf{FATT} \cite{fatt2021} is used to extract metadata and fingerprints such as JA3 \cite{ja32021} and HASSH \cite{hassh2021} from captured packets.
JA3 is a method for \enquote{creating SSL/TLS client fingerprints} whereas HASSH is a network fingerprinting standard that is used to identify specific client and server \ac{ssh} implementations.
In addition, it features live network traffic.
As noted by the author, FATT is based on a python wrapper for tshark, namely pyshark, and thus has performance downturns.
T-Pot applies FATT on every request made on the host network.

\textbf{Spiderfoot} \cite{spiderfoot2021} is an open-source intelligence automation tool that helps to screen targets to get information about what is exposed over the Internet.
It can target different entities such as \ac{ip} address, domain, hostname, or network subnet.
In addition, it features more than 200 modules that can be integrated as an extension.
T-Pot uses it to scan defensively and thus not include any other module.

\textbf{Suricata} \cite{suricata2021} is \enquote{a high performance \ac{ids}, \ac{ips} and \ac{nsm} engine}.
T-Pot lets Suricata analyze and assess any request made on the host network.

\textbf{P0f} \cite{p0f2021} is a fingerprinting tool that uses passive traffic fingerprinting mechanisms to check \ac{tcp}/\ac{ip} communications.
T-Pot lets P0f passively check any request made on the host network.

\textbf{Endlessh} \cite{endlessh2021} is an \ac{ssh} server that sends an endless, random \ac{ssh} banner.
The key idea is to lock up \ac{ssh} clients that try to connect to the \ac{ssh} server.
It lowers the transaction speed by intentionally inserting delays.
Due to the established connection before the cryptographic exchange, this module does not require any cryptographic libraries.

\textbf{HellPot} \cite{hellpot2021} is an \enquote{endless honeypot}.
If someone connects to this honeypot, it results in a memory overflow.
Its key idea is to send an endless data stream to the attacker until its memory or storage runs out.

\begin{sidewaystable}[htbp]
    \centering
    \caption[Overview of honeypots of T-Pot]{
        Overview of all available honeypots of T-Pot with interaction level, port, and a short description.
        Ports are marked with either \ac{tcp} or UDP; if a port misses any definition, both \ac{tcp} and UDP are allowed.
    }
    \begin{tabularx}{\linewidth}{l|XlX}
        \toprule
        \textsc{Honeypots}                        & \multicolumn{3}{c}{}                                                                                                                                                                                                            \\
                                                  & \textbf{Port}                                                                                               & \textbf{Interaction-level} & \textbf{Description}                                                                 \\
        \hline
        ADBHoney \cite{adbhoney2021}              & 5555/TCP                                                                                                    & low                        & \ac{adb} protocol honeypot                                                           \\
        Cisco ASA \cite{cymmetria2018}            & 5000/UDP, 8443/TCP                                                                                          & low                        & honeypot for CVE-2018-0101\cite{CVE-2018-0101} detection                             \\
        Citrix honeypot \cite{citrixhoneypot2020} & 443/TCP                                                                                                     & low                        & detects and logs CVE-2019-19781\cite{CVE-2019-19781} scans and exploitation attempts \\
        Conpot \cite{conpot2021}                  & 80, 102, 161, 502, 623, 1025, 2404, \newline 10001, 44818, 47808, 50100                                     & low                        & industrial honeypot for \ac{ics} and \ac{scada}                                      \\
        Cowrie \cite{cowrie2021}                  & 2222, 23                                                                                                    & high                       & \ac{ssh} and Telnet honeypot                                                         \\
        DDoSPot \cite{ddosspot2021}               & 1112/TCP                                                                                                    & low                        & log and detect UDP-based \ac{ddos} attacks                                           \\
        Dicompot \cite{dicompot2021}              & 1112/TCP                                                                                                    & medium                     & honeypot for the \ac{dicom} protocol                                                 \\
        Dionaea \cite{dionaea2021}                & 21, 42, 69/UDP, 8081, 135, 443, 445, \newline 1433, 1723, 1883, 1900/UDP, \newline 3306, 5060/UDP, 5061/UDP & low                        & capture malware copies                                                               \\
        Elasticpot \cite{elasticpot2021}          & 9200                                                                                                        & low                        & honeypot for Elasticsearch                                                           \\
        Glutton \cite{glutton2021}                & NFQ                                                                                                         & medium                     & MitM proxy for \ac{ssh} and TCP                                                      \\
        Heralding \cite{heralding2021}            & 21, 22, 23, 25, 80, 110, 143, 443, \newline 993, 995, 1080, 5432, 5900                                      & low                        & credential catching honeypot                                                         \\
        HoneyPy \cite{honeysap2021}               & 7, 8, 2048, 2323, 2324, 4096, 9200                                                                          & low                        & extendable honeypot                                                                  \\
        HoneySAP \cite{honeysap2021}              & 3299/TCP                                                                                                    & low                        & honeypot for SAP services                                                            \\
        Honeytrap \cite{honeytrap2021}            & NFQ                                                                                                         & medium                     & captures attacks via unknown protocols                                               \\
        IPPHoney \cite{ipphoney2021}              & 631                                                                                                         & low                        & \ac{ipp} honeypot                                                                    \\
        Mailoney \cite{mailoney2021}              & 25                                                                                                          & low                        & SMTP honeypot                                                                        \\
        MEDpot \cite{medpot2021}                  & 2575                                                                                                        & low                        & \ac{fhir} honeypot                                                                   \\
        RDPY \cite{rdpy2021}                      & 3389                                                                                                        & low                        & Microsoft \ac{rdp} honeypot                                                          \\
        SNARE/TANNER \cite{snare2021}             & 80                                                                                                          & low                        & web application honeypot                                                             \\
        \bottomrule
    \end{tabularx}
    \label{tab:overview-honeypots}
\end{sidewaystable}

\section{Results}
\label{sec:honeypots-heicloud}

The T-Pot has been deployed for three weeks (from \nth{26} of September to \nth{16} of October) and collected in total \numprint{607747} attacks.
Overall, RDPY ($46.08\%$), Honeytrap ($33.23\%$), and Cowrie ($12.42\%$) received most of the attacks with a total amount of \numprint{540398} attacks.
\autoref{fig:overview-attacks} shows the distribution of honeypot attacks.
The total numbers are based on \autoref{tab:overview-honeypots-attacks}.

\begin{figure}[htbp]
    \centering
    \includegraphics[width=\textwidth]{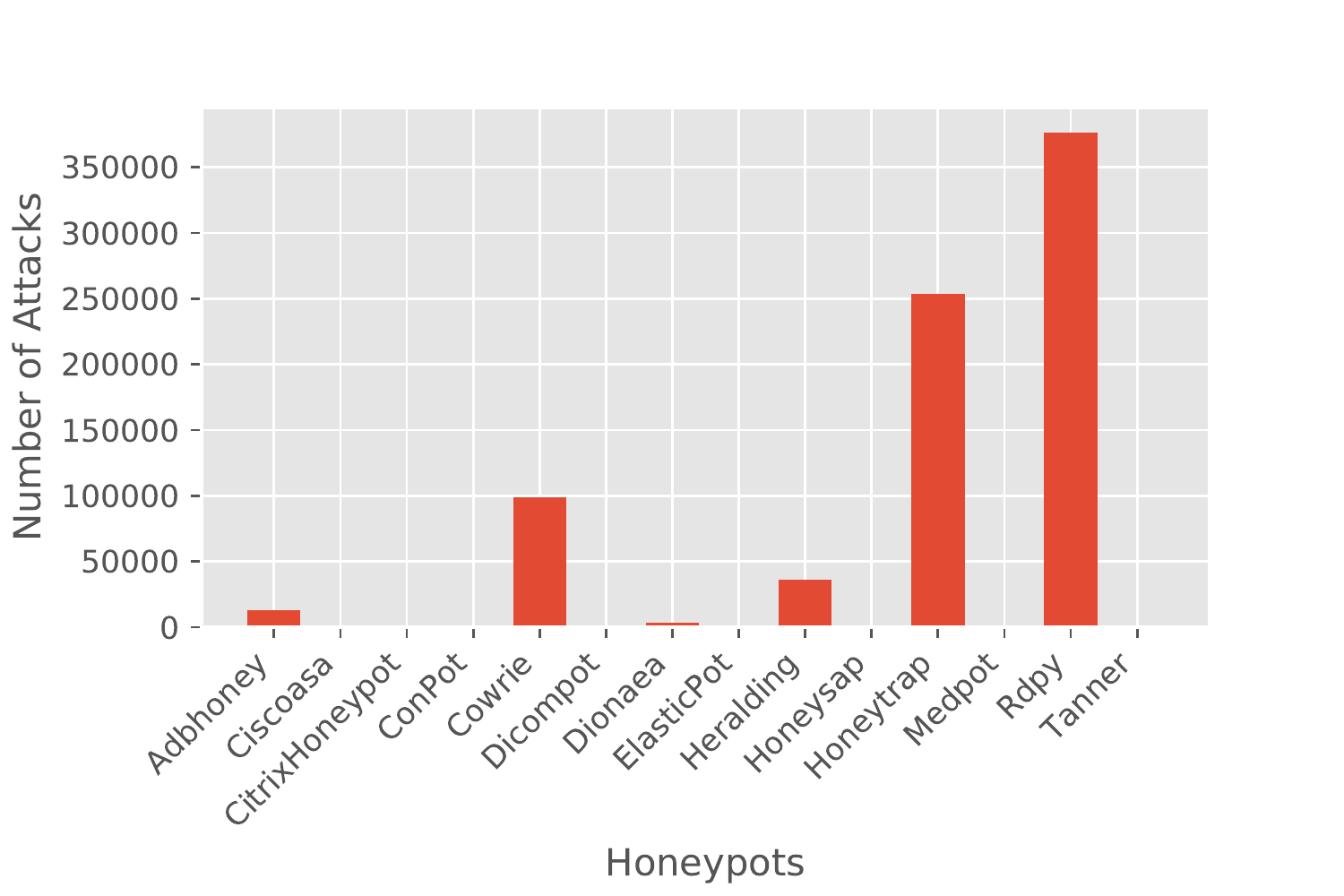}
    \caption[Distribution of honeypot attacks]{
        Distribution of honeypot attacks.
        Timestamp; \nth{26} of September to \nth{16} of October.
        A description of each honeypot can be found in section \ref{subsubsec:honeypots}.
    }
    \label{fig:overview-attacks}
\end{figure}

% ====================================================================================================================
% Dionaea
What is striking is the large disparity between the previously mentioned attacks on \ac{aws}, \ac{gcp}, and Azure.
Especially with the honeypot Dionaea, it is unclear why only \numprint{2368} attacks have been performed.
$96\%$ of \ac{ip} addresses connected to Dionaea are known attackers, and $70\%$ were acquired on port 81, unofficially known for Tor routing.
Neither any malware nor suspicious payload could be identified.
An assumption is that the packets run through a static filter.
Heidelberg has a centralized stateless firewall, indicating that specific ports or protocols are excluded.
A \textsc{nmap} \ac{tcp} SYN scan (\verb|nmap -sS -A 10.20.30.123|) has been performed to prove this assumption that ports are excluded.
The result clearly shows that port 139 for \ac{smb} is filtered, although the access security explicitly allows it.
The stateless firewall runs in front of heiCLOUD and filters many ports, including $113$.
Based on this, it can be assumed that most of the attacks on Dionaea are carried out via \ac{smb}, which would explain the total number of attacks.
The administrator of the university firewall had been consulted to exclude the T-Pot instance to validate if the actual number is even higher without any packet filter in front of it.
Respectively, no stateless packet filter has been applied to the T-Pot for three weeks (\nth{2} of December until \nth{23} of December).
It could identify a drastic increase in Dionaea attacks with a total number of \numprint{213053}.
Overall, $93\%$ of all attacks are on the \ac{smb} protocol followed by many database protocols such as MongoDB and MSSQL.
This confirms the assumption that a higher total number of attacks would be the result without the packet filter in front of the instance.

Comparing the number with \citet{Kelly2021} it shows that Dionaea attacks surpass every other cloud provider.
However, Dionaea attacks will not be included in later results because usually, a server is not allowed to be excluded from the university firewall.
Only for this research purpose to assess the effect of the packet filter has an exclusion been granted.

\begin{figure}[htbp]
    \centering
    \includegraphics[width=\textwidth]{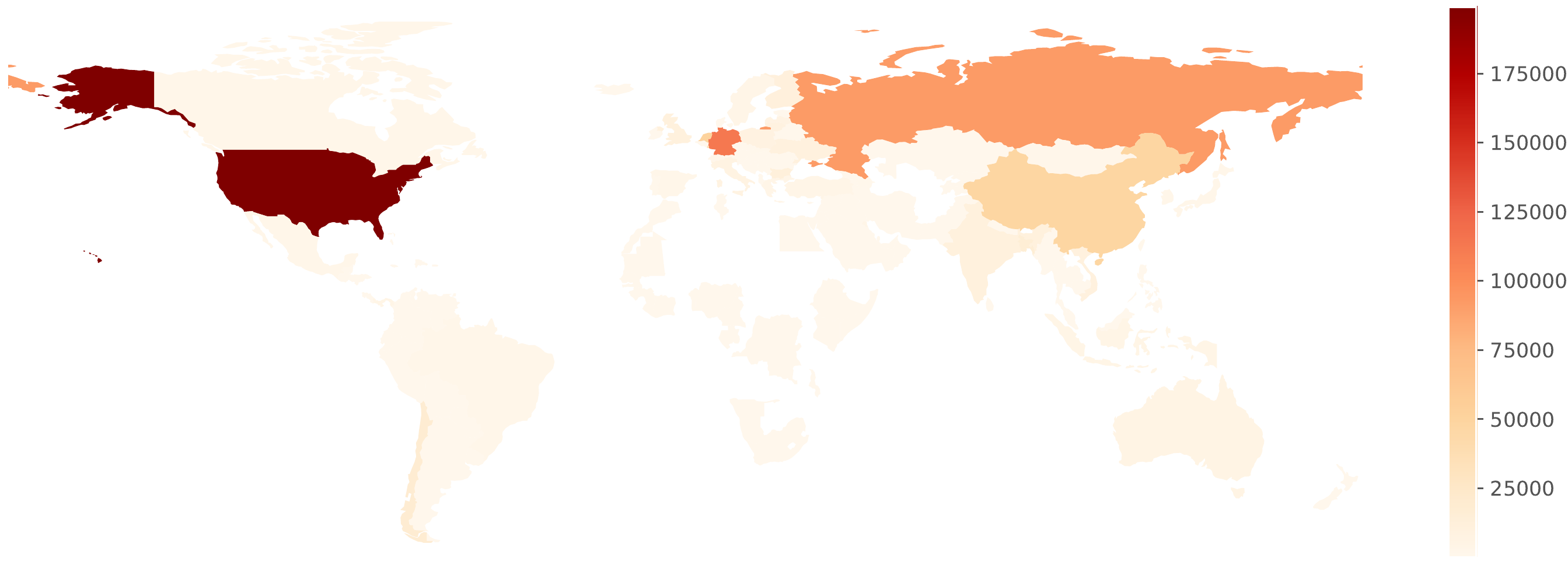}
    \caption[Attack distribution of T-Pot]{
        Attack distribution of T-Pot.
        The USA, Russia, China, and Germany are the most attacking countries.
        Timestamp; \nth{26} of September to \nth{16} of October.
    }
    \label{fig:attack-distribution}
\end{figure}

% ====================================================================================================================
% Location
Logstash uses GeoLite2 to resolve the source \ac{ip} address with information such as location, \ac{asn}, continent code, country name, and \ac{as} organization.
\autoref{fig:attack-distribution} indicates the geographical location of connections acquired to any honeypot.
Most attacks are originated from the United States, Germany, Russia, and China.
Large security scans of DFN or \ac{belwue} pushes Germany to second place; therefore, Germany can be considered negligible.
On the contrary, the geographical location of an \ac{ip} address merely indicates the true origin.
Due to technologies like \ac{vpn} or Tor, the last known node of an \ac{ip} address could be spoofed, and thus as stated by \citet{Kelly2021}, would remain insufficient to use.
Hence, no one should rely on geographical information.

\begin{figure}[htbp]
    \centering
    \includegraphics[width=\textwidth]{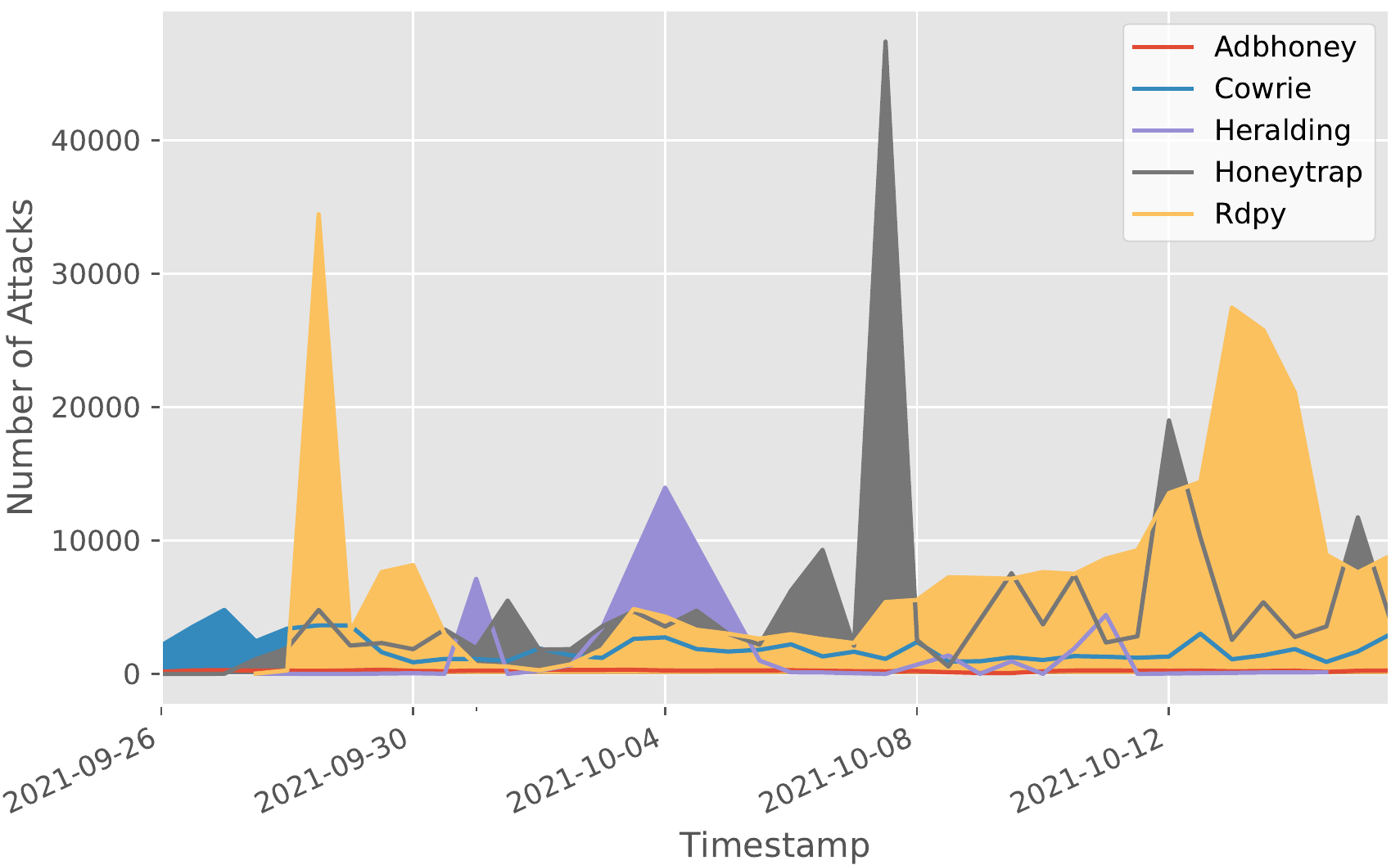}
    \caption[Attack histogram of T-Pot]{
        Attack histogram of T-Pot.
        Only the five most attacked honeypots are considered.
        Timestamp; \nth{26} of September to \nth{16} of October.
        A description of each honeypot can be found in section \ref{subsubsec:honeypots}.
    }
    \label{tpot-overview-histogram}
\end{figure}

Attacks are not equally distributed among all honeypots, and thus, different protocols and applications receive more attention than others.
\autoref{tpot-overview-histogram} shows the timeline of attacks that are executed on our instance separated by honeypots.
RDPY, Honeytrap, and Cowrie are the most attacked honeypots.
The high peak of Honeytrap in the middle indicates a full \textsc{nmap} scan from Germany that has been done to get an insight of the packet filtering at the Heidelberg University.
It identifies a bias towards remote desktop protocol attacks, shell-code exploitations, and commands to retrieve information about the \acs{cpu}, scheduled tasks (\verb|cat /proc/cpuinfo|, or \verb|crontab|), or privilege escalation.

\begin{figure}[htbp]
    \centering
    \includegraphics[width=\textwidth]{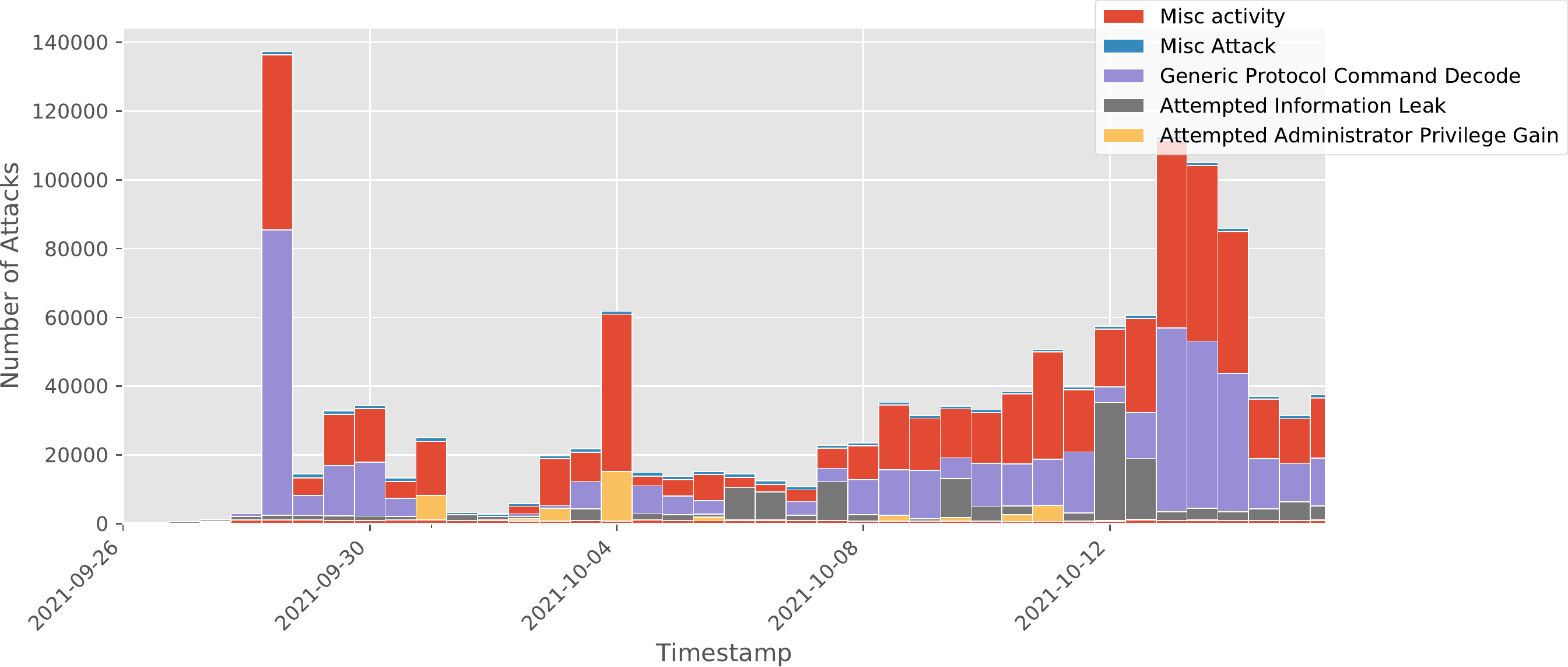}
    \caption[Suricata results of T-Pot]{
        Suricata results of T-Pot.
        Displays the five most listed alert categories.
        Timestamp; \nth{26} of September to \nth{16} of October.
    }
    \label{fig:suricata-results}
\end{figure}

% ====================================================================================================================
% Suricata
Suricata registered several alerts and \acsp{cve}.
The vast majority of alerts are \ac{rdp}-related policies, \ac{vnc} authentication failures, and \textsc{nmap} scans.
Most used vulnerabilities are
\begin{enumerate*}[label=(\roman*)]
    \item CVE-2001-0540\cite{CVE-2001-0540} which is a memory leak in terminal servers in Windows NT and Windows 2000 causing a denial of service (memory exhaustion) by malformed \ac{rdp} requests,
    \item CVE-2006-2369\cite{CVE-2006-2369} which is a RealVNC vulnerability allowing hackers to bypass authentication, and
    \item CVE-2012-0152\cite{CVE-2012-0152} which enables attackers for \ac{rdp} in Microsoft Windows Server 2008 R2 and R2 SP1 and Windows 7 Gold and SP1 to cause a denial of service by sending a series of crafted packets
\end{enumerate*}.
As derived from \autoref{fig:suricata-results}, the T-Pot has not received many attacks in the first week.
Starting from the 28th of September, the number of alerts is skyrocketing.
This would indicate that bots crawl \ac{ip} address ranges to find new machines and probe them.
Interestingly, zero-day exploits like the Apache vulnerability \cite{CVE-2021-42013} that came with version $2.49.0$ got registered in CVE on the \nth{6} of October and immediately recognized by Suricata on the \nth{15} of October.
Attackers could perform a remote code execution using path traversal attacks when the \ac{cgi} scripts of Apache are enabled.
The logs could trace back similar attacks like \verb|/cgi-bin/.\%2e/\%2e\%2e/bin/sh| until the \nth{7} of October, leaving an even smaller time frame to adapt to new exposures.
This shows how fast bots adapt to new vulnerabilities to compromise more systems.

\begin{figure}[htbp]
    \centering
    \includegraphics[width=\textwidth]{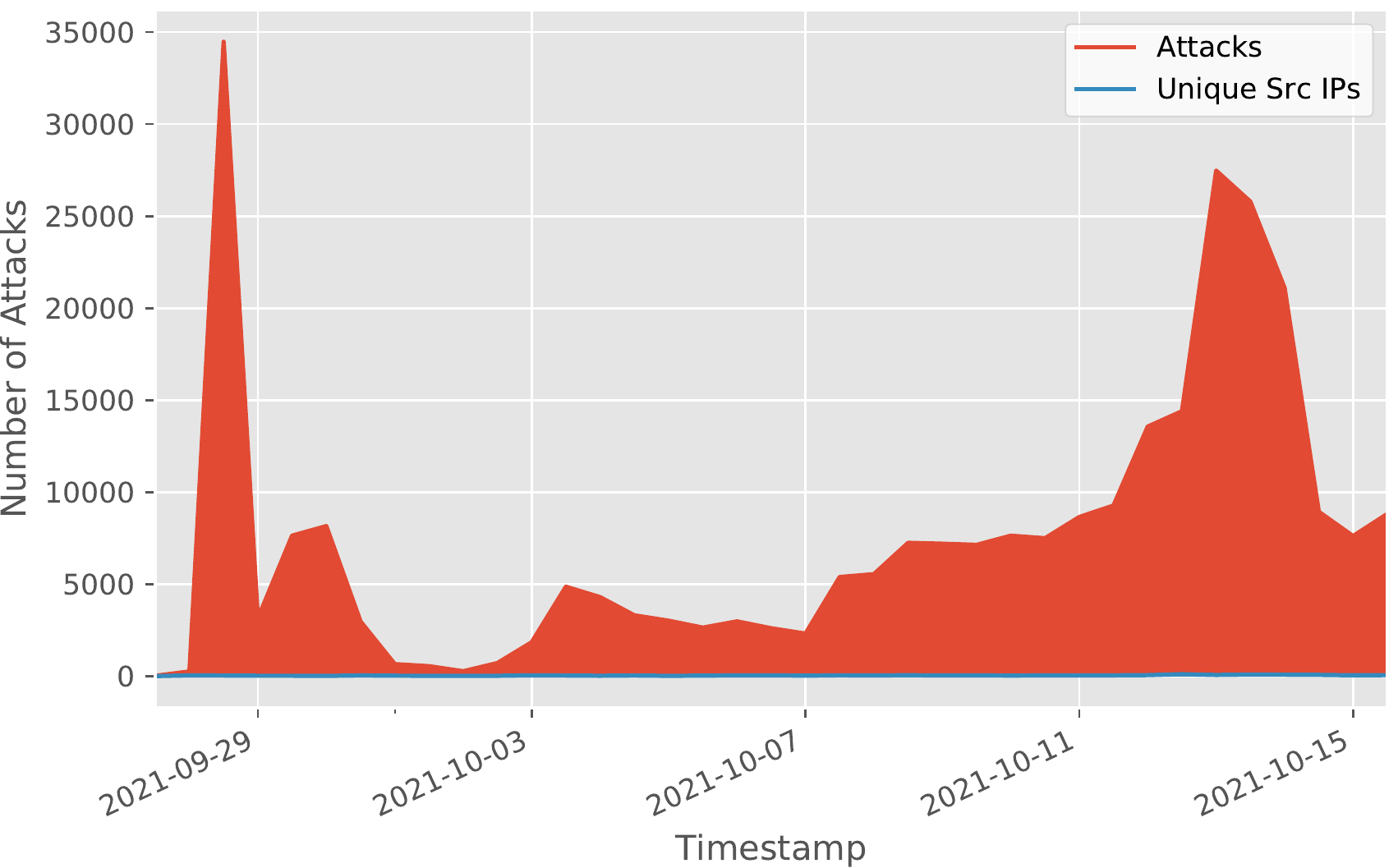}
    \caption[RDPY results of T-Pot]{
        RDPY attacks are separated into attacks and unique source \ac{ip} addresses.
        Timestamp; \nth{26} of September to \nth{16} of October.
        A description of the honeypot can be found in section \ref{subsubsec:honeypots}.
    }
    \label{fig:rdpy-results}
\end{figure}

% ====================================================================================================================
% RDPY
The results from RDPY in \autoref{fig:rdpy-results} backups the assumption that attacks originate from bots.
It shows that only a small margin represents unique source \ac{ip} addresses.
The rest of the attacks result in either a bad reputation, bot, crawler, or known attacker.
\autoref{fig:suricata-results} shows the distribution of alert categories that Suricata identified.
Respectively, misc activities sum up to roughly $1.5$ million entries, \ac{rdp} related alerts account for two-thirds of it.
Several \ac{rdp} attacks from 2021 back to 2001 had been executed on the T-Pot.
Respectively, CVE-2012-0152 and CVE-2001-0540 coincide with the ones \citet{Kelly2021} claim.

\begin{figure}[htbp]
    \centering
    \includegraphics[width=\textwidth]{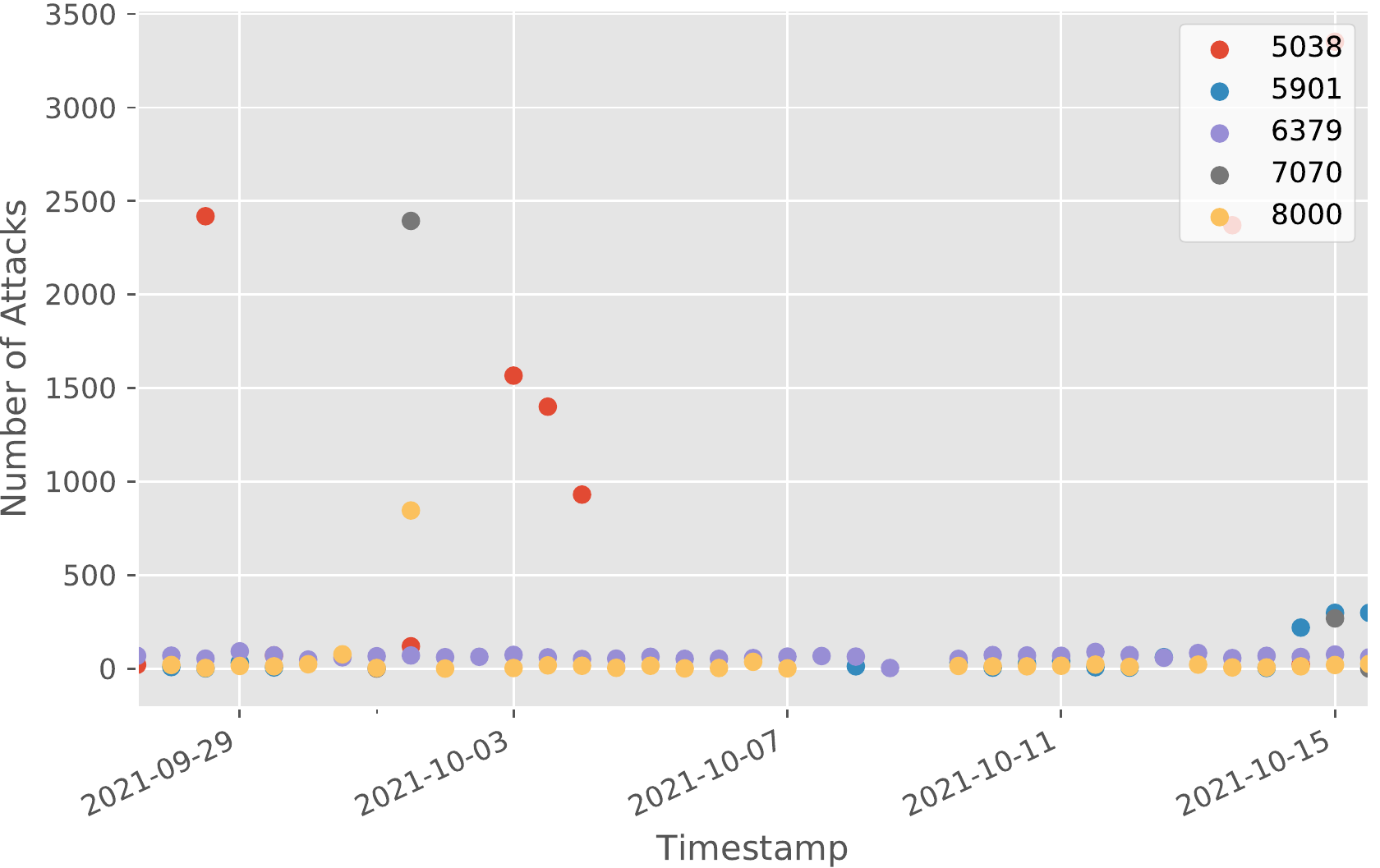}
    \caption[Honeytrap results of T-Pot]{
        Honeytrap results of T-Pot.
        Timestamp; \nth{26} of September to \nth{16} of October.
        A description of the honeypot can be found in section \ref{subsubsec:honeypots}.
    }
    \label{fig:honeytrap-results}
\end{figure}

% ====================================================================================================================
% Honeytrap
For NFQ related attacks, Honeytrap could identify three major services that are not provided by default.
Honeytrap functions as a honeypot to provide a service on ports that are not specified by default.
NFQ intercepts incoming \ac{tcp} connections during the \ac{tcp} handshake, and Honeytrap provides a service for it.
Most of these interceptions are made on
\begin{enumerate*}[label=(\roman*)]
    \item port 5038, which is used by a machine learning database called MLDB,
    \item port 5905, which an Intel Online Connect Access uses on Windows machines, and
    \item port 7070 which is used by Apple's QuickTime streaming server (RTSP)
\end{enumerate*}.
Nearly all ports attacks focused on \ac{rdp} connection attempts (\verb|Cookie: mstshash=Administr|).
However, $94\%$ of all connected \ac{ip} addresses on Honeytrap are resolved as known attackers.

\begin{figure}[htbp]
    \centering
    \includegraphics[width=\textwidth]{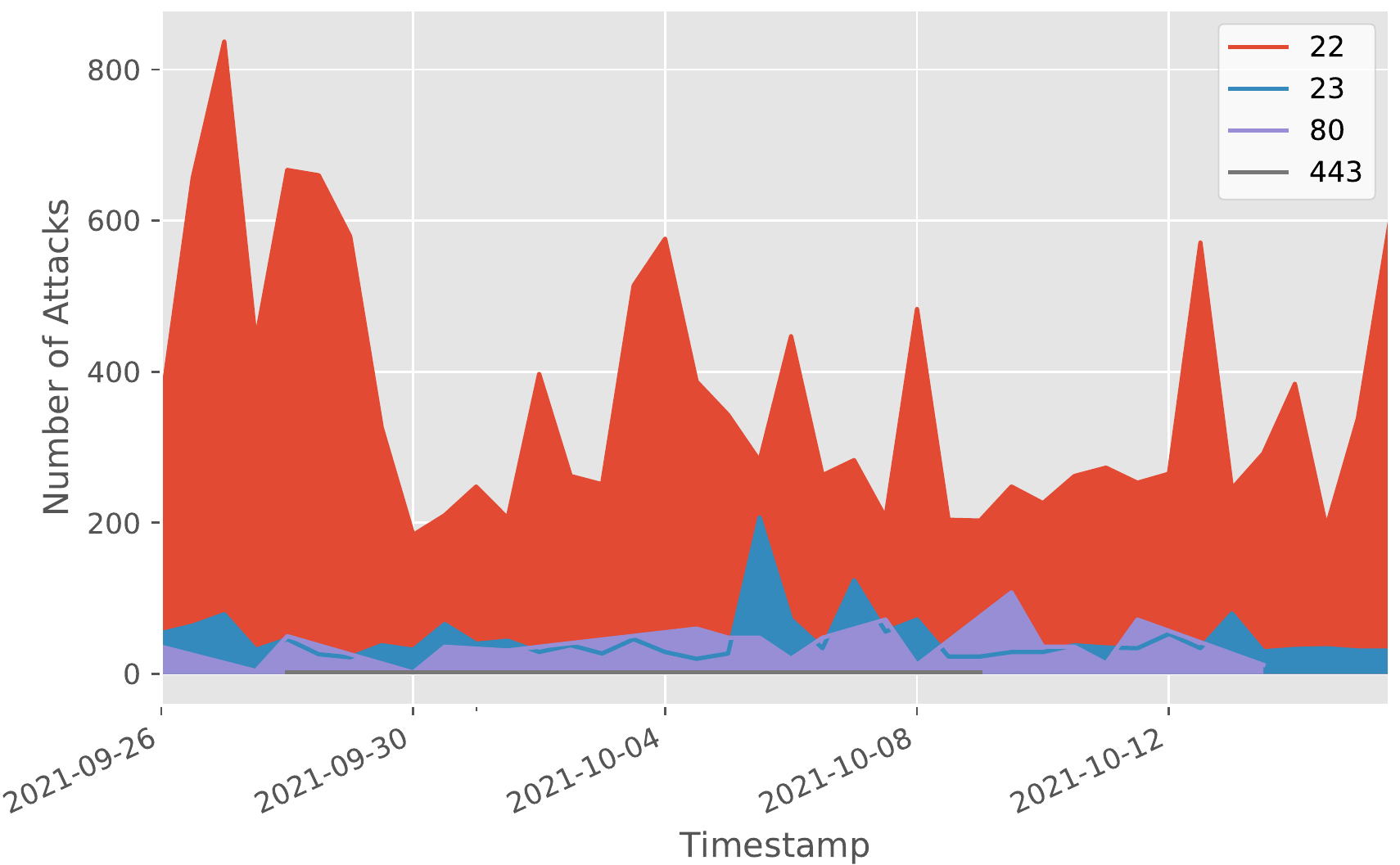}
    \caption[Cowrie results of T-Pot]{
        Cowrie results of T-Pot.
        Timestamp; \nth{26} of September to \nth{16} of October.
        A description of the honeypot can be found in section \ref{subsubsec:honeypots}.
    }
    \label{fig:cowrie-results}
\end{figure}

% ====================================================================================================================
% Cowrie
The third most compromised honeypot is Cowrie, with a strong bias towards \ac{ssh} and \ac{ftp}.
\autoref{fig:cowrie-results} shows all attacks executed on Cowrie separated by their port.
Respectively, \ac{ssh} port $22$ is the most considered port, resulting in high use for privilege escalation.
Besides using default credentials to log in (\verb|username: root|, \verb|password: root|, see \autoref{fig:cowrie-credentials} for top 10 credentials), adversaries used various commands to retrieve any information about the host system (\verb|nproc;uname -a|, \verb|cat /proc/cpuinfo|).
A unique information gathering attack could be identified that has been widely used on the T-Pot.
\autoref{lst:code-exploitation} shows all shell commands that are executed.
Attackers try to gain knowledge about running processes on the system (\verb|/bin/busybox|).
Interestingly, crypto mining attacks are getting more attractive to criminals.
For example, XMRig has been the most downloaded malware for cryptocurrency mining.
Some adversaries even executed complex tailored shell commands to exploit the host machine as a crypto miner (\autoref{lst:crypto-exploitation}).
It is not surprising that such attacks gain attraction concerning the current time.
Attackers could exploit machines for crypto mining in order to earn more money.
This looks more appealing than acquiring mining machines and hijacking electricity from surrounding apartments.

\begin{figure*}[htbp]
    \centering

    \begin{subfigure}[b]{0.49\textwidth}
        \centering
        \includegraphics[width=\textwidth]{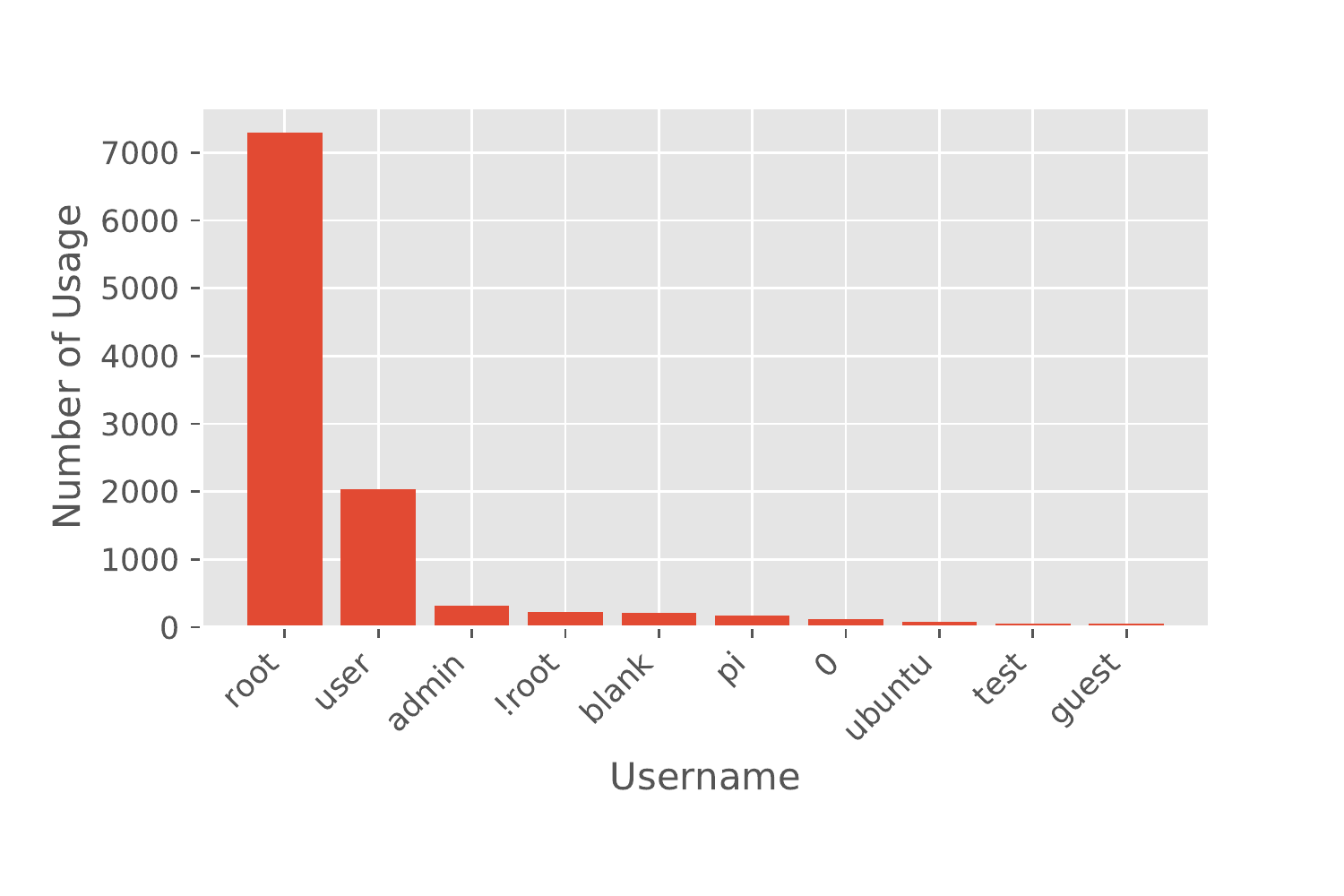}
        \caption{Cowrie username credentials}
        \label{fig:tpot-cowrie-username}
    \end{subfigure}
    \hfill
    \begin{subfigure}[b]{0.49\textwidth}
        \centering
        \includegraphics[width=\textwidth]{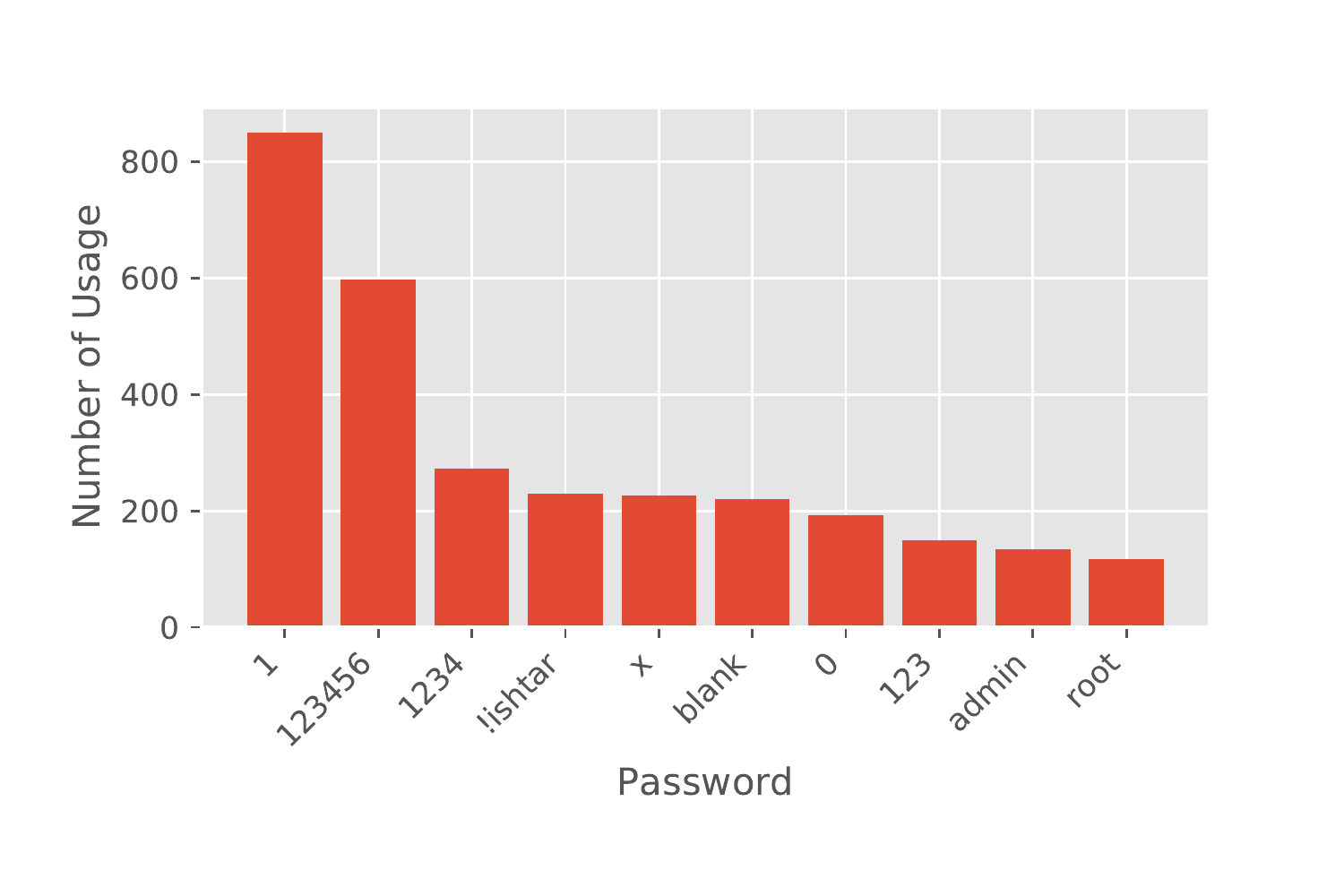}
        \caption{Cowrie password credentials}
        \label{fig:tpot-cowrie-password}
    \end{subfigure}
    \caption[Cowrie top 10 credentials on T-Pot]{
        Cowrie top 10 credentials used on T-Pot.
        Timestamp; \nth{26} of September to \nth{16} of October.
        A description of the honeypot can be found in section \ref{subsubsec:honeypots}.
    }
    \label{fig:cowrie-credentials}
\end{figure*}

% ====================================================================================================================
% P0f
P0f identified different Windows versions and Linux distributions in conjunction with various \ac{ssh} clients to compromise the T-Pot.
Like \citet{Kelly2021} presented, Windows 7 or 8 and Windows NT Kernel are the most used \ac{os} with $81\%$.
Unfortunately, disguising OS fingerprinting activities account for $84\%$ of all fingerprints.
Lastly, the results are cleaned up and all IPs from DFN and \acs{belwue} are excluded.
Both scan frequently and check if any vulnerability exists.
This distorts the findings, and thus, they have been filtered based on their subnet addresses.
However, the results show no notable changes.
The total number of attacks was hardly influenced by it.
This indicates that these scans do not greatly interfere with the findings.

% ====================================================================================================================
% Conclusion
On average, heiCLOUD has received $55.83\%$ more than Azure, \ac{gcp}, and \ac{aws}.
Attacks on Cowrie, RDPY, and Honeytrap are the most compromised honeypots.
In contrast to \citet{Kelly2021}, Dionaea and Glutton used to be the most considered honeypots for adversaries.
It can be assumed that attacks by bots had increased significantly since last year when \citet{Kelly2021} did their research.
Respectively, one unresolved question is if other cloud providers filter their network traffic.
It would explain the major difference between Heidelberg University and the big tech companies.
The cause for such an increase remains doubtful.
One explanation could root back to the Corona pandemic and the skyrocketing increase in home office activities.
Related to that is a higher usage in screen sharing software.
Considering the BSI\footnote{The Federal Office for Information Security is responsible for managing communication security for the German government.
Each year they publish a report for recent cybersecurity threats.}, report for cybersecurity 2021 \cite{bsi2021}, they revealed an increase of attack surfaces during the pandemic.
Respectively, the IT infrastructure could not keep up with this fast change and widen the company's attack surface.
Their conclusion overlays our assumption that attackers took advantage and increased their activities.
This phenomenon shows that nearly all attacks originate from bots that scan through \ac{ip} address ranges.
In total, $73\%$ of all \ac{ip} addresses are unresolved.
The known attacker reputation represents the largest part of resolved \ac{ip} addresses with $23\%$.
Fortunately, such reputations could technically be filtered by an organization's firewall and would lower the chance of an exploit.
Interestingly, after three weeks, the number of attacks originating from China decreased to almost zero percent.
This might indicate that the honeypot has been exposed, and further attacks represent a risk of revealing their compromises.
However, this assumption cannot be confirmed due to the lax geographical reliability of \ac{ip} addresses.

Our results emphasize the importance of honeypots.
It gives a proper security measure of an IT infrastructure and helps to identify potential leaks or vulnerabilities.
Moreover, it shows that T-Pot helps detect recent bot activities and gives an outlook on the newest trends of attacks.

\begin{sidewaystable}[htbp]
    \centering
    \caption[Overview of attacks on heiCLOUD, AWS, GCP, and Azure]{
        Overview of attacks on heiCLOUD, \ac{aws}, \ac{gcp}, and Azure.
        Only the top 10 most attacked honeypots are considered.
        \enquote{-} entails that a honeypot is not part of the top 10.
        The red and green arrows indicate whether heiCLOUD received more or fewer attacks than the other cloud providers on average .
    }
    \begin{tabularx}{\linewidth}{l|rrc|rr|rr|rr}
        \toprule
        \textsc{Honeypots} & \multicolumn{3}{c}{BASIS}              & \multicolumn{6}{c}{\textsc{comparison}}                                                                                                                                                                 \\
                           & \multicolumn{3}{c|}{\textsc{heiCLOUD}} & \multicolumn{2}{c|}{\textsc{\ac{aws}}}       & \multicolumn{2}{c|}{\textsc{GC}} & \multicolumn{2}{c}{\textsc{Azure}}                                                                                         \\
                           & \textbf{Number}                        & \textbf{Pct.}                           &                                  & \textbf{Number}                    & \textbf{Pct.} & \textbf{Number}   & \textbf{Pct.} & \textbf{Number}   & \textbf{Pct.} \\
        \hline
        ADBHoney           & \numprint{9302}                        & $1.65\%$                                & $\color{green}{\uparrow}$        & \numprint{413}                     & $0.17\%$      & \numprint{2497}   & $0.43\%$      & \numprint{442}    & $0.13\%$      \\
        Cisco ASA          & \numprint{674}                         & $0.11\%$                                & $\color{green}{\uparrow}$        & \numprint{260}                     & $0.10\%$      & \numprint{750}    & $0.13\%$      & \numprint{134}    & $0.04\%$      \\
        Citrix honeypot    & \numprint{1121}                        & $0.18\%$                                &                                  & -                                  & -             & -                 & -             & -                 & -             \\
        Conpot             & \numprint{615}                         & $0.10\%$                                &                                  & -                                  & -             & -                 & -             & -                 & -             \\
        Cowrie             & \numprint{75511}                       & $11.97\%$                               & $\color{red}{\downarrow}$        & \numprint{4503}                    & $1.81\%$      & \numprint{297818} & $51.25\%$     & \numprint{9,012}  & $2.64\%$      \\
        DDoSPot            & \numprint{0}                           & $0\%$                                   &                                  & -                                  & -             & -                 & -             & -                 & -             \\
        Dicompot           & \numprint{22}                          & $0\%$                                   &                                  & -                                  & -             & -                 & -             & -                 & -             \\
        Dionaea            & \numprint{2368}                        & $0.40\%$                                & $\color{red}{\downarrow}$        & \numprint{228075}                  & $91.91\%$     & \numprint{162570} & $27.98\%$     & \numprint{308102} & $90.42\%$     \\
        Elasticpot         & \numprint{385}                         & $0.06\%$                                &                                  & -                                  & -             & -                 & -             & -                 & -             \\
        Glutton            & \numprint{0}                           & $0\%$                                   & $\color{red}{\downarrow}$        & \numprint{11878}                   & $4.79\%$      & \numprint{84375}  & $15.52\%$     & \numprint{17256}  & $5.06\%$      \\
        Heralding          & \numprint{35680}                       & $4.34\%$                                & $\color{green}{\uparrow}$        & \numprint{1885}                    & $0.76\%$      & \numprint{12,255} & $2.11\%$      & \numprint{3370}   & $0.99\%$      \\
        HoneyPy            & \numprint{0}                           & $0\%$                                   & $\color{red}{\downarrow}$        & \numprint{172}                     & $0.07\%$      & \numprint{2149}   & $0.37\%$      & \numprint{497}    & $0.15\%$      \\
        HoneySAP           & \numprint{15}                          & $0\%$                                   &                                  & -                                  & -             & -                 & -             & -                 & -             \\
        Honeytrap          & \numprint{201949}                      & $32.01\%$                               &                                  & -                                  & -             & -                 & -             & -                 & -             \\
        IPPHoney           & \numprint{0}                           & $0\%$                                   &                                  & -                                  & -             & -                 & -             & -                 & -             \\
        Mailoney           & \numprint{0}                           & $0\%$                                   & $\color{red}{\downarrow}$        & \numprint{720}                     & $0.29\%$      & \numprint{9419}   & $1.62\%$      & \numprint{146}    & $0.04\%$      \\
        MEDpot             & \numprint{2}                           & $0\%$                                   &                                  & -                                  & -             & -                 & -             & -                 & -             \\
        RDPY               & \numprint{280040}                      & $49.15\%$                               & $\color{green}{\uparrow}$        & \numprint{100}                     & $0.04\%$      & \numprint{7916}   & $1.36\%$      & \numprint{1463}   & $0.43\%$      \\
        SNARE/TANNER       & \numprint{63}                          & $0.02\%$                                & $\color{red}{\downarrow}$        & \numprint{138}                     & $0.06\%$      & \numprint{1367}   & $0.24\%$      & \numprint{313}    & $0.09\%$      \\
        \hline
        \textsc{In total}  & \numprint{607747}                      & $100\%$                                 &                                  & \numprint{248144}                  & $100\%$       & \numprint{581116} & $100\%$       & \numprint{340735} & $100\%$       \\
        \bottomrule
    \end{tabularx}
    \label{tab:overview-honeypots-attacks}
\end{sidewaystable}

\section{Discussion}

One downside of T-Pot is the static hostname representation of Cowrie.
It always returns \verb|#1 SMP Debian 3.2.68-1+deb7u1| (\verb|uname -a|) as hostname information, leaving a tiny footprint when bots crawl through the web.
A random choice of hostname information could harden Cowrie from being exposed.
Next, if attackers scan open ports on T-Pot, it might be suspicious when many ports with services are open.
From a technical perspective, bots could check this state if it is uncommon and thus, exclude T-Pot from being probed.
However, T-Pot includes reasonable preventions like a random hostname and scheduled tasks.
Another major drawback is the latest endeavor to detect honeypots on the transport level.
As recently investigated by \citet{vetterl2020}, detecting honeypots is becoming easier due to a fatal flaw in the underlying protocol implementation.
\citet{vetterl2020} states that attackers always try to prevent their methods, exploits, and tools from being divulged.
Therefore, detecting honeypots before attacking them strongly motivates black hats.
\Autoref{chap:fingerprinting} will present a way to avoid such fingerprint activities with the honeypot Cowrie.

% ====================================================================================================================
% Listings
\begin{figure}[htbp]
\begin{lstlisting}[language=bash, caption={[Cowrie attack to gather various information about the system]Cowrie attack to gather various information about the system.}, label={lst:code-exploitation}]
enable
system
shell
sh
cat /proc/mounts; /bin/busybox $PROCESS_NAME
cd /dev/shm; cat .s || cp /bin/echo .s; /bin/busybox $PROCESS_NAME
tftp; wget; /bin/busybox $PROCESS_NAME
dd bs=52 count=1 if=.s || cat .s || while read i; do echo $i; done < .s
while read i
/bin/busybox $PROCESS_NAME
rm .s; exit
\end{lstlisting}
\end{figure}

\begin{figure}[htbp]
\begin{lstlisting}[language=bash, caption={[Cowrie attack to exploit the host machine as a crypto miner]Cowrie attack to exploit the host machine as a crypto miner.}, label={lst:crypto-exploitation}]
mkdir -p /home/osmc/.ssh/
echo ssh-rsa  $RSA_KEY >> /home/osmc/.ssh//authorized_keys
echo '<cmd7uname>'; uname -a
echo '</cmd7uname><cmd7uptime>'; uptime
echo '</cmd7uptime><cmd7w>'; w
echo '</cmd7w><cmd7who>'; who
echo '</cmd7who><cmd7last>'; last
echo '</cmd7last><cmd7lastlog>'; lastlog
echo '</cmd7lastlog><cmd7authkey>'; cat /home/osmc/.ssh//authorized_keys
echo '</cmd7authkey><cmd7lshome>'; ls -la /home
echo '</cmd7lshome><cmd7passwd>'; cat /etc/passwd
echo '</cmd7passwd><cmd7shadow>'; sudo -n cat /etc/shadow
echo '</cmd7shadow><cmd7psfaux>'; ps -faux
echo '</cmd7psfaux><cmd7netstat>'; netstat -npta
echo '</cmd7netstat><cmd7arpan>'; /usr/sbin/arp -an
echo '</cmd7arpan><cmd7ifconfig>'
/usr/sbin/ifconfig
echo '</cmd7ifconfig><cmd7localconf>'; cat /home/ethos/local.conf
echo '</cmd7localconf><cmd7remoteconf>'
cat /home/ethos/remote.conf
echo '</cmd7remoteconf><cmd7rclocal>'
cat /etc/rc.local
echo '</cmd7rclocal><cmd7claymorestub>'; cat /home/ethos/claymore.stub.conf
cat /hive-config/rig.conf; cat /hive-config/wallet.conf
cat /hive-config/vnc-password.txt
echo '</cmd7claymorestub><cmd7claymorezstub>'
cat /home/ethos/claymore-zcash.stub.conf
echo '</cmd7claymorezstub><cmd7sgminerconf>'
cat /var/run/ethos/sgminer.conf
echo '</cmd7sgminerconf><cmd7iptables>'
sudo -n iptables -S  && sudo -n iptables -t nat -S
echo '</cmd7iptables><cmdcrontab>'; crontab -l; echo '</cmdcrontab>'
exit
\end{lstlisting}
\end{figure}

  }
  {
  }

  \IfFileExists{chapters/concept}{
    \chapter{Catching Attackers in Restricted Network Zones}
\label{chap:concept}

The T-Pot identified a flood of threats when it was available on the Internet.
However, capacious networks have separated compartments, and services are usually not directly available without any protection.
Zoning is a well-known method to segment a network.
Heidelberg University applies zoning, and thus, it is an interesting question if an attacker probes services outside or within the network.
This chapter presents a concept that uses a honeypot-like detection tool to detect any dubious packets in the network.
It shows that attacks occurred in a restricted network zone of the Heidelberg University's internal network and contributed to an adaption of the stateless firewall.
Thus, improving the security of the network.

\section{University Network}

Honeypots that are accessible via the Internet receive a broad range of attacks.
As \citet{Spitzner2003} noted, a honeypot is not strictly bound to run in a \ac{dmz} or a network with direct Internet access.
The correct location has to be chosen based on the goals of the honeypot.
For example, one goal could be to catch attackers behind a perimeter firewall to reveal leaks or vulnerabilities.
As described in the chapter before, the honeypot was broadly available on the Internet, and attackers could probe it easily.
It collected on average \numprint{29840} attacks per day, resulting in a total amount of \numprint{607747} attacks.
Zoning a network into logical groups mitigates the risk of an open network.
Thus, the T-Pot would receive significantly fewer attacks in a controlled network zone.
A network infrastructure is segmented into the same communication security policies and security requirements.
For example, the Canadian government created its baseline for infrastructures, called Baseline Security Architecture Requirements for Network Security Zones in the Government of Canada (ITSG-22) \cite{csec2021}.
The four most common zones are:
\begin{enumerate*}[label=(\roman*)]
    \item Public Zone (PZ), which is entirely open,
    \item Public Access Zone (PAZ), which interacts as an interface between the PZ and internal services,
    \item Operation Zone (OZ), which processes sensitive information, and
    \item Restricted Zone (RZ), which includes business-critical services
\end{enumerate*} \cite{csec2021}.
A network zone restricts access and controls data communication flows \cite{csec2021}.

\begin{figure}[htbp]
    \centering
    \includegraphics[width=\textwidth]{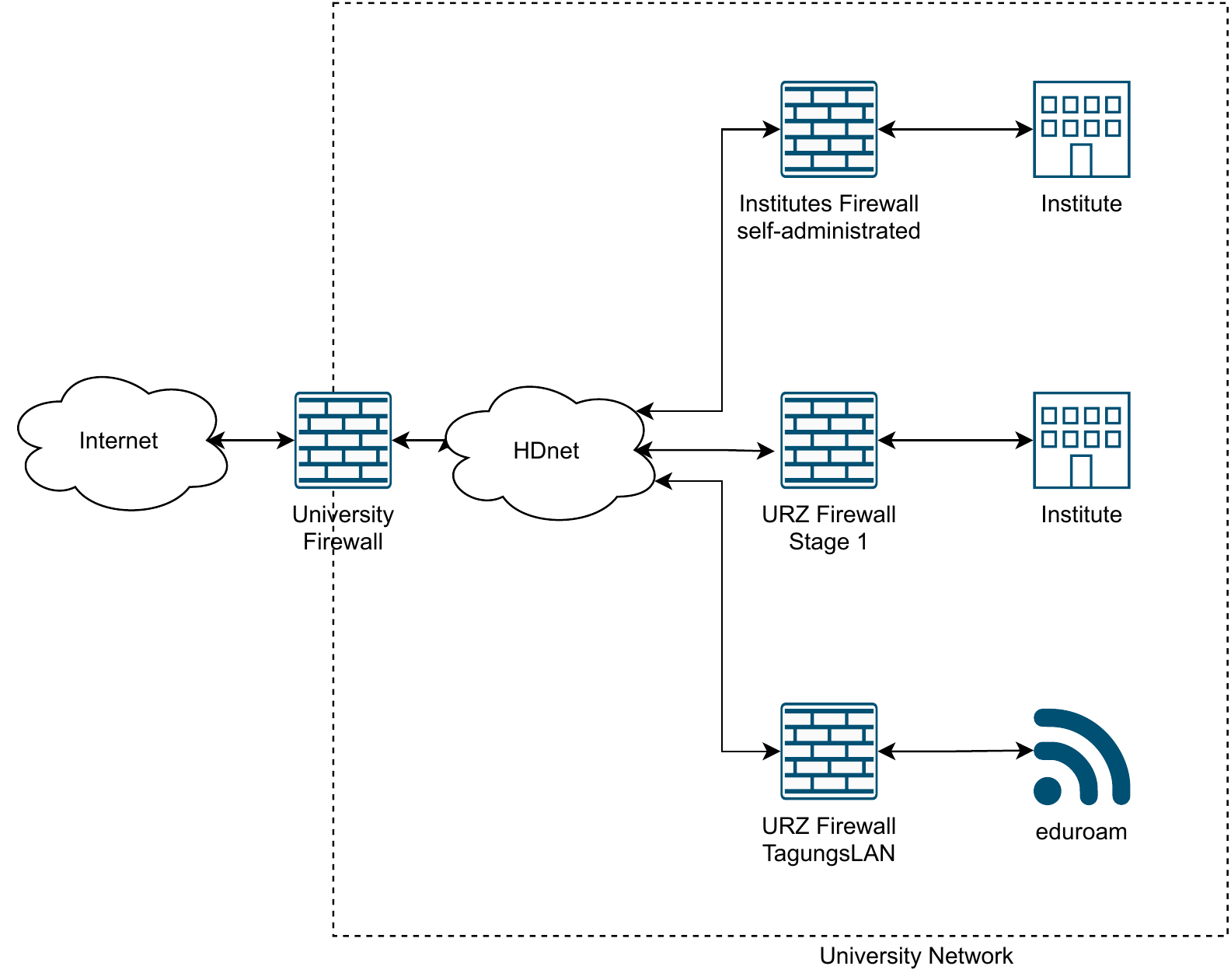}
    \caption[Draft of the University network]{
        Draft of the University network.
        The main doorkeeper is the university firewall.
        The HDnet is the internal network allowing institutes to communicate with each other.
    }
    \label{fig:university-network}
\end{figure}

The network at the Heidelberg University includes a central stateless firewall (\ac{acl}) that enfolds all institutes.
It entails a default blacklist that blocks certain services (such as \ac{smtp} or \ac{snmp}) and a stateless filter provided by BelWÜ.
Each institute can either use a pre-defined stateless firewall provided by the University Computing Center Heidelberg or use a self-administrated firewall inside the network.
\autoref{fig:university-network} outlines the association between these components.
The internal \enquote{HDnet} enables the communication between institutes without leaving the internal network.
Institute firewalls can be set up by each institute and are self-administrated.
They do have the possibility to use \acs{soho} routers\footnote{A \acl{soho} router is a broadband router used in small offices and home offices environments.} to disconnect certain network zones from the network.
It is recommended to configure the global \ac{acl} as a fallback solution in case of any downtime.
The University Computing Center Heidelberg offers stateless firewalls for router interfaces or \acp{vlan}.
This stateless firewall whitelists certain services and splits up into four stages.
Each stage can be individually activated per router interface.
Its key value is to maintain baseline security to avoid misconfigurations and port scans.
\autoref{tab:overview-security-zone} outlines these stages including the \ac{ip} address range.
Before applying one of these zones, the respective network has to oblige to client \ac{ip} addresses below \ipAddress{10.20.30.123/24}.
In addition, \ipAddress{10.20.30.123} is allocated for the gateway.
A network must adhere to these obligations if it applies to any pre-defined stages.

\begin{table}[htbp]
    \centering
    \caption[Overview of firewall stages]{
        Overview of firewall stages at the Heidelberg University.
        As an example, it applies the rules to subnet \ipAddress{10.20.30.123/24}.
        Rules are applied to any subnet.
    }
    \begin{tabularx}{\linewidth}{l|XX}
        \toprule
        \textsc{Name} & \textsc{Description}                      & \textsc{range}                     \\
        \hline
        Stage 0       & Filters broadcast communication           & \ipAddress{10.20.30.123-123/24}    \\
                      & No filtering                              & \ipAddress{10.20.30.123-123/24}  \\
        \hline
        Stage 1       & Allows common network protocol            & \ipAddress{10.20.30.123-123/24}   \\
                      & Allows services                           & \ipAddress{10.20.30.123-123/24} \\
        \hline
        Stage 3       & Internet access only via internal proxies & \ipAddress{10.20.30.123-123/24}   \\
        \hline
        Stage 4       & Only internal network communication       & \ipAddress{10.20.30.123-123/24}   \\
        \bottomrule
    \end{tabularx}
    \label{tab:overview-security-zone}
\end{table}

An interesting question is if attackers have access to restricted zones at the Heidelberg University.
It arises during the research of T-Pot if an adversary would try to probe any hosts in the internal university network.
In order to detect such events, a honeypot-like packet detection application is presented that helps identify any threats in a network.
In addition, it offers to deploy multiple instances and collect their data at a centralized instance.

\section{Honeypot-like Connection Detection Tool}

Recording and investigating connection attempts assimilates new honeypots.
Respectively, a new honeypot-like detection tool called MADCAT will be presented.
MADCAT has been developed by the BSI and helps to log any connection attempt being made on the host machine.
The acronym MADCAT stands for \textit{Mass Attack Detection Connection Acceptance Tools}.
It works as a honeypot-like detection application with a low-interaction level.
Its key idea is to log every connection attempt and further process it to retrieve credentials or shell exploitation.
\autoref{fig:madcat-architecture} gives an insight into how MADCAT works.
It runs on an Ubuntu distribution, either $18.04$ or $20.04$, and has been tested on Ubuntu $18.04$.
It processes packets from any interface that has been configured.
As an example, it could process Ethernet and wireless packets.
MADCAT itself consists of six independent modules for \ac{tcp}, \ac{udp}, \ac{icmp}, and raw packets that communicate with each other through a pipeline.
A module analyzes packets and logs the results in a queue.
In addition, \ac{udp} and \ac{tcp} offer a proxy to tunnel packets to another service.
Every 5 seconds \ac{tcp} postprocessor reads the newly arrived \ac{tcp} packets and processes them accordingly.
It resolves packets to log data, including source \ac{ip} address, protocol, and event type.
The enrichment processor is the final process step.
Its purpose is to log all queue-written packets in a specified format for further analysis.
The key idea of MADCAT is to get an insight into whether attackers have access to a particular network.
In contrast to T-Pot, the concept does not know what specific attacks are operated on the honeypot.
Instead, it ensures that no one else than authorized users has access.
Especially in high confidential areas, no attacker should be capable of sending even a single packet to a host in the network.
The vast range of honeypots does not provide tracking packets on a detailed level.

\begin{figure}[htbp]
    \centering
    \includegraphics[width=\textwidth]{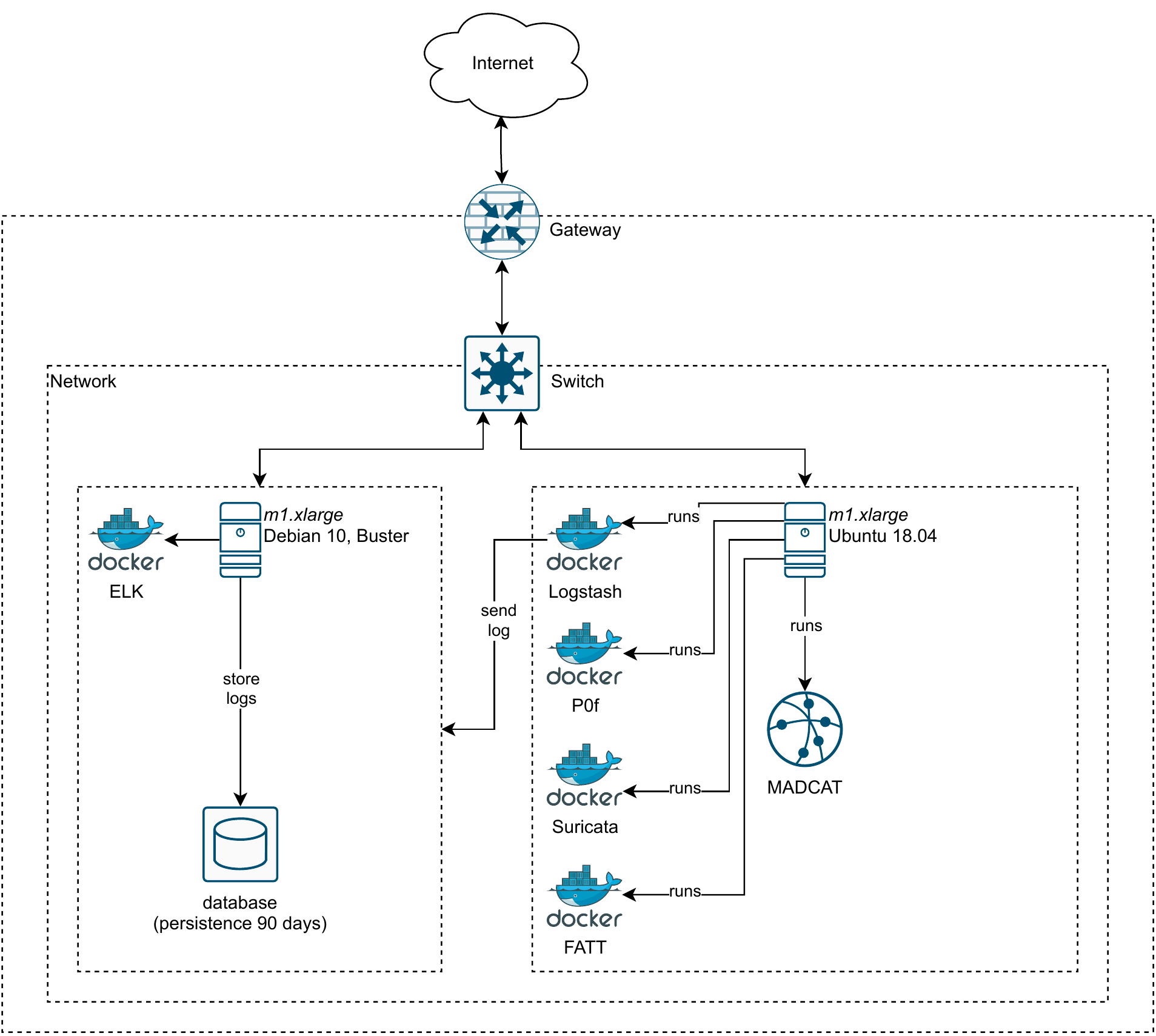}
    \caption[Concept to detect connection attempts]{
        Concept to detect connection attempts.
        It has been drafted to work in various scenarios.
        Kibana and Elasticsearch are deployed in heiCLOUD.
    }
    \label{fig:heicat-concept}
\end{figure}

In addition, a T-Pot instance will be deployed to have comparison data to the new concept.
It focuses on the \ipAddress{10.20.30.123/24} and \ipAddress{10.20.31.123/16} subnet.
The \ipAddress{10.20.30.123/24} subnet is used within University Computing Center Heidelberg building.
Every client in the building has a compelling connection in this subnet.
Otherwise, an Internet connection would not be feasible.
The subnet \ipAddress{10.20.31.123/16} connects clients to \enquote{eduroam}\footnote{The eduroam is an international Wi-Fi internet access point for researchers.}.
Like the four stages of the institute firewall, the \enquote{eduroam} network, also called \enquote{Tagungslan}, builds various permits into the subnet.
One essential difference is that services like \ac{smtp} and \ac{http} are not allowed, so attackers cannot deploy traps for users.
Moreover, each client is encapsulated in its subnet, which disables communication to other clients.
The instances are located in the building with \ac{ip} addresses \ipAddress{10.20.30.123} and \ipAddress{10.20.30.123}.
\autoref{fig:heicat-concept} outlines the concept using MADCAT and a separate instance to visualize our data.
The first instance with \ac{ip} address \ipAddress{10.20.30.123} provides Kibana and Elasticsearch to visualize and crawl logs.
The honeypot with \ac{ip} address \ipAddress{10.20.30.123} consists of MADCAT in conjunction with P0f, Suricata, and FATT.
Like T-Pot, it uses Logstash to forward data to Elasticsearch.
One benefit is the centralized approach to store data.
This allows to deploy more instances to randomly collect data from other zones.

\begin{figure}[htbp]
    \centering
    \includegraphics[width=\textwidth]{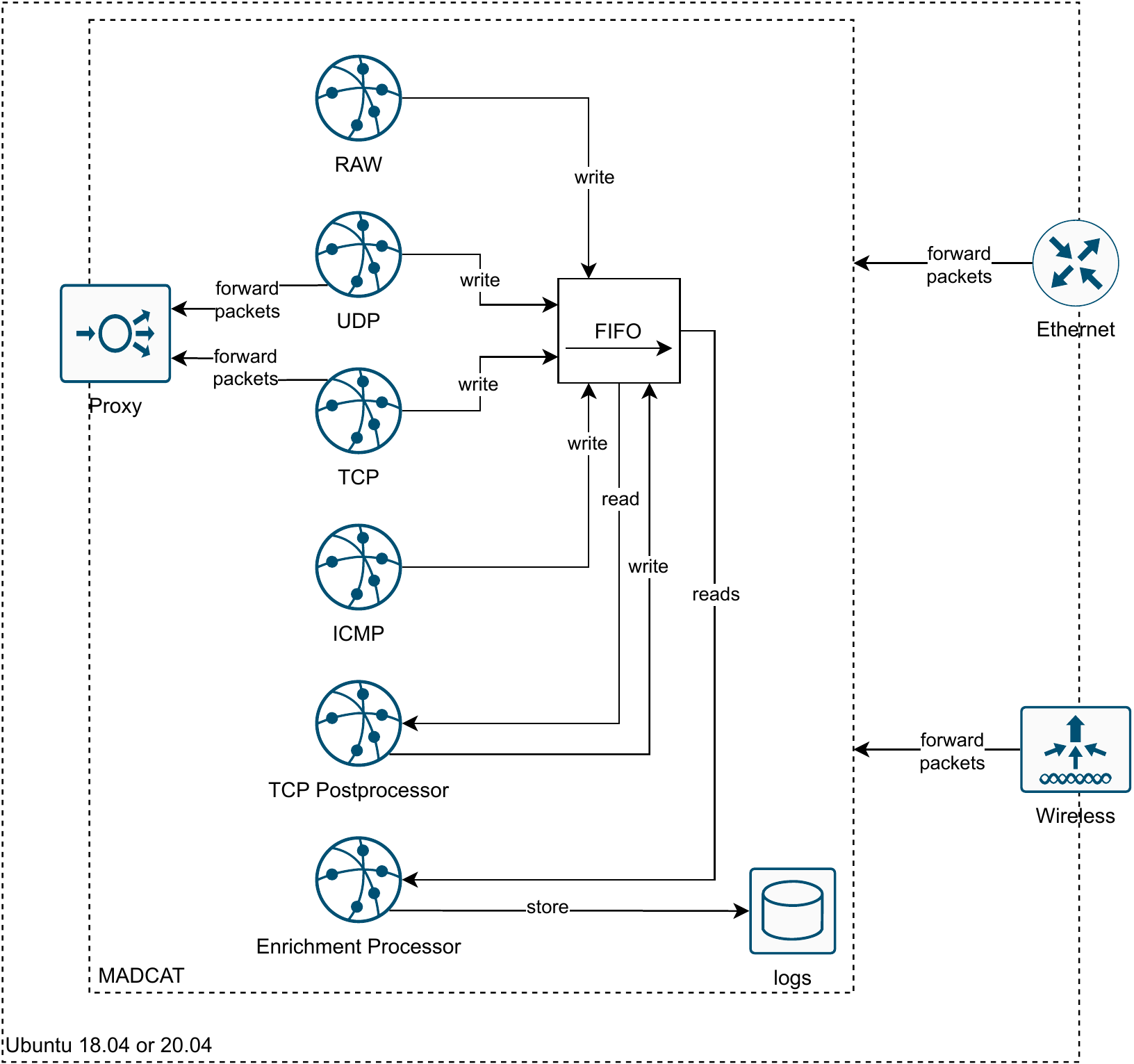}
    \caption[Visualization of the MADCAT packet flow]{
        Visualization of the MADCAT packet flow starting at the network interface.
        The Ethernet and wireless interface forwards packets to the desired module.
    }
    \label{fig:madcat-architecture}
\end{figure}

\section{Results}

MADCAT (\nth{28} of October till \nth{18} of November) and T-Pot (\nth{16} of November till \nth{7} of December) have been deployed for three weeks.
All instances had a connection to both subnets.
First, the results obtained in the subnet \ipAddress{10.20.30.123/24} will be presented, closing up with the ones claimed in the eduroam network.

In total, MADCAT received \numprint{35372} packets.
Overall, the modules \ac{tcp} ($66.62\%$) and raw ($33.26\%$) received the majority of all connection attempts.
The minority with less than one percent are suspicious packets with individual \ac{tcp} flags like reset or syn set. 
On the contrary, it could not identify any harmful activity based on these packets.
Overall, ConPot ($56.98\%$), Honeytrap ($31.35\%$), and Dionaea ($7.09\%$) received most of the attacks with a total number of $437$.
Interestingly, it could identify \ac{snmp} connections that are used by print servers to discover printers.

\begin{figure}[htbp]
    \centering
    \includegraphics[width=\textwidth]{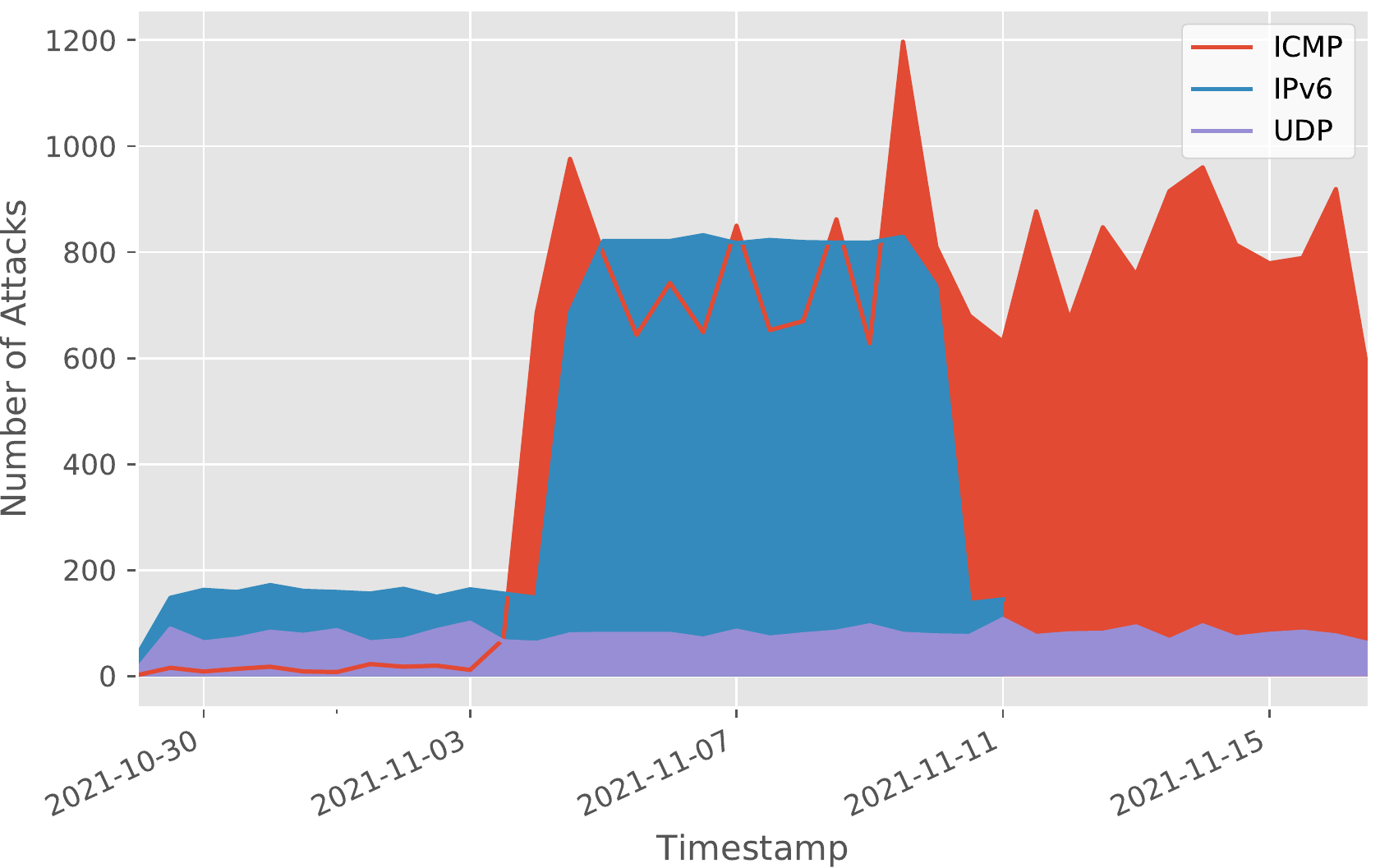}
    \caption[Protocol distribution of MADCAT]{
        Protocol distribution of MADCAT. \ac{icmp}, IPv6, and \ac{udp} are the most used protocols.
        Timestamp; \nth{28} of October to \nth{18} of November.
    }
    \label{fig:madcat-protocols}
\end{figure}

% ====================================================================================================================
% Madcat
\autoref{fig:madcat-protocols} shows the protocol distribution indicating a high amount of \ac{icmp} and IPv6 packets.
Only $11.59\%$ of all \ac{ip} address reputations could be resolved, splitting up into known attacker ($11.26\%$), mass scanner ($0.14\%$), bad reputation ($0.12\%$), and tor exit node ($0.08\%$).
Focusing on \ac{tcp} packets, $88.3\%$ are known attackers with source port 113 as the primary target.
The port 113 is officially known as the \ac{ident}\cite{rfc1413} used for identification/authorization on a remote server such as \ac{pop}, \ac{imap}, and \ac{smtp}.
A potential leak that allows adversaries to send \ac{ident} requests to the network could be spotted by comparing the results with the stateless firewall settings.
Decoding the payload of these \ac{tcp} packets shows that attackers instead used this port to get an \ac{smb} connection than deploying \ac{ident} protocol attacks.
It identified attempts to acquire an SSH session using \ac{smb} and \ac{sip} connection attempts and various \ac{http} requests.
For example, two payloads that have been sent to the instance show probing actions.
\autoref{lst:sip-exploitation} outlines a \ac{sip} probe that checks if any \ac{voip} service is active by answering the request packet.
Next, \autoref{lst:smb-exploitation} shows an \ac{smb} probe trying to achieve the same.
The \ac{ip} address reputation could help answer if a real user or an attacker sends these packets.
Both \ac{ip} addresses in this example were resolved as a known attacker; thus, it identified them as a probe packet before executing their attack.
A vital security interest in port 113 is negligible; however, the concept helps to detect such leaks, especially when stateless firewalls are the main doorkeeper for packets.

\begin{figure}[htbp]
    \centering
    \includegraphics[width=\textwidth]{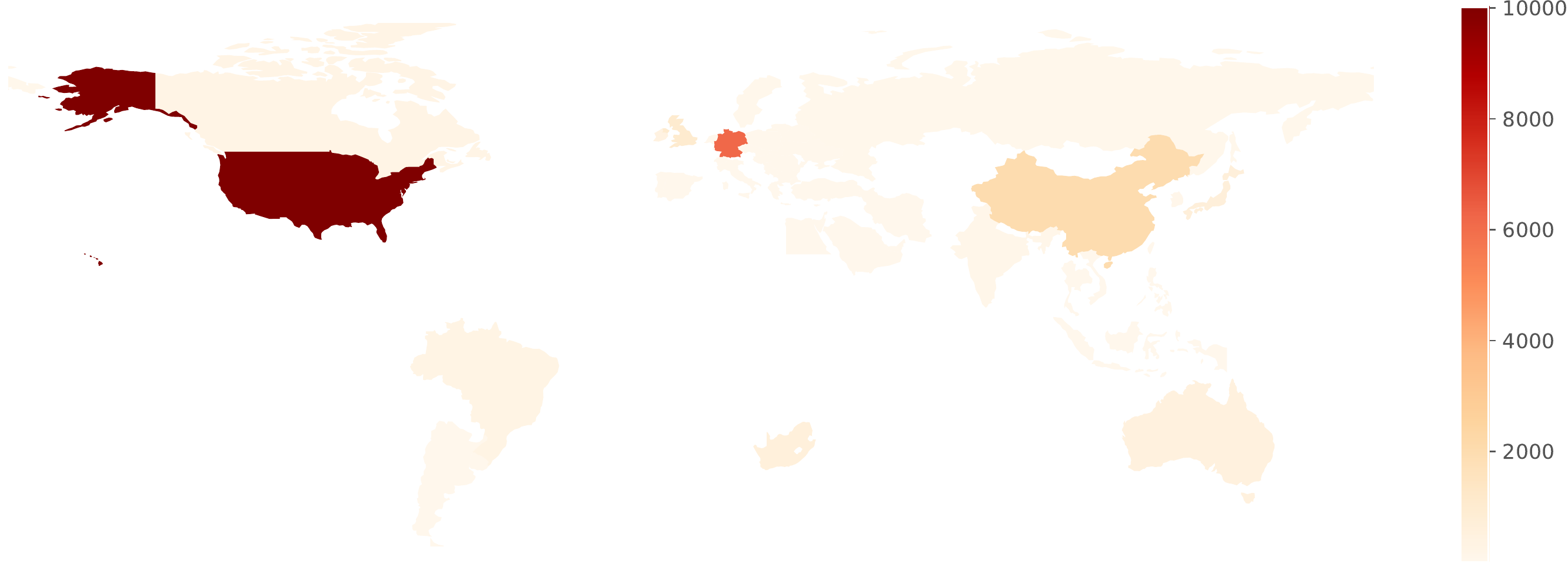}
    \caption[Attack distribution of MADCAT]{
        Attack distribution of MADCAT.
        The USA, Russia, and China are the most attacking countries.
        Timestamp; \nth{28} of October to \nth{18} of November.
    }
    \label{fig:madcat-attack-distribution}
\end{figure}

% ====================================================================================================================
% Location
\autoref{fig:madcat-attack-distribution} shows the attack distribution indicating the origin of an \ac{ip} address.
Most of the connections originate from the United States, Germany, and China.
As shown beforehand in \autoref{chap:cloud-security}, geographical information only outlines the last known location of a node.
Like the results in heiCLOUD, it can be assumed that this information is not reliable as an indicator of where attacks occur.
Nevertheless, it is interesting to see where the last node originated from.

\begin{figure}[htbp]
    \centering
    \includegraphics[width=\textwidth]{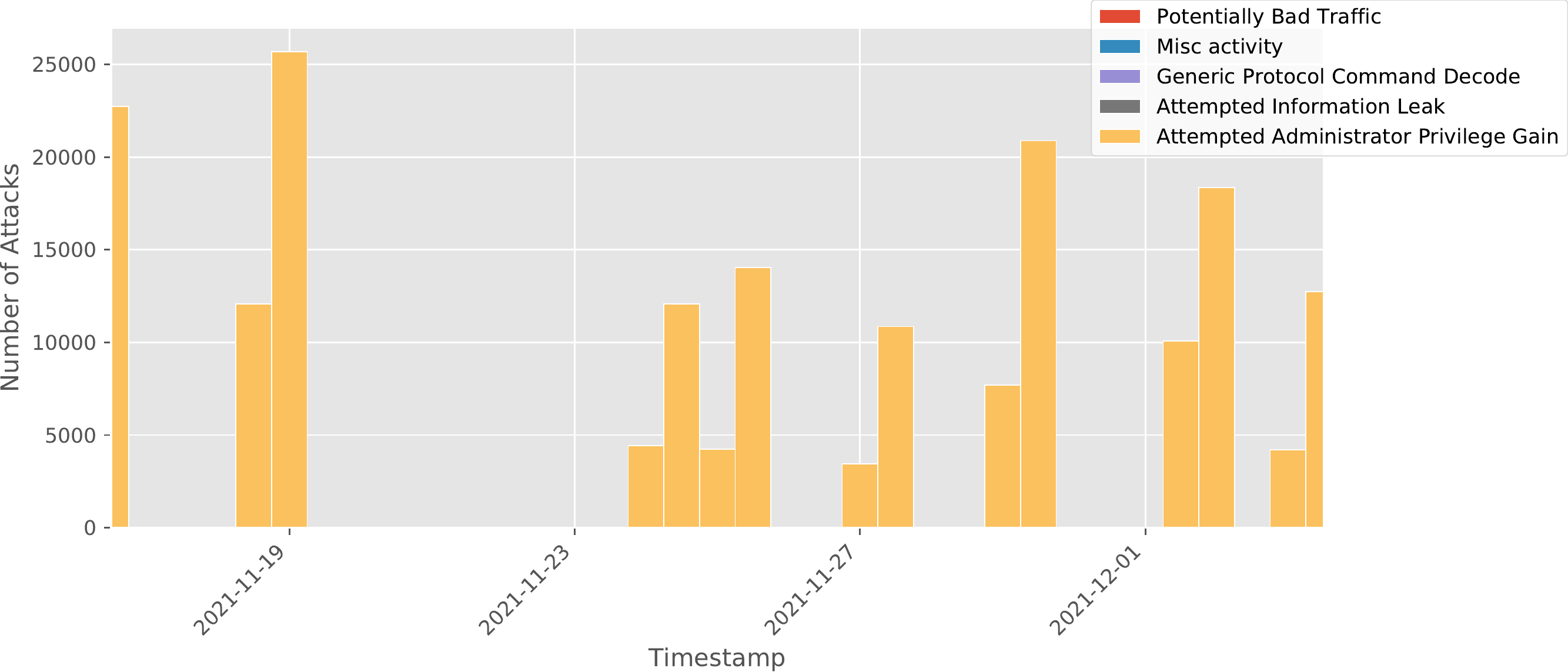}
    \caption[Suricata results of T-Pot]{
        Suricata results of T-Pot.
        Timestamp; from \nth{16} of November to \nth{7} of December.
    }
    \label{fig:suricata-distribution}
\end{figure}

% ====================================================================================================================
% Suricata
Suricata identified odd behaviors in the network (\autoref{fig:suricata-distribution}).
In total, it detected \numprint{292953} alerts and \acsp{cve}.
Besides minor alerts like \textsc{nmap} scans, Suricata registered alerts in \ac{snmp} requests, \ac{tcp} stack, and Wind River VxWorks.
CVE-2020-11899 \cite{CVE-2020-11899} accounts nearly $73.35\%$ with a total number of \numprint{214879}.
This \acs{cve} is one of 19 others forming the \textit{Ripple20} vulnerability in the low-level \ac{tcp}/\ac{ip} library developed by Treck, Inc.
One of the Track \ac{tcp}/\ac{ip} stack tasks is to reassemble fragmented packets.
Whenever a fragmented packet arrives, the stack tries to validate the total length in the \ac{ip} header.
If the total length is not correct, it trims the data.
However, this leads to inconsistency, and thus, resulting in a buffer overflow when someone sends fragmented packets through a tunnel.
A detailed description of the vulnerability can be found in \cite{ripple20}.
An adversary could send malformed IPv6 packets that cause an Out-of-bounds Read, resulting in potential remote code execution.
Only \ac{tcp}/\ac{ip} stack versions until $6.0.1.66$ are affected by this vulnerability.
Nevertheless, the tremendous alerts show the importance of adapting the IPv6 permits.
The second most recorded vulnerability with the highest score is CVE-2002-0013 \cite{CVE-2002-0013} that allows remote attackers to cause a denial of service or gain privileges in the SNMPv1 protocol.
The root cause for the \acs{cve} alert is the usage of the default public community for broadcast requests instead of configuring a private community with mandatory authentication.
To compromise \ac{snmp}, attackers have to have access to the network.
However, the university firewall blocks \ac{snmp} port 161 and 162 for \ac{tcp} and \ac{udp}, thus, restricting any access from outside.
If adversaries plan to deploy an attack on the \ac{snmp} protocol, they need to have a connection to the internal network.
Acquiring such a connection is rather hard to accomplish without any credentials.
On the contrary, all connection attempts registered by the concept have been made within the network, and they do reflect a normal \ac{snmp} communication.
Lastly, Wind River VxWorks 6.9.4 and vx7 in CVE-2019-12263 \cite{CVE-2019-12263} cause a buffer overflow due to the underlying \ac{tcp} component that results in a race condition.
Each connection attempt with CVE-2019-12263 is originated from Russia.
Hence, the assumption is that the source \ac{ip} address maliciously intended to send an urgent flag.
For the other \acsp{cve}, the \ac{ip} reputation could not be resolved.

\begin{figure}[htbp]
\begin{lstlisting}[language=bash, caption={[MADCAT connection attempt to exploit SIP connection]MADCAT connection attempt to exploit \ac{sip} connection. Received on the \nth{16} of November. \ac{ip} reputation: known attacker. Location Germany.}, label={lst:sip-exploitation}]
OPTIONS sip:nm SIP/2.0 Via: SIP /2.0/TCP nm;
branch=foo From: <sip:nm@nm>;
tag=root To: <sip:nm2@nm2> Call-ID: 50000 CSeq: 42 OPTIONS Max-Forwards: 70 Content-Length: 0 Contact: <sip:nm@nm> Accept: application/sdp
\end{lstlisting}
\end{figure}

\begin{figure}[htbp]
    \centering
    \includegraphics[width=\textwidth]{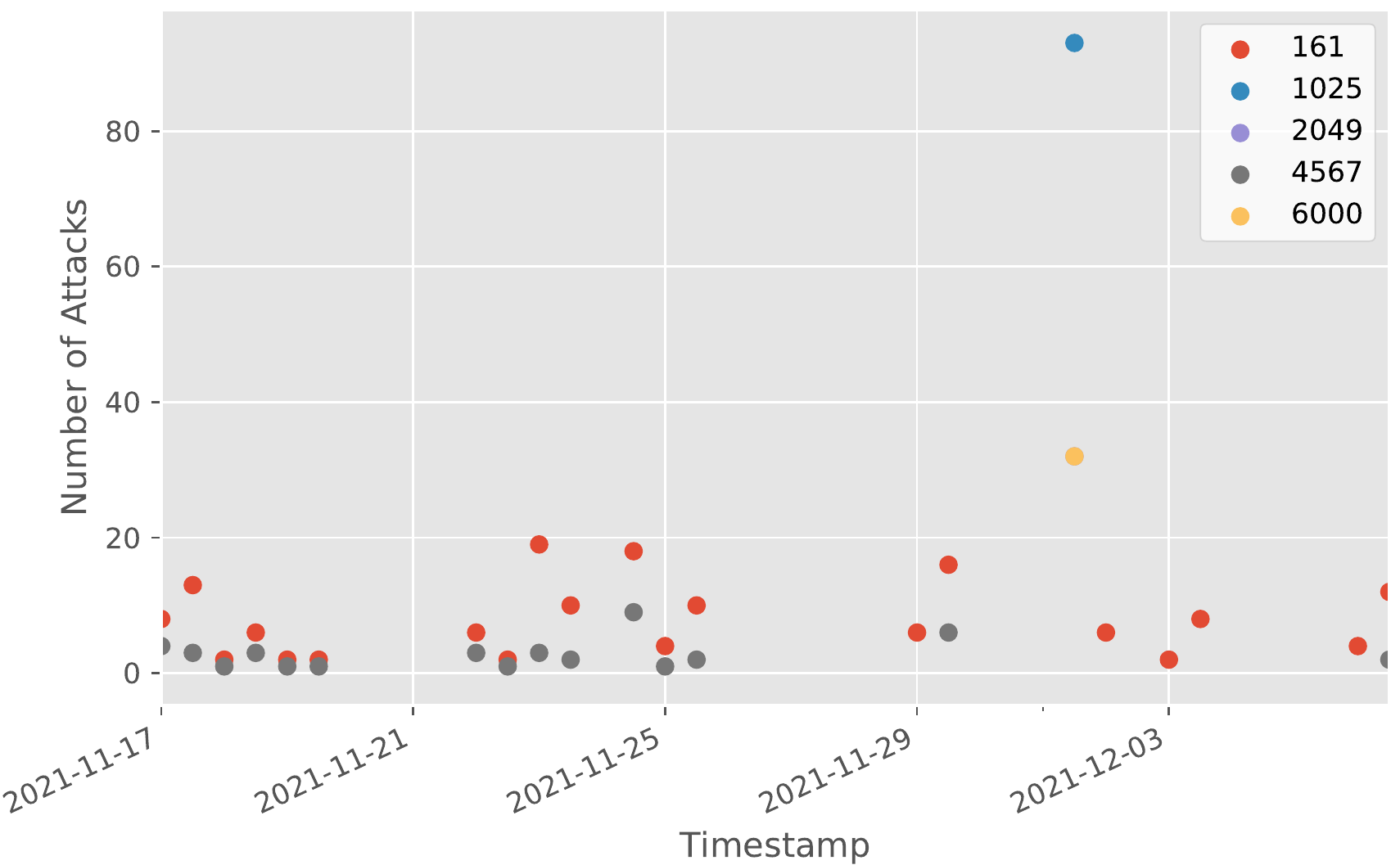}
    \caption[Attack port histogram of T-Pot]{
        Attack port histogram of T-Pot.
        Timestamp; from \nth{16} of November to \nth{7} of December.
    }
    \label{fig:port-histogram}
\end{figure}

% ====================================================================================================================
% T-Pot
Results from the T-Pot instance are exiguous, and in short, no real attacks such as shell exploitation have been performed.
All connection attempts originated from Germany within the same network and are made on ports $161$ and $4567$.
Conpot registered minor SNMPv2 Get, SNMPv1 Get, and GetNext requests.
A possible attack vector could be an \ac{snmp} reflection/amplification attack.
As previously discussed, the assumption is that devices within the network have a misconfigured printer and send broadcast requests frequently to find the machines.
This \ac{snmp} requests affiliate with day-to-day traffic in an internal network, and thus, are not suspicious.
The second most attacked honeypot is Honeytrap which received numerous packets on different ports, whereas $39\%$ evince an empty payload.
All of these received packets have a resolved \ac{ip} address in the subnet \ipAddress{10.20.30.123/16}.
It remains unclear if these connections are malicious or are acquired by accident.
Investigating the payload of outliers does not confirm the assumption of a vicious intention.
Thus, declaring these results as negligible.
Overall, most of the connection attempt received by the instance are from these \ac{ip} addresses: \ipAddress{10.20.30.123}, \ipAddress{10.20.30.123}, and \ipAddress{10.20.30.123}.

\begin{figure}[htbp]
\begin{lstlisting}[language=bash, caption={[MADCAT connection attempt to exploit SMB connection]MADCAT connection attempt to exploit \ac{smb} connection. Received on the \nth{16} of November. \ac{ip} reputation: known attacker. Location Germany.}, label={lst:smb-exploitation}]
PC NETWORK PROGRAM 1.0 MICROSOFT NETWORKS 1.03 MICROSOFT NETWORKS 3.0 LANMAN1.0 LM12X002 Samba NT LANMAN 1.0 NT LM 0.12.
\end{lstlisting}
\end{figure}

Lastly, the results from the eduroam network are considered.
Neither T-Pot nor MADCAT could identify any significant behavior for three weeks.
Unlike the subnet \ipAddress{10.20.30.123/24}, the honeypot did not register any suspicious packets, \ac{tcp} flags, or other \acp{cve}.
In retrospect, the eduroam configuration has been shown to work as designed.
Thus, the client seemed to be encapsulated from others and received no other packets.

Besides the subtle output it has received, the results have given an insight into the value of honeypots in a restricted network zone.
For Heidelberg University, using honeypots to evaluate their stateless firewall has never been considered.
The initial concept has shown that it delivered minor findings in the subnet \ipAddress{10.20.30.123/24} with stage 1 firewall.
As a result, the port 113 used for the \ac{ident} protocol will be removed in the future to reduce the attack surface, thus, contributing to the firewall definition.
Overall, the two instances received numerous packets containing interesting payloads.
Compared to the T-Pot, which has been used in heiCLOUD, results are as expected delicate, and data analysis turns out to be more detailed.
The statement from \citet{Spitzner2003} that honeypots only receive little input and nearly every input is suspicious matches the results only halfway.
As shown beforehand, the results are dramatically little; however, only a few requests seemed suspicious.
Nonetheless, the initial question of whether attackers have access to the restricted network zone at the Heidelberg University has been answered.

\section{Discussion}

This chapter has shown that honeypots help find potential leaks in restricted network zones.
Though, it remains questionable if the concept can deliver accurate results.
The instance has been running for three weeks in the two different subnets.
The honeypot has to be detected as a vulnerable target to deliver meaningful data.
However, it could not detect any large scans on the instance; thus, it is very likely that either an attacker could not find the instance or no one had any access.
In the eduroam network, large scans are negligible due to the firewall permits.
It can be assumed that the results are accurate and do not show any discrepancy.
Considering the subnet with stage one institute firewall, it identified attacks on port 113, resulting in an adaption of the stage one permits.
On the contrary, it could not register any other odd packets on other ports.
A detailed investigation could resolve whether the honeypot is available to attackers.
A misconfiguration of the university firewall has been detected which proves this assumption.

In December, from the \nth{21} to the \nth{23}, a misconfiguration of the university firewall resulted in a flood of attacks.
In total, the T-Pot instance received \numprint{46328} attacks in three days.
It turns out that five ports were open during that time, allowing attackers to probe the instance (\autoref{fig:tpot-misconfig-port-histogram}).
The most attacked honeypots are RDPY ($58.58\%$), Honeytrap ($24.53\%$), and Cowrie ($11.69\%$).

\begin{figure}[htbp]
    \centering
    \includegraphics[width=\textwidth]{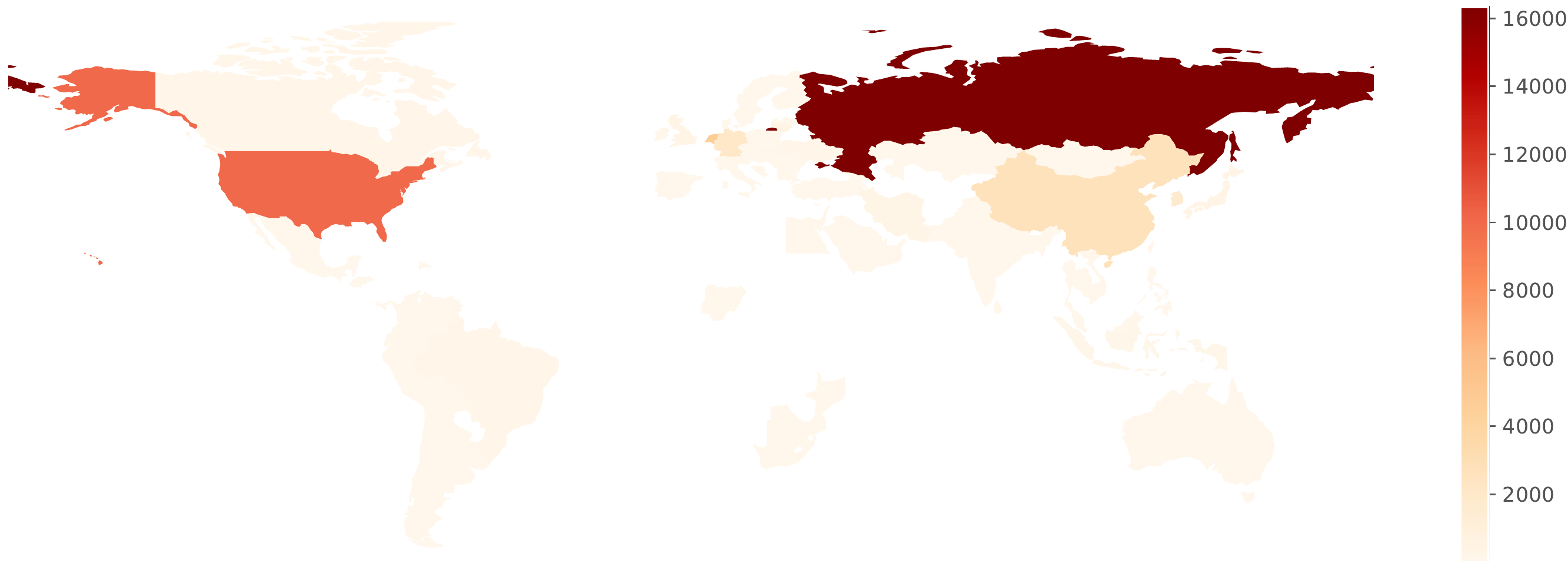}
    \caption[Attack distribution of T-Pot]{
        Attack distribution of T-Pot.
        USA, Russia, China, and Germany are the most attacking countries.
        Timestamp; from \nth{21} of December to \nth{23} of December.
    }
    \label{fig:tpot-misconfig-map}
\end{figure}

Like the geographical information of other honeypots, most of the connections originate from the United States, Russia, and China (\autoref{fig:tpot-misconfig-map}).
These similarities indicate a bias of the origin even though the location information is not reliable.

\begin{figure}[htbp]
    \centering
    \includegraphics[width=\textwidth]{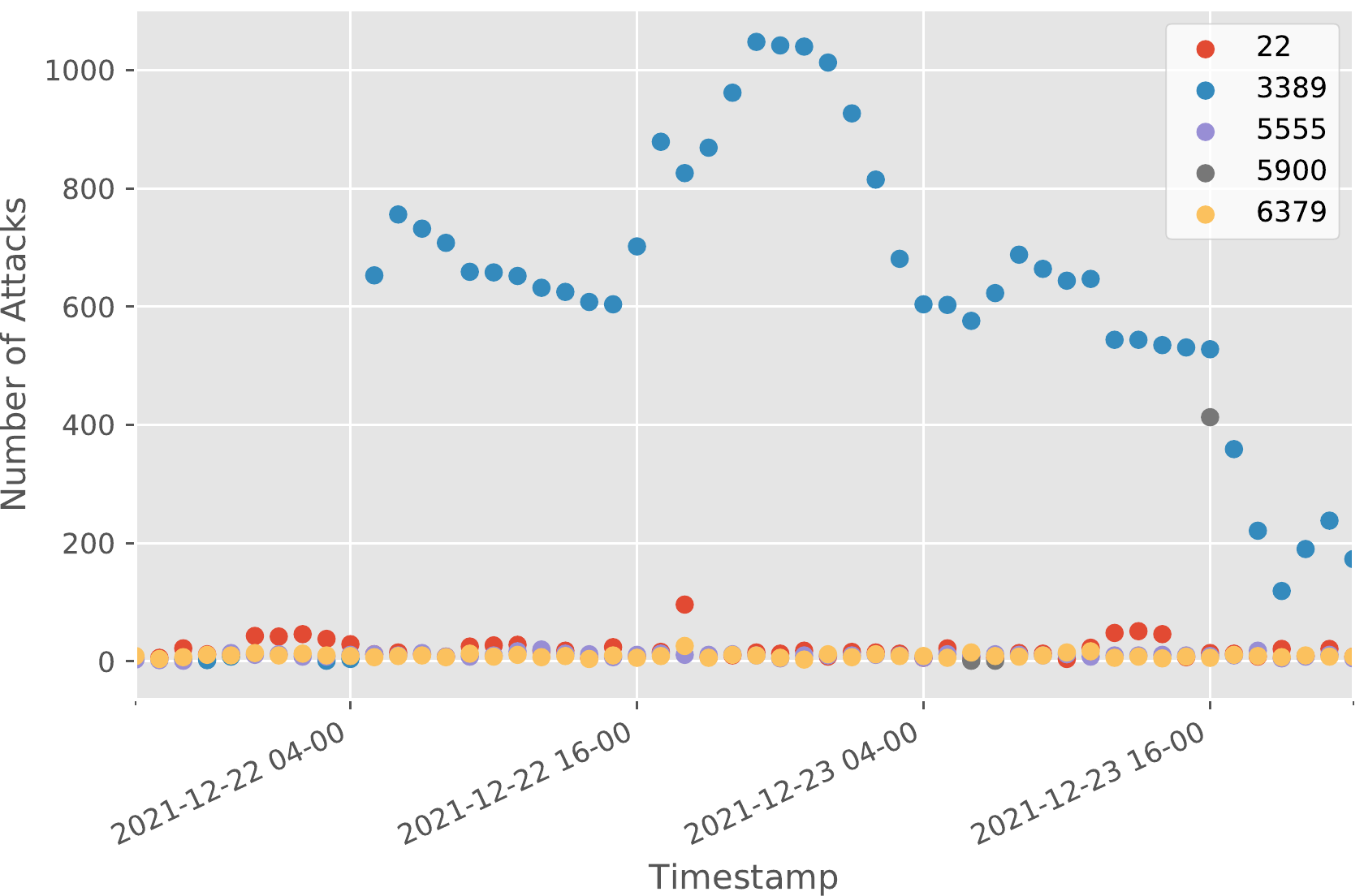}
    \caption[Attack port histogram of T-Pot]{
        Attack port histogram of T-Pot.
        Timestamp; from \nth{21} of December to \nth{23} of December.
    }
    \label{fig:tpot-misconfig-port-histogram}
\end{figure}

On RDPY and Honeytrap, many connection attempts on various ports have been made.
Based on the Suricata results, adversaries tried to gain administrator privileges.
For Cowrie, attackers tried to log in and execute commands by brute force or guesswork.
Moreover, the latest crypto-mining malware has been used, which resembles the findings of other honeypots.
These results overlap strongly with those obtained by the T-Pot instance in heiCLOUD.

The firewall administrator stated that the misconfiguration was fixed on the \nth{23} of December, resulting in a decrease in attacks on the T-Pot instance.
These results have successfully answered our discussion of whether an attacker could detect the host machines at the university building.
It clearly shows that attackers scan these \ac{ip} address ranges and send malicious packets whenever they can.

  }
  {
  }

  \IfFileExists{chapters/experiment}{
    \chapter{Mitigate Fingerprint Activities of Honeypots}
\label{chap:fingerprinting}

\epigraph{There is a generic weakness in the current generation of low- and medium-interaction honeypots because of their reliance on off-the-shelf libraries to implement large parts of the transport layer.}{\textit{Alexander Vetterl}}

Detecting honeypots before launching attacks helps to avoid the disclosure of information.
\Autoref{chap:cloud-security} has shown that bot activities are on the rise, and more attacks than ever have been launched.
However, the vast majority of attacks have been identified to be repetitive.
This chapter conducts two experiments on whether it is possible to fingerprint honeypots.
First, it reproduces the findings from \citet{vetterl2020} to prove the initial question if any fingerprint activity is feasible.
Consequently, it presents a concept to disguise Cowrie and verify this assumption with an experiment.

\section{OpenSSH}
\label{sec:openssh}

OpenSSH is one of the most used applications that enables SSH.
Before proceeding with generic weaknesses of honeypots, a short intermezzo about OpenSSH is given.

\begin{figure}[htbp]
    \centering
    \includegraphics{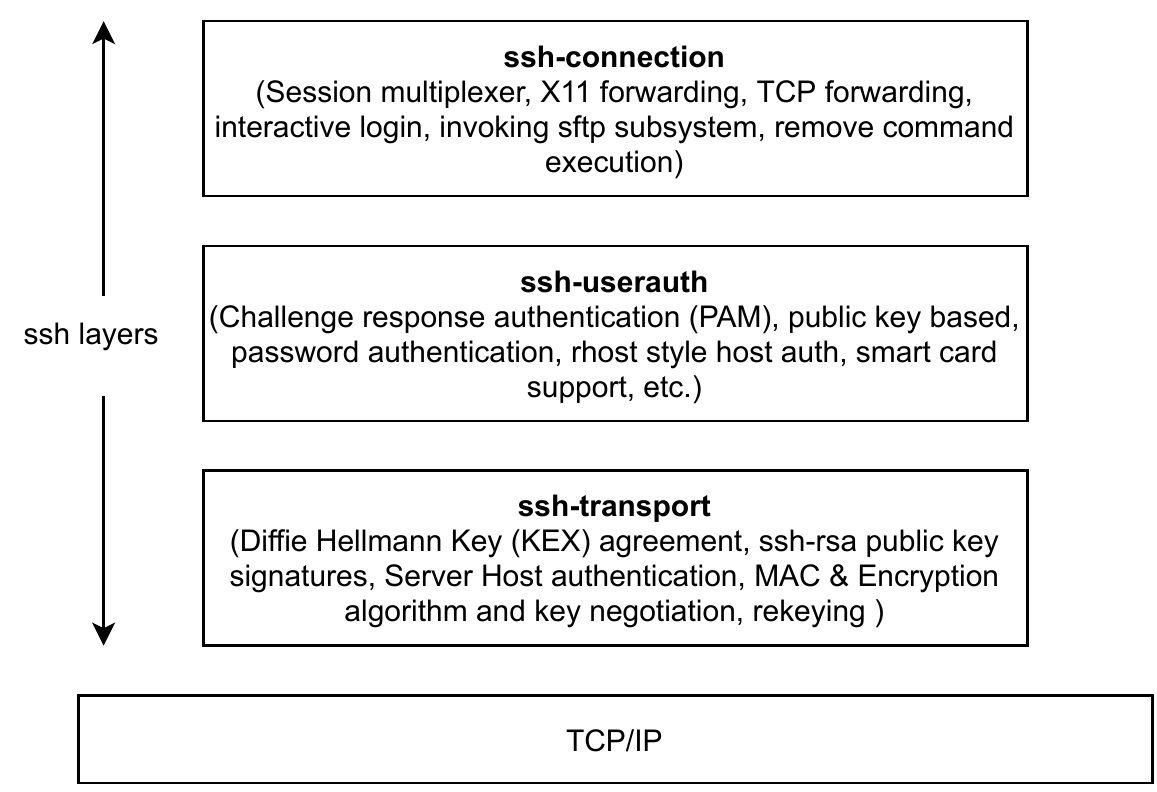}
    \caption[OpenSSH architecture]{
        OpenSSH architecture (derived from \cite{openssh2007}).
        The ssh-transport layer builds the foundation for the other layers on top.
        In addition, each layer lists examples of functionalities that it supports. 
    }
    \label{fig:openssh-architecture}
\end{figure}

OpenSSH consists of three major layers, namely \verb|ssh-connection|, \verb|ssh-userauth|, and \verb|ssh-transport| (\autoref{fig:openssh-architecture}) \cite{openssh2007}.
The last layer is the most important because it provides the basic functionalities for crypto operations, such as key exchange and encryption.

The first layer is responsible for authenticating the user to the SSH daemon, namely sshd.
Based on two-way authentication, the client authenticates the SSH daemon with the help of the \verb|ssh-transport| \cite{openssh2007}.
Finally, a secure connection is established, and the key exchange is done.
The next step is to authenticate the user of the client.
It offers authentication methods such as username/password, public key, or smart-card authentication \cite{openssh2007}.
If the \verb|ssh-userauth| layer is successful, it will establish a secure channel through the \verb|ssh-connection| layer \cite{openssh2007}.
Each session is handled in a so-called channel.

The \verb|ssh-connection| layer handles multiple sessions simultaneously over a single \verb|ssh-userauth| layer with the TCP/IP layer below \cite{openssh2007}.
It is responsible for executing arbitrary commands, forwarding X11 connections, establishing VPN tunnels, and more.

In addition, OpenSSH has built-in features such as keeping alive messages and redirecting stdin to \textit{/dev/null} for specialized X11 windows \cite{openssh2007}.

\autoref{fig:openssh-flow-diagram} outlines a sample session between a client and a server.
The key exchange initialization is the first message between them to negotiate all ciphers and keys for communication.
For this chapter, no other than the key exchange initialization message will be considered.

\section{Preliminary Work}

Attackers have a strong motivation to reveal honeypots before launching an attack.
Without any protection, attackers would disclose their methods, and thus, newly developed attacks would become useless.
As shown in \autoref{chap:cloud-security}, attackers do try to get information about the host system.
\citet{vetterl2020} discussed various methods of fingerprinting; however, executing commands in a shell and examining the response leaves precarious information to the honeypot itself.
His technical report evaluated methods to detect honeypots at the transport level.
As stated, the value of a honeypot would be merely zero if detection on transport level would work.
He presents fingerprinting methods for SSH, Telnet, and HTTP/Web.
Due to the complexity of each method, this section focuses on SSH fingerprinting using the honeypot Cowrie.

The idea to detect SSH honeypots is to look for deviations in the response.
Therefore, \citet{vetterl2020} sends a set of probes $P = \{P_1, P_2, \dots, P_n\}$ to a given set of implementations of a network protocol $I = \{I_1, I_2, \dots, I_n\}$ and stores the set of responses $R = \{R_1, R_2, \dots, R_n\}$.
He calculated the cosine similarity coefficient $C$ for the given set of responses.
The goal is to find the best $P_i$ where the sum of $C$ is the lowest.
\autoref{fig:draft-cosine-similarity} presents these steps.

The cosine similarity outputs the similarity between vectors of numerical attributes.
It is widely used in text semantics to measure the similarity of sets of information such as two sentences.
\citet{vetterl2020} outlines that it can be used in \enquote{traffic analysis to find abnormalities and to measure domain similarity}.
Mathematically, it computes the angle between two vectors.
For each set of information $A$, we create a vector $D_A$.
Referring to the use case with SSH, we use the response from the server as information $A$.
If $\theta$ is the angle between $D_A$ and $D_B$, then:

\begin{equation} \label{eq:cosine-similarity}
    \cos \theta = \frac{D_A \cdot D_B}{\|D_A\| \|D_B\|}
\end{equation}

where \enquote{$\cdot$} is the dot product obtained by:

\begin{equation}
    D_A \cdot D_B = \sum_{i=1}^{n} (D_{A_i} \times D_{B_i})
\end{equation}

and $\|D_A\|$ (resp. $\|D_B\|$) is the Euclidean norm, obtained by $\sqrt{\sum_{i=1}^{n} D_{A_i}^2}$ (resp. $\sqrt{\sum_{i=1}^{n} D_{B_i}^2}$).
The values of vectors are non-negative.
The similarity between items is the value $\cos \theta$, $\cos \theta = 1$ indicates equality.

\begin{figure}[htbp]
    \centering
    \includegraphics{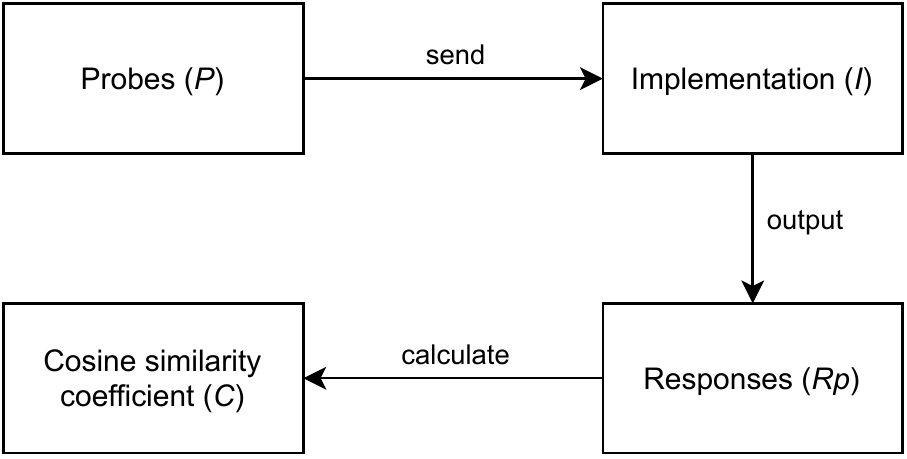}
    \caption[Process to obtain the cosine similarity coefficient]{
        Process to obtain the cosine similarity coefficient (derived from \cite{vetterl2020}).
    }
    \label{fig:draft-cosine-similarity}
\end{figure}

In order to find the best $P_i$ for SSH, \citet{vetterl2020} first created different SSH version strings based on the format: \verb|SSH-protoversion-swversion SP comment crlf|.
He used different lower and upper case variations, 12 different \verb|protoversions| ranging from $0.0$ to $3.2$, \verb|swversion| set to \verb|OpenSSH| or empty string, \verb|comment| set to \verb|FreeBSD| or empty string, and \verb|crlf| to either \verb|\r\n| or empty string.
In total, summing up to \numprint{192} client version strings.
Second, he created different \verb|SSH2_MSG_KEXINIT| packets with 16 key-exchange algorithms, two host key algorithms, 15 encryption algorithms, 5 \ac{mac} algorithms, and three compression algorithms.
In total, he sent \numprint{58,752} key exchange initialization messages.
Combining them with the \numprint{192} client versions, he ended up sending \numprint{157925376} packets.
The version string \verb|SSH-2.2-OpenSSH \r\n| and the \verb|SSH2_MSG_KEXINIT| packet including \textit{ecdh-sha2-nistp521} as the key-exchange algorithm, \textit{ssh-dss} as host key algorithm, \textit{blowfish-cbc} as encryption algorithm, \textit{hmac-sha1} as mac algorithm, and \textit{zlib@openssh.com} as compression algorithm, with the wrong padding, resulting in the lowest cosine similarity coefficient $C$.
\autoref{lst:ssh-debug} shows the SSH debug information with the modified version string and key exchange message.

\begin{figure}
\begin{lstlisting}[language=bash, caption={[Example OpenSSH connection with probed SSH packet]OpenSSH connection attempt with probed SSH packet. All non-essential debug information have been removed to lay emphasis on the modified key exchange initialization.}, label={lst:ssh-debug}]
Local version string SSH-2.2-OpenSSH 
SSH2_MSG_KEXINIT sent
SSH2_MSG_KEXINIT received
kex: algorithm: ecdh-sha2-nistp521
kex: host key algorithm: ssh-dss
kex: server->client cipher: blowfish-cbc@openssh.com MAC: <implicit> compression: zlib@openssh.com
kex: client->server cipher: blowfish-cbc@openssh.com MAC: <implicit> compression: zlib@openssh.com
\end{lstlisting}
\end{figure}

\autoref{tab:cosine-similarity} has been derived from \citet{vetterl2020} to present his results of the cosine similarity of OpenSSH, Twisted, and Cowrie.
Twisted has been added to have an example with an older SSH honeypot.
As seen, it differs fundamentally from OpenSSH.
At most, it scores $0.52$ whereas various OpenSSH versions start at $0.98$.
The number of hosts significantly decreases with a cosine similarity score of $0.90$ and higher.
Cowrie responses are not too far away from OpenSSH, with an average of $0.80$.
However, scanning through the web with a minimum score of $0.90$ and higher would exclude all honeypots.
Thus, distinguishing Cowrie from OpenSSH with SSH packets is a feasible method.
Moreover, \citet{vetterl2020} performed an Internet-wide scan, and detected \numprint{758} Kippo and \numprint{2021} Cowrie honeypots.
These results show that the values of honeypots would decrease to zero when fingerprinting activities are used.

\begin{figure}[htbp]
    \centering
    \includegraphics{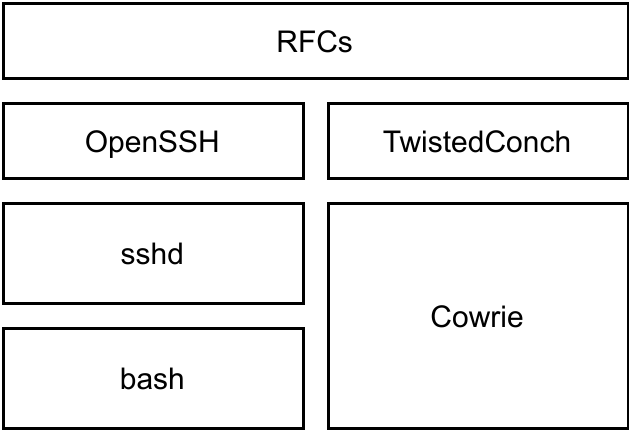}
    \caption[Architecture of OpenSSH and Cowrie]{
        Architecture of OpenSSH and Cowrie.
        OpenSSH and TwistedConch have subtle protocol differences (derived from \cite{vetterl2020}).
    }
    \label{fig:cowrie-openssh}
\end{figure}

\citet{vetterl2020} states that current low- and medium-interaction honeypots have a generic weakness due to the underlying off-the-shelf libraries.
Cowrie is based on TwistedConch\footnote{\href{https://github.com/twisted/twisted/commit/4e3b22afe1f76b360733b65d6b835b7aaae6deb6}{TwistedConch 27.0.1} on GitHub}, a Python 2/3 library that implements the SSH protocol.
Any bash command and its response are tweaked by Cowrie, and thus, resulting in a discrepancy to OpenSSH.
For example, Cowrie version $1.1.0$ missed \verb|tftp|\footnote{Trivial File Transfer Protocol (TFTP) is a lockstep File Transfer Protocol} that later came with version $1.2.0$.
Therefore, it is a continuous struggle to add new commands to avoid early disclosures of Cowrie.

\autoref{fig:cowrie-openssh} shows the difference between OpenSSH and Cowrie.
Both have to fulfill the RFC4250 \cite{rfc4250} which defines the protocol.
OpenSSH and TwistedConch implement the RFC requirement.
As an example, \citet{vetterl2020} found that Cowrie used to have random bytes for the key exchange initialization packet\footnote{Each packet consists of the packet and padding length, the \ac{mac}, a payload, and random padding.}.
With respect to RFC4253 \cite{rfc4253} that defines the \ac{bpp} of SSH, the random padding is used to solidify the total length of the packet to be a multiple of the cipher block size.
The RFC in section 6 defines that the padding consists of 4 random bytes.
Based on the statement of the OpenSSH authors, random bytes have been changed to \verb|NULL| characters due to no security implications.
Thus, an adversary could have detected a Cowrie honeypot with a single key exchange initialization packet.
Nowadays, Cowrie adapted itself to have \verb|NULL| characters as padding to mitigate such an exploit.
However, these subtle differences give adversaries precautionary information and influence the cosine similarity coefficient.

\begin{table}[htbp]
    \caption[Overview of the cosine similarity of OpenSSH, Cowrie, and Twisted]{
        Overview of the cosine similarity of OpenSSH, Cowrie, and Twisted.
        Twisted has been added to have a comparison to an older honeypot.
    }
    \begin{tabular}{lc|cccccccccc}
    \toprule
                   &   & A & B      & C      & D      & E      & F      & G      & H      & I      & J      \\
    \hline
    OpenSSH 6.6    & A & - & $0.98$ & $0.98$ & $0.94$ & $0.94$ & $0.42$ & $0.78$ & $0.79$ & $0.79$ & $0.79$ \\
    OpenSSH 6.7    & B &   & -      & $0.98$ & $0.98$ & $0.98$ & $0.41$ & $0.80$ & $0.81$ & $0.81$ & $0.80$ \\
    OpenSSH 6.8    & C &   &        & -      & $0.96$ & $0.96$ & $0.42$ & $0.78$ & $0.79$ & $0.79$ & $0.79$ \\
    OpenSSH 7.2    & D &   &        &        & -      & $0.98$ & $0.42$ & $0.80$ & $0.80$ & $0.80$ & $0.80$ \\
    OpenSSH 7.5    & E &   &        &        &        & -      & $0.42$ & $0.78$ & $0.79$ & $0.79$ & $0.79$ \\
    \\
    \cline{1-2} \cline{8-12}
    \\
    Twisted 15.2.1 & F &   &        &        &        &        & -      & $0.50$ & $0.51$ & $0.51$ & $0.52$ \\
    \\
    \cline{1-2} \cline{9-12}
    \\
    Cowrie 96ca2ba & G &   &        &        &        &        &        & -      & $0.98$ & $0.98$ & $0.98$ \\
    Cowrie dc45961 & H &   &        &        &        &        &        &        & -      & $0.99$ & $0.99$ \\
    Cowrie dbe88ed & I &   &        &        &        &        &        &        &        & -      & $0.99$ \\
    Cowrie fd801d1 & J &   &        &        &        &        &        &        &        &        & -      \\
    \bottomrule
    \end{tabular}
    \label{tab:cosine-similarity}
\end{table}

\section{Experiment 1: Reproduce Vetterl et al.'s findings}

First, the reproduction of the outdated OpenSSH library that \citet{vetterl2020} used will be investigated.
In his work, he used the version $7.5P1$, which deviates from the latest version $8.8P1$.
Older versions rely on OpenSSL $1.0.2$, including outdated algorithms and functions.
For the \verb|SSH2_MSG_KEXINIT| packet, the encryption algorithm \textit{blowfish-cbc} is outdated and has been removed with version $7.6P1$.
Building the version $7.5P1$ requires the libraries \textit{libssl} ($1.0.2$), \textit{libssl-dev} ($1.0$), \textit{libssh-dev} ($0.7.3-2$), and \textit{libssh-4} ($0.9.6-1$).
All of these libraries are outdated and have been removed from any Debian installation.
Using the latest versions of these libraries results in missing encryption algorithms and host key algorithms.
Thus, replacing the libraries is a necessary task.
It is required to download the libraries, remove the current versions, and install the outdated ones.
The version $7.5P1$ allows modifying the key exchange initialization message proposal in a single file.
On the contrary, this has been removed starting from version $7.6P1$.
After compiling the application, its behavior has been tested with a Debian $11$ Buster and a Debian Jessie $9$ Docker image.
Both are new machines with no other installed packages than the SSH daemon.
Debian $11$ uses the latest OpenSSH version, whereas Jessie is at $6.7P1$.
These environments help to uniquely identify variations in the protocol version.

\begin{figure}[htbp]
\begin{lstlisting}[language=bash, caption={[OpenSSH connection attempt with probed message]OpenSSH connection attempt for version $7.5P1$ and $8.8P1$ with probed key exchange initialization message. All non-essential debug information have been removed to lay emphasis on the modified key exchange initialization.}, label={lst:ssh-openssh}]
OpenSSH_7.5p1, OpenSSL 1.0.2u  20 Dec 2019
Local version string SSH-2.2-OpenSSH
SSH2_MSG_KEXINIT sent
SSH2_MSG_KEXINIT received
kex: algorithm: ecdh-sha2-nistp256
kex: host key algorithm: ssh-dss
Unable to negotiate with ::1 port 22: no matching cipher found. Their offer: aes128-ctr,aes192-ctr,aes256-ctr,aes128-gcm@openssh.com,aes256-gcm@openssh.com,chacha20-poly1305@openssh.com
\end{lstlisting}
\end{figure}

\autoref{lst:ssh-openssh} shows the connection attempt with the adjusted version string and \verb|SSH2_MSG_KEXINIT| packet.
Both Debian machines return the same response.
Using the outdated version $7.5P1$, it results in an incompatibility.
The return message outlines that \textit{blowfish-cbc} is not supported anymore.
OpenSSH kept the encryption algorithm usable for compatibility reasons for clients until $7.6P1$.
Later patches removed the \textit{blowfish-cbc} from the application; thus, a reproduction of \citet{vetterl2020} remains not feasible with the latest version.
Testing it with version $7.3P1$ that has been compiled on the machine results in a successful connection attempt.
\citet{vetterl2020} does not outline any expected response of OpenSSH; thus, it can be assumed that a connection attempt would have been successful due to the existing ciphers during that time.
Adapting the version $8.8P1$ with \textit{chacha20-poly1305} instead of \textit{blowfish-cbc} for the encryption algorithm results in a successful connection attempt.
Therefore, the key exchange initialization has been adapted to use \textit{chacha20-poly1305} as encryption algorithm instead.
Next, the DSA host key algorithms are marked as too weak and are not included automatically during the key exchange initialization.
Using \textit{ssh-dss} requires the extra flag \verb|-oHostKeyAlgorithms=+ssh-dss|.
In order to avoid weak algorithms, the \textit{ssh-ed25519} host key algorithm is used, and the response has been promising to probe instances.
So far, the key exchange initialization packet with \textit{ecdh-sha2-nistp521} as key exchange algorithm, \textit{ssh-ed25519} as host key algorithm, \textit{chacha20-poly1305} as encryption algorithm, \textit{hmac-sha1} as mac algorithm, and \textit{zlib@openssh.com} as compression algorithm have been successfully tested on the two Debian instances.

\begin{figure}[htbp]
\begin{lstlisting}[language=bash, caption={[Cowrie connection attempt with probed message]Cowrie connection attempt with probed key exchange initialization message. All non-essential debug information have been removed to lay emphasis on the modified key exchange initialization.}, label={lst:ssh-cowrie}]
OpenSSH_8.8p1, OpenSSL 1.1.1l  24 Aug 2021
Local version string SSH-2.2-OpenSSH 
SSH2_MSG_KEXINIT sent
Bad packet length 1349676916.
ssh_dispatch_run_fatal: Connection to 10.20.30.123 port 22: message authentication code incorrect
\end{lstlisting}
\end{figure}

The most interesting question remains about Cowrie's response deviation.
\citet{vetterl2020} claims that it results in a \verb|bad version *| exception.
Cowrie has fixed this issue in the meantime, and thus, it does not leak vulnerable information anymore.
For the experiment, the default Cowrie implementation version $v.2.3.0$\footnote{\href{https://github.com/cowrie/cowrie/commit/555ff10d95f6239d9d6efee8a2d05def316ab144}{Cowrie v2.3.0} on GitHub} of the T-Pot instance is used.
\autoref{lst:ssh-cowrie} outlines the connection attempt.
Unambiguously, Cowrie results in a \verb|bad packet length *| exception, and thus, deviates fundamentally from an Open\-SSH response.
The underlying off-the-shelf library TwistedConch checks if a packet is within \numprint{1048576} bytes ($1$ MB) (\autoref{lst:twistedconch}).
Any packet that exceeds that threshold causes this exception, which results in a loss of connection for the client.
This static check is performed when Cowrie tries to get the request packet.
It remains dubious why TwistedConch has added it whenever a packet has to be returned.
In the RFC4253, the minimum packet size is $5$ bytes whereas maximum packet size is set to \numprint{32768} bytes (\numprint{256} KB).
Debugging Cowrie shows that the exception occurs during the version string validation (\autoref{lst:cowrie-version-string}, line 16).
The server validates if the version string matches the allowed versions $1.99$ and $2.0$.
Any higher or lower version will result in a \verb|Protocol major versions differ.\n| exception by calling the function \verb|_unsupportedVersionReceived|.
This response would match the behavior of OpenSSH.

Therefore, the version strings $1.0$, $1.99$, $2.0$, and $2.2$ have been tested on Cowrie and OpenSSH.
As a result, Cowrie's \verb|bad packet length *| exception occurs when the version does not match the expected one.
This result diverges from OpenSSH, as only versions under $1.99$ lead to the same exception as Cowrie.
For any higher version, the connection can be established successfully.
It can be assumed that Cowrie has an error in validating the version string.
Debugging Cowrie shows that the method to return the \verb|Protocol major versions differ.\n| exception is called, but the client does not receive this message.
Hence, the assumption is that the underlying library TwistedConch is responsible for the incorrect message.

% Cosine similarity coefficient, show the value
Calculating the cosine similarity coefficient of both responses shows that the coefficient with $0.46$ is lower than the results from \citet{vetterl2020}.
In his study, the coefficient between Cowrie and OpenSSH was on average $0.80$.
Different implementation approaches to reproduce his results have been considered.
The standard implementation to retrieve the coefficient returned the best result with $0.46$.
Moreover, a soft cosine similarity with English word vectors from \citet{mikolov2018advances} has been used; however, it did not improve the result.
In summary, the same response could not be reproduced.
Nevertheless, it shows that both responses have similarities.

In conclusion, these are the protocol deviations that \citet{vetterl2020} has presented in his technical report.
Thus, this section could successfully recreate his findings by detecting Cowrie on the transport level.
Adversaries who modify their SSH client to send the specific version string and key exchange initialization message could detect Cowrie honeypots and stop further activities.

\begin{figure}[htbp]
\begin{lstlisting}[language=python, caption={[TwistedConch packet length validation]TwistedConch packet length validation. Line 3 validates if the packet length is not greater than $1$ MB. If this check is not successful, the client receives a bad packet length exception.}, label={lst:twistedconch}]
def getPacket(self):
    ...
    if packetLen > 1048576: # 1024 ** 2
        self.sendDisconnect(DISCONNECT_PROTOCOL_ERROR,
                            'bad packet length %s' % packetLen)
    return
    ...
\end{lstlisting}
\end{figure}

\section{Attempt to Disguise Cowrie}

Cowrie has to be tweaked to hide its generic weakness.
Fixing the significant flaws in Cowrie to avoid early detection remains an ephemeral patch.
The continued use of libraries that reimplement the behavior of OpenSSH leads attackers to try to find subtle protocol differences and exclude any host machine that deviates.
Such approaches could be achieved by arbitrary Internet-wide scanning and calculating the cosine similarity coefficient.
Thus, the value of honeypots would decrease to almost zero.
Therefore, a new solution is required to disguise SSH honeypots.
\citet{vetterl2020} presented a solution to use OpenSSH as an intermediary instance between the attacker and Cowrie.
Unfortunately, this solution is outdated, and newer versions contain significant changes in structure and functions.
The concept is based on \citet{vetterl2020} solution, but due to newer versions available, the solution has to be updated to the latest version.
By default, OpenSSH itself cannot act as an intermediary; therefore, it is necessary to customize the latest version to enable this feature.
\autoref{fig:cowrie-fix} visualizes the flow of SSH packets between an attacker and Cowrie.
Cowrie is hidden in the background, and it is only accessible via the loopback address \ipAddress{127.0.0.1} on port 65522.
The updated daemon is exposed to the Internet, and it is accessible via \ipAddress{10.20.30.123} on port 22.
Each connection to OpenSSH is forwarded to the honeypot through a \ac{nat} rule\footnote{iptables -t nat -A PREROUTING -p tcp --dport 22 -j REDIRECT --to-port 65222}.
Accordingly, an attacker should not be able to detect Cowrie through response deviations.

\begin{figure}[htbp]
    \centering
    \includegraphics{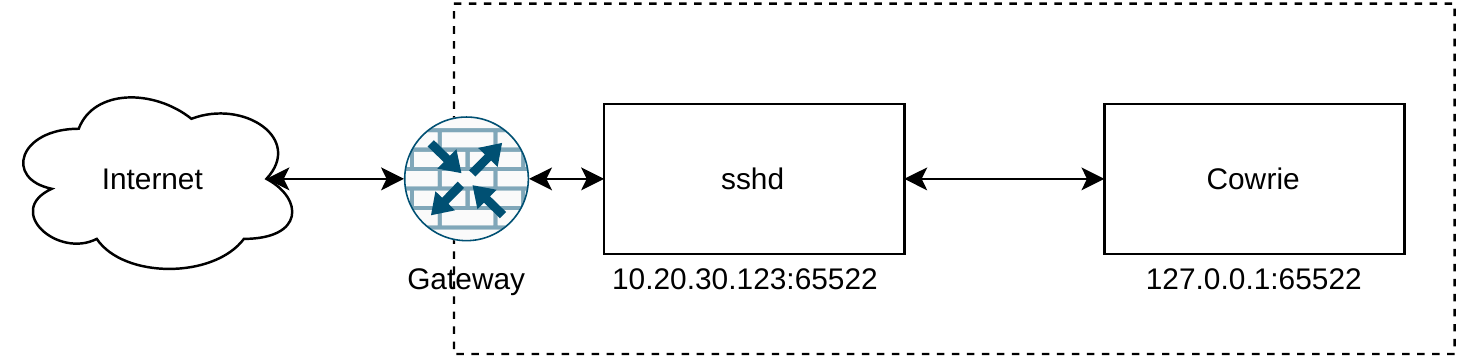}
    \caption[Architecture of OpenSSH and Cowrie]{
        Architecture of OpenSSH and Cowrie (derived from \cite{vetterl2020}).
        A \ac{nat} rule forwards the communication from port 22.
        Only the SSH daemon is accessible from extern.
    }
    \label{fig:cowrie-fix}
\end{figure}

For instance, the latest OpenSSH version $8.8P1$\footnote{\href{https://github.com/openssh/openssh-portable/commit/bf944e3794eff5413f2df1ef37cddf96918c6bde}{OpenSSH 8.8P1} on GitHub} is used.
The implementation is based on \citet{vetterl2020} version $6.3P1$\footnote{\href{https://github.com/amv42/sshd-honeypot/commit/f58830161002baec9d3ed218e78ddb06b0d40a23}{sshd-honeypot} on GitHub}.
As mentioned beforehand, due to major differences between both versions, a smooth transition is unattainable without modifications.
Fortunately, the basic idea to morph OpenSSH into an intermediary instance stays the same.

In total, the connection and user authentication layer has to be modified.
These are the following steps required to change the SSH daemon:

\begin{itemize}
    \item User authentication layer: permit any connection to communicate to Cowrie without an authentication running in front of it.
    \item Connection layer: create a separate channel to communicate with the attacker that forwards the packets to Cowrie.
    \item Connection layer: handle the communication with Cowrie in a new channel separated from others.
\end{itemize}

The first step is to tweak the authentication to permit any session to forward an incoming connection to Cowrie (\autoref{lst:openssh-auth}).
Initially, it checks each session to see if the chosen authentication method returns true.
In order to skip the authentication process, the server must return true for any client that tries to connect to the honeypot.
Therefore, the authentication method has to be overridden in the main method and the allowed user method that checks if the user is permitted to log in.
The authentication process validates if a connection to Cowrie is successful and returns true.
In case of failure, the authentication would fail, resulting in a loss of connection for the client.
Next, the \textit{libssh} library expects a different integer for a successful authentication; therefore, the result is converted to the expected format.
The allowed user method is changed to return true for any user trying to connect to the honeypot.
Cowrie continues the authentication process and communicates with the attacker.

Second, the communication has to be forwarded to the honeypot (\autoref{lst:openssh-channel}).
In OpenSSH, communications are handled in channels as seen beforehand in \autoref{sec:openssh}.
Technically, the daemon opens a SOCKS connection for each session to communicate with the client.
SOCKS is a network protocol to exchange packets between servers and clients.
The SSH daemon needs a separate channel to store the attacker's session and forward packets to communicate with Cowrie.
The channel is implemented in version $6.3P1$ and can be used in $8.8P1$ with minor adaptions.
The method validates if the Cowrie channel is open and writes the new packets into the buffer.
In the main method, when the daemon is started, the channel is created, and a connection to the running Cowrie instance is opened to forward a new session.
If Cowrie is unavailable, the startup will fail; thus, it has to run prior to the SSH daemon.

Lastly, the server loop responsible for connecting the client to the correct port must be modified.
It puts direct TCP/IP connections in the respective channel.
The connection layer handles multiple sessions simultaneously over a single user authentication layer.
Without this adaption, Cowrie would not receive any packet.
The function in \autoref{lst:openssh-serverloop} handles these connections.
For instance, it checks if TCP forwarding is allowed and if the port of Cowrie is defined.
Then, it connects the current session to the respective port.
The SSH daemon has to be adapted to start and set up the channel in the main method at startup.
In addition, the configuration has to be extended to configure the daemon to set the Cowrie IP address and the port.

After compiling the version, a brief test proved a valid connection to the SSH daemon.

\section{Experiment 2: Avoid fingerprinting of Cowrie}

The last experiment to conclude this chapter is to test if the concept helps to disguise Cowrie and avoid fingerprint activities based on a custom local string version and key exchange initialization message.

For instance, \citet{vetterl2020} original $6.3P1$ \textit{sshd-honeypot} and the newly implemented version $8.8P1$ will be used for this experiment.
The forwarding communications are handled by an unmodified Cowrie version $2.3.0$ running in a Docker environment.
The version $6.3P1$ has been tested in heiCLOUD on a Debian 9 Jessie distribution, whereas the version $8.8P1$ with our latest adaption has been tested on a Debian 10 Buster.
In addition, both versions are validated locally in an encapsulated environment.
The clients to test the two concepts are a standard OpenSSH $8.8.P1$ and the modified version with custom local version string and key exchange initialization message to fingerprint honeypots.

The standard client's requests do not result in a bad packet length exception for both servers.
This behavior reflects an original SSH daemon communication and represents a successful test.
The requests from the modified client are successful on the latest version, whereas the older server $6.3P1$ had problems with new encryption and host key algorithms.
A successful connection to the original server from \citet*{vetterl2020} could be recreated by using the $7.3P1$ version.
This version has been used to verify Cowrie beforehand.
The concept can forward any related packet to Cowrie and hide the generic weakness of TwistedConch.
Therefore, whether Cowrie can be disguised to prevent any fingerprint activities with the help of OpenSSH has been answered successfully.
This section can confirm this assumption based on the reproduction and implementation of the concept.
On the other side, Cowrie receives the connection and log information (\autoref{lst:log-cowrie}).
However, one downside is the connection loss due to timeout restrictions.
This issue is a minor bug and could be fixed in the future.

In conclusion, this experiment has shown that the initial idea of hiding Cowrie in the background and directing the communication through OpenSSH prevents fingerprint activities of an adversary.
In addition, it has shown that protocol implementations change rapidly to adapt to new security standards, leading to outdated honeypots.

\section{Discussion}

Depending on the interaction level, honeypots will always deviate from production instances.
As seen in the two experiments beforehand, detecting a generic weakness is doable in a respective time, as well as mitigating it.
Thus, finding and fixing the weaknesses of honeypots becomes a continuous cycle.
However, this chapter also outlined the importance of the libraries that were used.
TwistedConch is the bottleneck of Cowrie, and it is updated\footnote{Based on the lastest GitHub commit of the Python library} frequently.
Libraries that reimplement protocols have to be always up-to-date.
In conclusion, such libraries should be chosen carefully to avoid bugs that leave harmful information to attackers.

\begin{figure}[htbp]
    \centering
    \includegraphics{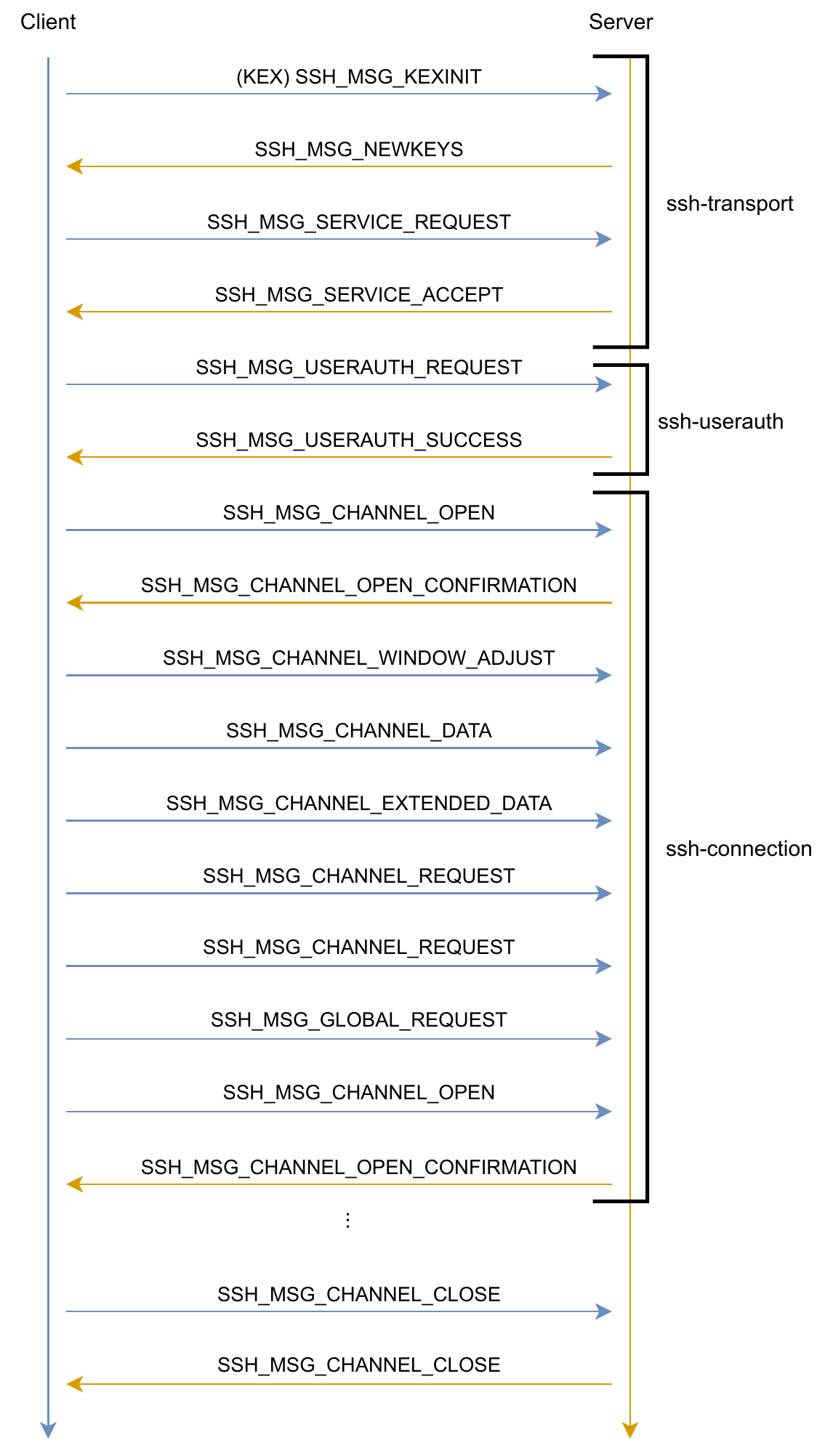}
    \caption[OpenSSH sample session flow diagram]{
        OpenSSH sample session flow diagram (derived from \cite{openssh2007}).
        In addition, the right side indicates the layers that are responsible for handling the messages.
    }
    \label{fig:openssh-flow-diagram}
\end{figure}

% ====================================================================================================================
% Listings
\begin{figure}[htbp]
\begin{lstlisting}[language=python, caption={[Cowrie version string validation]Cowrie version string validation. It tweaks the same results as OpenSSH in line 16.}, label={lst:cowrie-version-string}]
def _unsupportedVersionReceived(self, remoteVersion: bytes) -> None:
    """
    Change message to be like OpenSSH
    """
    self.transport.write(b"Protocol major versions differ.\n")
    self.transport.loseConnection()

def dataReceived(self, data: bytes) -> None
    ...
    if not self.gotVersion:
        ...
        self.otherVersionString = self.buf.split(b"\n")[0].strip()
        ...
        # Checks if the version string has a correct format
        m = re.match(br"SSH-(\d+.\d+)-(.*)", self.otherVersionString)
        if m is None:
            ...
            self.transport.write(b"Invalid SSH identification string.\n")
            self.transport.loseConnection()
            return
        else:
            ...
            # Checks if version string is either 1.99 or 2.0
            if remote_version not in self.supportedVersions:
                self._unsupportedVersionReceived(self.otherVersionString)
                return
            ...
        ...
    ...
\end{lstlisting}
\end{figure}

\begin{figure}[htbp]
\begin{lstlisting}[language=c, caption={[Tweaked OpenSSH authentication]Tweaked OpenSSH authentication to connect to Cowrie. Only the essential code parts to change the authentication method have been added.}, label={lst:openssh-auth}]
int
auth_password(struct ssh *ssh, const char *password)
{
    Authctxt *authctxt = ssh->authctxt;
    /* Send the request to Cowrie */
    int rc;
    rc = authenticate_password(authctxt->user, password);
    authctxt->valid = 1;
    /* libssh returns different values compared to OpenSSH, for SSH_AUTH_SUCCESS=0 returns 1 */
    if (rc == 0)
    {
        finish_connection_setup();
        return 1;
    }
    else
    {
        return 0;
    }
    ...
}
int authenticate_password(const char *username, const char *password)
{
    int rc = -1;
    /* No logins if we could not connect to Cowrie */
    if (ssh_client_conns1[0].error != 1)
    {
        rc = ssh_userauth_password(ssh_client_conns1[0].initial_session, username, password);
    }
    return rc;
}
int
allowed_user(struct ssh *ssh, struct passwd * pw)
{
    return 1;
}
\end{lstlisting}
\end{figure}

\begin{figure}[htbp]
\begin{lstlisting}[language=c, caption={[Tweaked OpenSSH channel]Tweaked OpenSSH channel to connect to Cowrie. Only the essential code parts to change the authentication method have been added.}, label={lst:openssh-channel}]
static int
channel_handle_wfd(struct ssh *ssh, Channel *c,
   fd_set *readset, fd_set *writeset)
{
    ...
    // Implement channel logic to forward data to Cowrie
    int nbytes;
    char buffer[65507] = {0};
    ssh_client_conns1[0].rfd = c->rfd;
    ssh_client_conns1[0].wfd = c->wfd;
    ssh_client_conns1[0].efd = c->efd;
    // Check the connection to Cowrie, if not, close the sshd-client connection
    if (ssh_channel_is_open(channel_rw1.channel_data) &&
           !ssh_channel_is_eof(channel_rw1.channel_data))
    {
        // Read data from the channel (Cowrie)
        nbytes = ssh_channel_read_nonblocking(channel_rw1.channel_data, buffer, sizeof(buffer), 0);
        if (nbytes > 0 && ssh_client_conns1[0].got_command != 1 && ssh_client_conns1[0].subsystem_req != 1)
        {
            write(ssh_client_conns1[0].wfd, buffer, nbytes);
        }
        else if (nbytes > 0 && ssh_client_conns1[0].got_command == 1)
        {
            sshbuf_putf(&c->input, buffer, nbytes);
        }
    } else
    {
        if (ssh_client_conns1[0].counter_disconnect == 0)
        {
            ssh_client_conns1[0].to_disconnect = 1;
        }
    }
    ...
}
\end{lstlisting}
\end{figure}

\begin{figure}[htbp]
\begin{lstlisting}[language=c, caption={[Tweaked OpenSSH server loop]Tweaked OpenSSH server loop to connect to Cowrie. Only the essential code parts to change the authentication method have been added.}, label={lst:openssh-serverloop}]
static Channel *
server_request_direct_tcpip(struct ssh *ssh, int *reason, const char **errmsg)
{
    ...
        ...
        /* Implement direct-TCP/IP forwarding */
        if (sshd_honey_options.tcpForwardingPort != 0)
        {
            /* Redirect to the host specified in sshd_config */
            c = channel_connect_to_port(
                    ssh, 
                    sshd_honey_options.tcpForwardingHost, 
                    sshd_honey_options.tcpForwardingPort,
                    "direct-tcpip", 
                    "direct-tcpip", 
                    reason,
                    errmsg
                );
        }
        else
        {
            /* Redirect to any host */
            c = channel_connect_to_port(ssh, target, target_port, "direct-tcpip", "direct-tcpip", reason, errmsg);
        }
    ...
    /* Make sure cowrie is aware of all requests (successful or not) */
    ssh_channel_open_forward(channel_rw1.channel_data_1,
                             target, target_port,
                             originator, originator_port);

    sprintf(ssh_client_conns1[0].target_ip, "%s", target);
    sprintf(ssh_client_conns1[0].target_port, "%d", target_port);
    ...
}
\end{lstlisting}
\end{figure}

\begin{figure}[htbp]
\begin{lstlisting}[language={}, caption={[Cowrie log information]Cowrie log information. The new connection from this experiment has been acquired. Cowrie fetched information about the local string version and kex message.}, label={lst:log-cowrie}]
New connection: 127.0.0.1:65522 [session: 2ca9a619ceb8]
Remote SSH version: SSH-2.0-libssh_0.9.6
SSH client hassh fingerprint: ....
kex alg=b'curve25519 -sha256' key alg= b'ssh-ed25519'
outgoing: b'aes256 -ctr' b'hmac-sha2-512' b'none'
incoming: b'aes256 -ctr' b'hmac-sha2-512' b'none'
\end{lstlisting}
\end{figure}

  }
  {
  }

  \IfFileExists{chapters/conclusion}{
    \chapter{Conclusion}

This thesis has shown that organizations can spot malicious activities using honeypot solutions.
The result in this thesis successfully answered the original question of whether honeypots contribute to a more secure infrastructure.
It can confirm this assumption based on its results in the cloud and in the university network.
The first approach was to collect data with the help of the T-Pot solution and compare them to a previous study of similar cloud providers.
It has shown that these activities increased significantly.
The universitys' cloud solution heiCLOUD has received more attacks than ever, putting it in the first place compared to other cloud providers.
It has seen various attacks in RDP, VoIP, and SSH.
The number of attacks related to cryptocurrencies is particularly striking, reflecting the current situation of highly traded GPUs.
In addition, the latest attacks like the Apache vulnerability in version $2.49.0$ could be traced back to very early stages, showing how fast attackers adapt to new vulnerabilities.
It is assumed that most of the executed attacks on the instance came from bots.

Next, this thesis has focused on the university's internal network and implemented a new concept to detect every single packet sent to a host machine.
The MADCAT solution, in conjunction with IDS tools, helped identify the open port 113 that has been used to deploy attacks.
It has shown that known attackers with an IP address originating from Russia have probed the instance, and as an assumption, further attacks would have been carried out.
In retrospect, this helped remove the port from the firewall's permits, thus improving the security at the Heidelberg University.
Any other suspicious behavior in the eduroam network could not be registered, proving that the firewall works as intented.

Moreover, honeypots like Cowrie have a fundamental flaw because they rely on off-the-shelf libraries.
These libraries often reimplement protocol behaviors like OpenSSH and add a subtle difference to the response.
On the contrary, this deviation of responses can be used to detect honeypots on the transport level.
Adversaries could spot honeypots before deploying any attack based on a cosine similarity coefficient, thus avoiding exposures to newly developed attacks.
The findings \citet{vetterl2020} claims in his work have been recreated by adapting OpenSSH $8.8P1$ and testing it on different Debian instances.
Due to outdated algorithms, the key exchange initialization message has been updated to work with the latest version.
It shows that the latest Cowrie version $2.3.0$ results in a bad packet length because the local version string does not match the expected ones of the underlying library TwistedConch.
This result deviates fundamentally from OpenSSH.
Lastly, an attempt to protect Cowrie from early exposure has been made by hiding it in the background and tunneling requests through a customized OpenSSH daemon.
This has successfully fixed the generic weakness of Cowrie so that connecting to Cowrie works without running into a bad packet length error.
The last chapter shows that honeypots are not flawless, and developers should be careful when deciding on additional libraries.

In conclusion, this thesis has presented concepts to catch attackers for different scenarios and shows that malicious activities have increased tremendously.
In addition, it has taken a deep dive into an edge-breaking study to detect honeypots on transport level and has disguised Cowrie to block such activities.
An interesting future study could involve the development of a generic method to fingerprint honeypots.
Future research could also examine other libraries that reimplement protocols to find generic weaknesses and deviations.
Ultimately, using honeypots as a security parameter has been proven promising for further implementation.

  }
  {
  }

\cleardoublepage
%\pagenumbering{Roman}
%\setcounter{page}{10}

\bibliographystyle{plainnat}

% Installation and configuration
%\begin{appendices}
%  \include{chapters/appendix.tex}
%\end{appendices}

\end{document}